\documentclass{elsarticle}

\usepackage{epsfig}
\usepackage{a4wide}
\usepackage{amssymb,amsmath}

\newcommand\as{\alpha_s}
\def\ep{\epsilon}
\def\beq{\begin{equation}}
\def\eeq{\end{equation}}
\def\beeq{\begin{eqnarray}}
\def\eeeq{\end{eqnarray}}
\def\cm{{\cal M}}
\def\kper{k_{\perp}}
\def\ktil{\widetilde k}
\newcommand\f[2]{\frac{#1}{#2}}
\newcommand{\la}{\langle}
\newcommand{\ra}{\rangle}
\def\Ph{{\hat P}}
\def\nn{\nonumber}
\def\lra{\leftrightarrow}
\def\hs{\hspace{.1mm}}

\begin{document}

\vspace{\baselineskip}

\title{Double-real radiation in hadronic top quark pair production
\\ as a proof of a certain concept}

    \author[aachen]{M. Czakon}

    \address[aachen]{
      Institut f\"ur Theoretische Teilchenphysik und Kosmologie,
      RWTH Aachen University,\\  D-52056 Aachen, Germany
    }

\cortext[thanks]{Preprint numbers: TTK-10-58, SFB/CPP-10-134}

\begin{abstract}

  \noindent

  Using the recently introduced \textsc{Stripper} approach to
  double-real radiation, we evaluate the total cross sections for the
  main partonic channels of the next-to-next-to leading order
  contributions to top quark pair production in hadronic collisions:
  $gg \rightarrow t\bar tgg$, $gg \rightarrow t\bar tq\bar q$, $q\bar
  q \rightarrow t\bar tgg$, $q\bar q \rightarrow t\bar tq'\bar q'$,
  $q' \neq q$. The results are given as Laurent expansions in
  $\epsilon$, the parameter of dimensional regularization, at a number
  of $m/E_{\mbox{\scriptsize CM}}$ values spreading the entire
  variation range, with $m$ the mass of the top quark and
  $E_{\mbox{\scriptsize CM}}$ the center-of-mass energy. We describe
  the details of our implementation and demonstrate its main properties:
  pointwise convergence and efficiency. We also prove the
  cancellation of leading divergences after inclusion of the
  double-virtual and real-virtual contributions. On a more technical
  note, we extended the double-soft current formulae to the case of
  massive partons.

\end{abstract}

\maketitle


\section{Introduction}

In \cite{Czakon:2010td} we have proposed \textsc{Stripper} (SecToR
Improved Phase  sPacE for real Radiation), a novel subtraction
scheme for the evaluation of double-real radiation contributions in
next-to-next-to leading order (NNLO) QCD calculations. As suggested by
the name, it is the phase space that acquires a special r\^ole in our
approach. Once it is suitably parameterized and decomposed,
Laurent expansions of arbitrary infrared safe observables can be
obtained without any analytic integration, by applying numerical
Monte Carlo methods. An important feature of \textsc{Stripper} is its
process independence. In the actual calculation, general subtraction
terms are combined with process dependent amplitudes. The simplicity of our
construction contrasted with the complexity of double-real radiation
singularity structure may lead to scepticism. The present publication
is meant to prove that \textsc{Stripper} delivers on its promises.

There are many other subtraction schemes for real radiation. At
next-to-leading order (NLO), most calculations are done with the
method of \cite{Catani:1996vz, Catani:2002hc}, but other approaches
are subject to active development \cite{Frixione:1995ms,
  Frederix:2009yq} (see also \cite{Nagy:1996bz}). At the
NNLO level, the situation is more 
complex. Much has been achieved with Sector Decomposition
\cite{Binoth:2000ps, Anastasiou:2003gr, Binoth:2004jv} and
Antenna Subtraction \cite{GehrmannDeRidder:2005cm, Daleo:2009yj,
  Glover:2010im, Boughezal:2010mc}, but there are more
specialised methods for colorless states \cite{Catani:2007vq} and, very recently, for
massive final states \cite{Anastasiou:2010pw}. General tools are also
being developed in \cite{Somogyi:2005xz, Somogyi:2006cz,
  Somogyi:2006db, Somogyi:2006da, Bolzoni:2010bt}. There were, of
course, many other proposals \cite{Weinzierl:2003fx, Kilgore:2004ty,
  Frixione:2004is}, which have not been completely developed.

To demonstrate the virtues of \textsc{Stripper}, we have chosen
hadronic top quark pair production, as it is of great phenomenological
relevance, but its theoretical description is still incomplete. The
current state of the art in the field is as follows
\begin{enumerate}

\item differential cross sections including complete off-shell effects
  and leptonic decays are known at NLO \cite{Denner:2010jp,
    Bevilacqua:2010qb} (until recently, they 
  were only available in the narrow-width approximation
  \cite{Bernreuther:2004jv, Melnikov:2009dn}); 

\item fixed order threshold expansions for the quark-annihilation and
  gluon-fusion channels are known at NNLO up to constants, both for
  the total cross section \cite{Beneke:2009ye}, and for the invariant mass
  distribution \cite{Ahrens:2009uz};

\item soft-gluon resummation for the previously mentioned channels is
  understood at the NNLL level for the total cross section
  \cite{Beneke:2009rj, Czakon:2009zw} (see also
  \cite{Moch:2008qy}), and selected differential distributions
  \cite{Ahrens:2010zv} (see also \cite{Kidonakis:2010dk});

\item mixed soft-gluon and Coulomb resummation is understood at the
  NNLL level as well \cite{Beneke:2010da};

\item two-loop virtual amplitudes are known analytically in the
  high-energy limit \cite{Czakon:2007ej, Czakon:2007wk}, for the
  planar contributions \cite{Bonciani:2009nb, Bonciani:2010mn},
  and fermionic contributions in the quark-annihilation channel
  \cite{Bonciani:2008az}; in the same channel, the full amplitude is known
  numerically \cite{Czakon:2008zk};

\item one-loop squared amplitudes are known analytically at the NNLO
  level \cite{Korner:2008bn, Kniehl:2008fd};

\item one-loop real-virtual (one additional massless parton in the
  final state) amplitudes have been obtained in the course of several
  projects connected to top quark pair production in association with
  jets \cite{Dittmaier:2007wz, Bevilacqua:2010ve, Melnikov:2010iu};

\item approximations of the one-loop real-virtual amplitudes are known
  in all singular limits involving only massless partons
  \cite{Bern:1994zx, Bern:1998sc, Kosower:1999rx, Bern:1999ry,
    Catani:2000pi} (this is needed for the evaluation of the
  real-virtual contributions).

\end{enumerate}

We are interested in an NNLO calculation. Schematically, the involved
partonic cross sections are a sum of three terms mentioned already
above
\begin{equation}
d\sigma_{t\bar t+X}^\text{NNLO} = d\sigma^{VV} + d\sigma^{RV} +
d\sigma^{RR} \; ,
\end{equation}
where $d\sigma^{VV}$ denotes the double-virtual
(two-loop and one-loop squared), $d\sigma^{RV}$ the real-virtual
(one-loop with one additional parton), and $d\sigma^{RR}$ the
double-real (tree-level with two additional partons) corrections.
We will ignore the need for collinear renormalization, which involves
lower order cross sections expanded to higher orders in $\epsilon$,
the parameter of dimensional regularization. These have
been derived (although not published) for the analysis of
\cite{Czakon:2008ii}.

Currently missing are the double-real and real-virtual
contributions. Once these are known, one can provide complete NNLO cross
sections beyond the known threshold expansions. The real-virtual contribution
has a simpler singularity structure than the double-real, and should
be obtainable once the soft-gluon current in the presence of heavy
quarks has been derived.

As far as the double-real contribution is concerned, there are many
partonic channels, which need to be considered in
principle. Nevertheless, current phenomenological applications require
the knowledge of the cross sections with gluon-gluon and
quark-anti-quark initial states. This leads to our choice of the
following channels
\begin{equation}
gg \to t\bar tgg , \; 
gg \to t\bar tq\bar q , \; 
q\bar q \to t\bar tgg , \; 
q\bar q \to t\bar tq'\bar q' \; .
\end{equation}
We will only consider the case $q' \neq q$, because the case of
identical quarks is expected to be numerically irrelevant. In the
present publication, we will provide numerical values for the cross
sections for the four channels as function of $m/E_\text{CM}$.

The paper is organized as follows. In the next Section, we will
discuss the phase space, its volume, parameterization and
decomposition. Subsequently, we will describe the derivation of the
subtraction terms, convergence and cancellation of leading
divergences. In Section 4, we will describe the technical details of
our implementation, demonstrate its efficiency and describe several
tests. Section 5 contains the results of numerical simulations for the
four chosen channels. Apart from the main text and Conclusions, the
publication consists of a number of Appendices. They contain the
collinear splitting functions, a discussion of the double-soft limit
in the presence of massive partons, another approach to the collinear
limit in the double-soft limit, the Born cross sections, and finally a
list of the software that we have used.


\section{Phase space}


\subsection{Volume}
\label{sec:volume}

\vspace{.2cm}

A numerical approach, as the one advocated here, poses
substantial problems, when assessing the correctness of both the
approach and implementation. We will, therefore, start by introducing
the only truly non-trivial integral that relates to our computation,
but can be evaluated entirely analytically: the volume of the phase
space.

We are interested in the following class of processes
\begin{equation}
a(p_1)+b(p_2) \rightarrow t(q_1)+\bar t(q_2)+c(k_1)+d(k_2) \; ,
\end{equation}
where $a,b$ are initial and $c,d$ final state partons, in particular
$ab = gg$ or $q\bar q$, and $cd = gg$ or $q'\bar q'$. Additionally
\begin{equation}
\;\;\;\; p_1^2=p_2^2=k_1^2=k_2^2 = 0 \; , \;\;\;\; q_1^2 = q_2^2 = m^2
\neq 0 \; , 
\end{equation}
and as usual
\begin{equation}
s = (p_1+p_2)^2 \; .
\end{equation}
Throughout this publication we will work in the partonic
center-of-mass system.

The phase space measure in $d$-dimensions is
\begin{equation}
\label{eq:ps1}
d \Phi_{4} = \frac{d^{d-1} k_1}{(2\pi)^{d-1} 2k_1^0} \frac{d^{d-1}
  k_2}{(2\pi)^{d-1} 2k_2^0} \frac{d^{d-1} q_1}{(2\pi)^{d-1} 2q_1^0}
\frac{d^{d-1} q_2}{(2\pi)^{d-1} 2q_2^0}   (2\pi)^d
\delta^{(d)}(k_1+k_2+q_1+q_2-p_1-p_2) \; ,
\end{equation}
with $d=4-2\epsilon$ as usual. We wish to evaluate the integral of
unity with this measure. The result will be given as a product of two
functions
\begin{equation}
\int d \Phi_{4} = P_4(s,\epsilon) \Phi(x,\epsilon) \; ,
\end{equation}
with
\begin{equation}
x = \frac{1-\beta}{1+\beta} \; , \;\;\;\; \beta =
\sqrt{1-\frac{4m^2}{s}} \; ,
\end{equation}
and $P_4$ the volume of the phase space in the purely massless case
\begin{equation}
P_4(s,\epsilon) = 2^{-11+6\epsilon}\pi^{-5+3\epsilon}
\frac{\Gamma^4(1-\epsilon)} {\Gamma(3-3\epsilon)\Gamma(4-4\epsilon)}
s^{2-3\epsilon} \; ,
\end{equation}
which can be easily obtained from the imaginary part of the three-loop
massless sunrise diagram \cite{GehrmannDe Ridder:2003bm}. By definition, the $\Phi$ function
must satisfy two boundary conditions
\begin{equation}
\Phi(1,\epsilon) = 0 \; , \;\;\;\; \Phi(0,\epsilon) = 1 \; .
\end{equation}
The first of the above equations is just the vanishing of the phase
space at threshold, which is located at $x = 1$. The second follows
from the normalization in the massless case, which in turn corresponds
to $x = 0$.

\begin{figure}[t]
\begin{center}
\includegraphics[width=.40\textwidth]{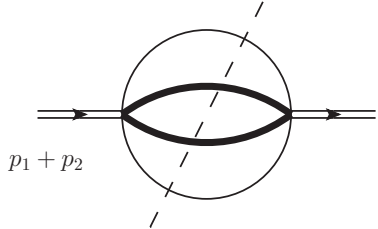}
\end{center}
\caption{Three-loop sunrise graph used to obtain the phase space
  volume. Thick lines are massive, whereas the dashed line denotes a
  cut, which in this case corresponds to the imaginary part of the
  integral.}
\label{fig:sunrise}
\end{figure}

In order to obtain $\Phi$, we will use the method of differential
equations \cite{Kotikov:1990kg, Remiddi:1997ny}. To this end, we
exploit again the fact that $\Phi$ 
is given, up to normalization, by the imaginary part of the three-loop
sunrise diagram, this time with two massless and two massive lines,
see Fig.~\ref{fig:sunrise}. Reducing the mass derivatives of the three
occurring master integrals using integration-by-parts identities, we
obtain the following two equations
\begin{eqnarray}
&&\frac{\partial}{\partial x} \left( \frac{(1+x)^4}{x^2}
  \Phi(x,\epsilon) \right) = \frac{(1+x)^3}{x^3} \left( 2(x-1) \;
  \epsilon \; \Phi(x,\epsilon) + (1+x)^3 \frac{\partial
    \Psi(x,\epsilon)}{\partial x} \right) \; , \\ \nonumber \\ &&
  \frac{\partial^2 \Psi(x,\epsilon)}{\partial x^2}
  -\frac{1-20x+3x^2}{x(1-x^2)}\frac{\partial
    \Psi(x,\epsilon)}{\partial x} -\frac{24}{x(1+x)^2}\Psi(x,\epsilon)
  = \nonumber \\ && \;\;\;\;\;\;\;\; -\epsilon \left(
  \frac{1-22x+x^2}{x(1-x^2)}\frac{\partial \Psi(x,\epsilon)}{\partial
    x}+\frac{24(2-\epsilon)}{x(1+x)^2}\Psi(x,\epsilon)
  +\frac{2(1-2\epsilon)}{(1+x)^4}\Phi(x,\epsilon) \right) \; ,
\end{eqnarray}
where we have only kept two of the master integrals, $\Phi$ and
$\Psi$. Notice that $\Psi$'s sole purpose is to provide a solvable
second order differential equation and a neat boundary condition,
which follows again from the vanishing of the phase space at threshold
\begin{equation}
\Psi(1,\epsilon) = 0 \; .
\end{equation}
For this reason, we do not even bother specifying its exact
definition.

With the three boundary conditions, the solution of the system of
differential equations is unique, and can be obtained recursively as a
series expansion in $\epsilon$. In principle, we need five terms of
the expansion corresponding to the five relevant terms of the
expansion of the cross section, ranging from $1/\epsilon^4$ to
$\epsilon^0$. As the expressions quickly become extremely lengthy, we
will only reproduce the first two, and give a high precision numerical
value at some benchmark point for the complete expansion\footnote{The
  analytic result for $\Phi(x)$ is attached to the electronic preprint
  version of this publication}. $\Phi$ is
given by
\begin{equation}
\Phi(x,\epsilon) = \sum_{i=0}^\infty \; \epsilon^i \; \Phi^{(i)}(x) \;
,
\end{equation}
with
\begin{eqnarray}
\Phi^{(0)}(x) &=& -\frac{48 x^2 H(-1,0,x)}{(x+1)^4}+\frac{24 x^2
  H(0,0,x)}{(x+1)^4}+\frac{12 \left(x^4+5 x^3+6 x^2+5 x+1\right) x
  H(0,x)}{(x+1)^6} \nonumber \\ && -\frac{x^5+23 x^4+\left(34+4 \pi
  ^2\right) x^3+\left(4 \pi ^2-34\right) x^2-23 x-1}{(x+1)^5} \; ,
\\ \nonumber \\ \Phi^{(1)}(x) &=& -\frac{672 x^2
  H(-1,-1,0,x)}{(x+1)^4}-\frac{96 x^2 H(-1,0,-1,x)}{(x+1)^4}+\frac{144
  x^2 H(-1,0,0,x)}{(x+1)^4} \nonumber \\ &&-\frac{480 x^2
  H(-1,0,1,x)}{(x+1)^4}+\frac{336 x^2 H(0,-1,0,x)}{(x+1)^4}+\frac{48
  x^2 H(0,0,-1,x)}{(x+1)^4} \nonumber \\ &&-\frac{72 x^2
  H(0,0,0,x)}{(x+1)^4}+\frac{240 x^2 H(0,0,1,x)}{(x+1)^4}-\frac{4
  \left(3 x^2+56 x+3\right) x H(0,0,x)}{(x+1)^4} \nonumber
\\ &&+\frac{8 \left(15 x^4+122 x^3+190 x^2+122 x+15\right) x
  H(-1,0,x)}{(x+1)^6} \nonumber \\ &&+\frac{24 \left(x^4+5 x^3+6 x^2+5
  x+1\right) x H(0,-1,x)}{(x+1)^6} \nonumber \\ &&+\frac{120
  \left(x^4+5 x^3+6 x^2+5 x+1\right) x H(0,1,x)}{(x+1)^6} \nonumber
\\ &&-\frac{4 \left(2 x^5+61 x^4+4 \left(50+\pi ^2\right) x^3+2
  \left(89+4 \pi ^2\right) x^2+\left(86+4 \pi ^2\right) x+13\right) x
  H(0,x)}{(x+1)^6} \nonumber \\ &&+\frac{\left(-2 x^5-46 x^4+4 \left(8
  \pi ^2-17\right) x^3+4 \left(17+8 \pi ^2\right) x^2+46 x+2\right)
  H(-1,x)}{(x+1)^5} \nonumber \\ &&-\frac{10 \left(x^5+23 x^4+34
  x^3-34 x^2-23 x-1\right) H(1,x)}{(x+1)^5} \nonumber \\ &&-\frac{2 x
  \left(\pi ^2 \left(18 x^4+43 x^3+8 x^2+43 x+18\right)+180 x (x+1)^2
  \zeta_3 \right)}{3 (x+1)^6} \; .
\end{eqnarray}
The $H$ functions are standard harmonic polylogarithms (HPL)
\cite{Remiddi:1999ew}. Their initial weight is 2, but the last term of the expansion
we are interested in, {\it i.e.} $\epsilon^4$, contains HPLs up to
weight six.

We can also obtain the behavior of $\Phi$ near threshold, either from
the differential equations, or from the actual solution. The result is
\begin{equation}
\label{eq:betaphase}
\Phi\left( \frac{1-\beta}{1+\beta}, \epsilon \right) = \frac{64}{315}
\beta^{9-10\epsilon} \Big( 1+{\cal O}(\epsilon) \Big) + {\cal O}\big(
\beta^{10} \big) \; .
\end{equation}

The numerical benchmark expansion is chosen at $x=1/2$
\begin{eqnarray}
\Phi\left(1/2,\epsilon\right)  &=& 0.00001122829901964763 \; \nonumber
\\ &+& 0.0001283543727784153 \; \epsilon \nonumber \\ &+&
0.0007325963156455679 \; \epsilon^2 \nonumber \\ &+&
0.002782712436211506 \; \epsilon^3 \nonumber \\ &+&
0.007910064621069109 \; \epsilon^4 \nonumber \\ &+& {\cal
  O}\big(\epsilon^5\big) \; .
\end{eqnarray}
One may wonder why it is not sufficient to test the implementation
close to the massless case and avoid the work that led to the above
expressions. The reason is that the presence of large logarithms of
the mass, up to $\log^4 (m^2/s)$, implies a high sensitivity of the
result to the value of the small mass. A point like $x=1/2$ has no
special properties and, therefore, no large cancellations are
expected.

Using the result for $\Phi(x)$ and its two derivatives in $x$, we have
a complete set of master integrals and can evaluate the integral of
any polynomial in scalar products of $p_1+p_2$, $k_{1,2}$ and
$q_{1,2}$. We will find it later useful to have the result for the
following integral
\begin{eqnarray}
&& \!\!\!\!\!\!\!\! \int d\Phi_4 \; \left( \frac{k_1 \cdot q_1}{s}
  \right)^2 = \frac{P_4(s,\epsilon)}{48 (\epsilon-1)^2 \left(12
    \epsilon^2-31 \epsilon+20\right) (1-x) (1+x)^4} \nonumber \\ &&
  \times \Big( (x+1) \left(x^2-1\right) x^2 \left(\epsilon^2 \left(6
  x^2-8 x+6\right)+\epsilon \left(-19 x^2+4 x-19\right)+2 \left(7
  x^2+3 x+7\right)\right) \; \Phi ''(x) \nonumber \\ && -(x+1) x
  \big(2 \epsilon^3 \left(5 x^4-66 x^3+82 x^2-66 x+5\right)+\epsilon^2
  \left(-47 x^4+526 x^3-210 x^2+510 x-35\right) \nonumber \\ &&
  +\epsilon \left(72 x^4-636 x^3-66 x^2-628 x+34\right)+4 \left(-9
  x^4+59 x^3+26 x^2+62 x-2\right)\big) \; \Phi '(x) \nonumber \\ && -2
  (x-1) \big(2 \epsilon^4 \left(x^4+60 x^3+26 x^2+60
  x+1\right)-\epsilon^3 \left(11 x^4+562 x^3+470 x^2+562 x+11\right)
  \nonumber \\ && +\epsilon^2 \left(22 x^4+933 x^3+1028 x^2+933
  x+22\right)-\epsilon \left(19 x^4+660 x^3+852 x^2+660 x+19\right)
  \nonumber \\ && +2 \left(3 x^4+85 x^3+122 x^2+85 x+3\right)\big) \;
  \Phi (x) \Big) \; .
\end{eqnarray}
The value at our benchmark point is
\begin{eqnarray}
&& \!\!\!\!\!\!\!\! \frac{1}{P_4(s,\epsilon)} \left. \int d\Phi_4 \;
  \left( \frac{k_1 \cdot q_1}{s} \right)^2 \; \right|_{x=\frac{1}{2}} =
\\ \nonumber && \Big( 
1.4553533
+ 16.673868 \; \epsilon
+ 95.377076 \; \epsilon^2
+ 363.03219 \; \epsilon^3
+ 1033.8027 \; \epsilon^4
+{\cal O}\big(\epsilon^5\big)
\Big) \times 10^{-9} \; .
\end{eqnarray}
%


\subsection{Parameterization of the massless system}

\vspace{.2cm}

We will now introduce a suitable parameterization of the phase
space. We will closely follow the lines of \cite{Czakon:2010td}, where the massless
system has been specified. In the next subsection, we will define a
parameterization for the heavy system.

Before we give the momentum vectors, let us note that we can always
choose them such that their $\epsilon$-dimensional components
vanish. This is due to the rotational invariance remaining in the
system as long as we only have three vectors $\vec p_1, \vec k_1, \vec
k_2$ (notice that $\vec p_2 = -\vec p_1$ by assumption). Therefore, we
will specify the vectors, as if they were purely
four-dimensional. The only consequence of the existence of the
additional degrees of freedom is the modified form of the phase space
measure, which is then sufficient to regulate all singularities. We
will also exploit rotational invariance and space inversion invariance
of the matrix elements (which can also be viewed as $d$-dimensional
rotation invariance) to restrict the momenta as follows
\begin{equation}
k_1^x = 0 \; , \;\;\;\; k_2^x > 0 \; .
\end{equation}
In the actual Monte Carlo simulation, the momenta should be rotated
randomly around the $z$-axis and the sign of $x$-axis should also be
chosen at random, in order to fill out the complete phase space.

With the above assumptions, let
\begin{eqnarray}
p_1^\mu &=& \frac{\sqrt{s}}{2} (1,0,0,1) \; ,  \nonumber \\ p_2^\mu
&=& \frac{\sqrt{s}}{2} (1,0,0,-1) \; ,  \nonumber \\ n_1^\mu &=&
\frac{\sqrt{s}}{2} \beta^2 (1,0,\sin\theta_1,\cos\theta_1) \; ,
\nonumber \\ n_2^\mu &=& \frac{\sqrt{s}}{2} \beta^2
(1,\sin\phi\sin\theta_2,\cos\phi\sin\theta_2,\cos\theta_2) \; ,
\nonumber \\ k_1^\mu &=& \hat{\xi}_1 \; n_1^\mu \; , \nonumber
\\ k_2^\mu &=& \hat{\xi}_2 \; n_2^\mu \; ,
\end{eqnarray}
where $\phi,\theta_{1,2} \in [0,\pi]$, and $n^\mu_{1,2}$ are
auxiliary vectors needed to define soft limits, whereas
$\hat{\xi}_{1,2}$ are used to parameterize the energies. Notice the
hats above the variables. We shall use them to denote all variables,
which are going to be transformed due to further phase space
decomposition. The angular variables are replaced by another set in
two steps.

We first define
\begin{eqnarray}
\label{eq:eta12}
\hat{\eta}_{1,2} &=& \frac{1}{2} (1-\cos\theta_{1,2}) \; , \\ \eta_3
&=& \frac{1}{2} (1-\cos\theta_3) \nonumber \\ &=& \frac{1}{2}
(1-\cos\phi\sin\theta_1\sin\theta_2-\cos\theta_1\cos\theta_2)
\nonumber \\ &=& \frac{1}{2}
(1-\cos(\theta_1-\theta_2)+(1-\cos\phi)\sin\theta_1\sin\theta_2) \;
, \label{eq:eta3first}
\end{eqnarray}
where $\theta_3$ is the relative angle between $\vec k_1$ and $\vec
k_2$, and by definition $\hat{\eta}_{1,2},\eta_3 \in [0,1]$. One of the
main ideas of the subtraction scheme \cite{Czakon:2010td} is to change variables in
such a way that all collinear limits be parameterized with just two
variables, $\hat{\eta}_{1}$ and $\hat{\eta}_2$. In order to do so, we
introduce
\begin{equation}
\zeta = \frac{1}{2} \frac{(1 - \cos(\theta_1 - \theta_2))(1 +
  \cos\phi)}{1 - \cos(\theta_1 - \theta_2) + (1 -
  \cos\phi)\sin\theta_1\sin\theta_2} \in [0,1] \; ,
\end{equation}
which can be inverted to give
\begin{equation}
\eta_3 =
\frac{(\hat{\eta}_1-\hat{\eta}_2)^2}{\hat{\eta}_1+\hat{\eta}_2-2\hat{\eta}_1
  \hat{\eta}_2-2(1-2\zeta)
  \sqrt{\hat{\eta}_1(1-\hat{\eta}_1)\hat{\eta}_2(1-\hat{\eta}_2)}} \;
. \label{eq:eta3second}
\end{equation}
Clearly, the collinear limits are now at $\hat{\eta}_1 = 0$,
$\hat{\eta}_2 = 0$, $\hat{\eta}_1 = 1$, $\hat{\eta}_2 = 1$ or
$\hat{\eta}_1 = \hat{\eta}_2$. While $\theta_{1,2}$ are obtained from
Eq.~(\ref{eq:eta12}), $\phi$ is given by solving
Eq.~(\ref{eq:eta3first}) and Eq.~(\ref{eq:eta3second}) 
\begin{equation}
\cos\phi =  \frac{1-(1-2 \hat{\eta}_1) (1-2 \hat{\eta}_2)-\frac{2
    (\hat{\eta}_1-\hat{\eta}_2)^2}{\hat{\eta}_1+\hat{\eta}_2-2
    \hat{\eta}_1 \hat{\eta}_2-2 (1-2 \zeta ) \sqrt{(1-\hat{\eta}_1)
      \hat{\eta}_1 (1-\hat{\eta}_2) \hat{\eta}_2}}}{4
  \sqrt{(1-\hat{\eta}_1) \hat{\eta}_1 (1-\hat{\eta}_2) \hat{\eta}_2}}
\; .
\end{equation}
Notice that
\begin{equation}
\label{implication:1}
\hat{\eta}_1 = \hat{\eta}_2 \; \Rightarrow \; \cos\phi = 1 \; ,
\end{equation}
whereas
\begin{equation}
\hat{\eta}_1 = 0 \vee \hat{\eta}_2 = 0 \vee \hat{\eta}_1 = 1 \vee
\hat{\eta}_2 = 1 \; \Rightarrow \; \cos\phi = 2\zeta-1 \; .
\end{equation}
The last statement is valid, when $\hat{\eta}_1 \neq \hat{\eta}_2$,
but seems to contradict
implication~(\ref{implication:1}). Fortunately, in the limiting cases
$\hat{\eta}_1=\hat{\eta}_2=0$ and $\hat{\eta}_1=\hat{\eta}_2=1$ the
momentum vectors do not depend on $\phi$.

The final set of parameters specifying the kinematics of the massless
partons is
\begin{equation}
\zeta, \; \hat{\eta}_1, \; \hat{\eta}_2, \; \hat{\xi}_1, \;
\hat{\xi}_2 \; .
\end{equation}
The first three are unrestricted within the range $[0,1]$, whereas the
energy variables belong to one of the two non-overlapping regions
(apart from a measure zero set) \cite{Czakon:2010td}
\begin{equation}
\label{eq:range}
\big\{(\hat{\xi}_1,\hat{\xi}_2): \;  0 \leq \hat{\xi}_1 \leq 1 \; , \;
0 \leq \hat{\xi}_2 \leq \hat{\xi}_1 \; \xi_{max}(\hat{\xi}_1) \big\}
\; ,
\end{equation}
\begin{equation}
\big\{(\hat{\xi}_1,\hat{\xi}_2): \;  0 \leq \hat{\xi}_2 \leq 1 \; , \;
0 \leq \hat{\xi}_1 \leq \hat{\xi}_2 \; \xi_{max}(\hat{\xi}_2) \big\}
\; ,
\end{equation}
where
\begin{equation}
\xi_{max}(\xi) = \min\left( 1, \;
\frac{1}{\xi}\frac{1-\xi}{1-\beta^2\eta_3\xi}\right) \leq 1 \; .
\end{equation}
Not only do these conditions guarantee that the massive states can
always be produced, but they are also suggestive of a decomposition of
the phase space, which we will perform later on.

Having specified the parameterization of the phase space, we can
rewrite the measure Eq.~(\ref{eq:ps1}) in the new variables. We split
it into two parts
\begin{equation}
d\Phi_4 = d\Phi_3(p_1+p_2;k_1,k_2) \; d\Phi_2(Q;q_1,q_2) \; ,
\end{equation}
with
\begin{equation}
Q = p_1+p_2-k_1-k_2 \; .
\end{equation}
$d\Phi_3(p_1+p_2;k_1,k_2)$ is not exactly the three-particle phase
space of $k_1,k_2$ and $Q$, because the only constraint that it is
subjected to is $Q^2 \geq 4m^2$. On the other hand,
$d\Phi_2(Q;q_1,q_2)$ is the two-particle phase space. We have
\begin{eqnarray}
\label{eq:ps3}
d\Phi_3(p_1+p_2;k_1,k_2) &=&
\frac{\pi^{2\epsilon}}{8(2\pi)^5\Gamma(1-2\epsilon)} s^{2-2\epsilon}
\beta^{8-8\epsilon}  \; (\zeta(1-\zeta))^{-\frac{1}{2}-\epsilon}
\nonumber \\ &\times& \; (\hat{\eta}_1(1-\hat{\eta}_1))^{-\epsilon}
(\hat{\eta}_2(1-\hat{\eta}_2))^{-\epsilon}
\frac{\eta_3^{1-2\epsilon}}{|\hat{\eta}_1-\hat{\eta}_2|^{1-2\epsilon}}
\; \hat{\xi}_1^{1-2\epsilon} \hat{\xi}_2^{1-2\epsilon} \nonumber
\\ &\times& d\zeta \; d\hat{\eta}_1 \; d\hat{\eta}_2 \; d\hat{\xi}_1
\; d\hat{\xi}_2 \; .
\end{eqnarray}
The first line above will be of no further concern, since we are only
going to perform variable changes on the subset
$\hat{\eta}_1,\hat{\eta}_2,\hat{\xi}_1,\hat{\xi}_2$. Therefore, we will
define
\begin{eqnarray}
d\mu_\zeta &=&
\frac{\pi^{2\epsilon}}{8(2\pi)^5\Gamma(1-2\epsilon)} s^{2-2\epsilon}
\beta^{8-8\epsilon}  \; (\zeta(1-\zeta))^{-\frac{1}{2}-\epsilon} \;
d\zeta \; = \; \mu_\zeta \; d\zeta \; ,
\\ \label{eq:mu1}
d\mu_{\eta\xi} &=&  (\hat{\eta}_1(1-\hat{\eta}_1))^{-\epsilon}
(\hat{\eta}_2(1-\hat{\eta}_2))^{-\epsilon}
\frac{\eta_3^{1-2\epsilon}}{|\hat{\eta}_1-\hat{\eta}_2|^{1-2\epsilon}}
\; \hat{\xi}_1^{1-2\epsilon} \hat{\xi}_2^{1-2\epsilon} \; d\hat{\eta}_1
\; d\hat{\eta}_2 \; d\hat{\xi}_1 \; d\hat{\xi}_2 \; ,
\end{eqnarray}
with $d\Phi_3 = d\mu_\zeta \; d\mu_{\eta\xi}$. Despite the splitting, $d\mu_{\eta\xi}$
depends on $\zeta$ through $\eta_3$.


\subsection{Parameterization of the massive system}

\vspace{.2cm}

In order to parameterize the massive system, we perform a boost to the
center-of-mass frame of $Q$. Denoting the momenta of the heavy quarks
in this frame by $q'_{1,2}$, we have
\begin{eqnarray}
q^0_i &=& \frac{Q^0 q'^0_i+\overrightarrow{Q} \cdot
  \overrightarrow{q'_i}}{\sqrt{Q^2}} \; , \nonumber
\\ \overrightarrow{q_i} &=& \overrightarrow{q'_i} + \left( q'^0_i +
\frac{\overrightarrow{Q} \cdot \overrightarrow{q'_i}}{Q^0+\sqrt{Q^2}}
\right) \frac{\overrightarrow{Q}}{\sqrt{Q^2}} \; , \;\;\;\; i=1,2 \; .
\end{eqnarray}
The problem that we face is that once three $(d-1)$-dimensional momenta
of the massless partons have been specified with
$\epsilon$-dimensional components vanishing, we do not have the
freedom to keep the latter components of $q'_{1,2}$ vanishing
anymore. The easiest solution would be to restrict the momenta of the
heavy quarks, which are always resolved after all, to lie in the four
physical dimensions. This would simplify the parameterization, but we
would loose the possibility to use the integrals derived in
Section~\ref{sec:volume}. Furthermore, to obtain finite partonic cross
sections, it is necessary to add collinear counterterms, which are
convolutions of splitting functions with lower order cross
sections. If we would like to use the results obtained in
\cite{Czakon:2008ii} for this purpose, we need the heavy quarks in
$d$-dimensions.

Let us, therefore, define the $q'_{1,2}$ momenta through three
spherical angles $\theta_Q, \phi_Q$ and $\rho_Q$.
\begin{eqnarray}
\begin{array}{rcrcl}
{q'_1}^0 &=& {q'_2}^0 &=& \frac{1}{2} \sqrt{Q^2} \; ,  \\ &&&&
\\ \overrightarrow{q'_1} &=& -\overrightarrow{q'_2} &=& \frac{1}{2}
\sqrt{Q^2+\beta^2-1} \\ &&&& \\ &&&\times& \left(\sin\rho_Q
\sin\phi_Q \sin\theta_Q \; \vec n^{(d-4)}, \cos\rho_Q \sin\phi_Q
\sin\theta_Q,\cos\phi_Q \sin\theta_Q,\cos\theta_Q \right) \; . 
\end{array}
\end{eqnarray}
In principle, the three angles should lie in the range
$[0,\pi]$. Nevertheless, we can assume $\phi_Q \in [0,2\pi]$ and
$\rho_Q \in [0,\pi/2]$, as long as we exploit the independence of the
results from the sign of $\vec n^{(d-4)}$. In fact, without loss of
generality, we can set
\begin{equation}
\vec n^{(d-4)} = (\vec 0^{(d-5)}, 1) \; ,
\end{equation}
and forget about the $(d-5)$-dimensional components. Thus we have to
work with five-dimensional vectors. We will soon see that the
contribution of those vectors, which have a non-vanishing fifth
dimension is suppressed by a power of $\epsilon$ as one would expect.

The two-particle phase space is now
\begin{eqnarray}
d\Phi_2(Q;q_1,q_2) &=&
\frac{(4\pi)^{\epsilon}\Gamma(1-\epsilon)}{8(2\pi)^2\Gamma(1-2\epsilon)}
\left( Q^2 \right)^{-\epsilon}\left( \sqrt{1-\frac{4m^2}{Q^2}}
\right)^{1-2\epsilon} (1-\cos^2\theta_Q)^{-\epsilon} \left(
\sin^2\phi_Q \right)^{-\epsilon} \nonumber \\ &\times& 
\frac{4^{1+\epsilon}\Gamma(-2\epsilon)}{\Gamma^2(-\epsilon)(1-\cos^2\rho_Q)^{1+\epsilon}}
\; d\cos\theta_Q \; d\phi_Q \; d\cos\rho_Q \nonumber \\
&=& \mu_2 \; d\cos\theta_Q \; d\phi_Q \; d\cos\rho_Q \; .
\end{eqnarray}
It depends on $\zeta,\hat\eta_{1,2},\hat\xi_{1,2}$ only through $Q^2$,
although the momentum vectors $q_{1,2}$ depend on each of these variables
independently. Close to threshold, where $Q^2 \approx s$, we recover
the behavior Eq.~(\ref{eq:betaphase})
\begin{equation}
\int d\Phi_4 \propto s^{2-3\epsilon} \beta^{9-10\epsilon} \; .
\end{equation}
More interestingly, however, the ratio
$\Gamma(-2\epsilon)/\Gamma^2(-\epsilon)$ is of the order $\epsilon$,
which means that we need a divergent contribution from the integral to
obtain a cross section in four dimensions. This is indeed guaranteed
by the following
\begin{equation}
\label{eq:5d}
\frac{4^{1+\epsilon}\Gamma(-2\epsilon)}{\Gamma^2(-\epsilon)(1-
  \cos^2\rho_Q)^{1+\epsilon}} = \delta(1-\cos\rho_Q) +
\frac{4^{1+\epsilon}\Gamma(-2\epsilon)}{\Gamma^2(-\epsilon)} \left[
  \frac{1}{(1-\cos^2\rho_Q)^{1+\epsilon}} \right]_+ \; ,
\end{equation}
where the ``+''-distribution is defined as
\begin{equation}
\int_0^1 d\cos\rho_Q \left[ \frac{1}{(1-\cos^2\rho_Q)^{1+\epsilon}}
  \right]_+ f(\cos\rho_Q) = \int_0^1 d\cos\rho_Q \frac{1}{(1-
  \cos^2\rho_Q)^{1+\epsilon}} \Big( f(\cos\rho_Q)-f(1) \Big) \; ,
\end{equation}
and the integrand on the right-hand side should be expanded in a
Taylor series in $\epsilon$. While we leave the discussion of the
implementation details to Section~\ref{sec:implementation}, we note
that we chose to use equation Eq.~(\ref{eq:5d}) to divide the phase
space into two contributions
\begin{eqnarray}
\label{eq:phi2splitting}
d\Phi_2 &=& d\Phi_2^{(d|\epsilon)} + d\Phi_2^{(\epsilon)} \nonumber \\
&=& \mu_2^{(d|\epsilon)} d\cos\theta_Q \; d\phi_Q + \mu_2^{(\epsilon)}
d\cos\theta_Q \; d\phi_Q \; d\cos\rho_Q \; ,
\end{eqnarray}
with
\begin{eqnarray}
\mu_2^{(d|\epsilon)} &=& \frac{(4\pi)^{\epsilon}\Gamma(1-
  \epsilon)}{8(2\pi)^2\Gamma(1-2\epsilon)} \left( Q^2
\right)^{-\epsilon}\left( \sqrt{1-\frac{4m^2}{Q^2}}
\right)^{1-2\epsilon} (1-\cos^2\theta_Q)^{-\epsilon} \left(
\sin^2\phi_Q \right)^{-\epsilon} \; , \\
\mu_2^{(\epsilon)} &=&
\frac{(16\pi)^\epsilon}{4(2\pi)^2\Gamma(-\epsilon)} \left( Q^2
\right)^{-\epsilon}\left( \sqrt{1-\frac{4m^2}{Q^2}}
\right)^{1-2\epsilon} (1-\cos^2\theta_Q)^{-\epsilon} \left(
\sin^2\phi_Q \right)^{-\epsilon} \nonumber \\ &\times& \left[
\frac{1}{(1-\cos^2\rho_Q)^{1+\epsilon}} \right]_+ \; .
\end{eqnarray}
$d\Phi_2^{(d|\epsilon)}$ would be the entire phase space, if we could
rotate the $\epsilon$-dimensional components away. One can expect that
the additional contribution from $d\Phi_2^{(\epsilon)}$ will be small
in practice. We will show later that this is indeed the case.

At this point, we would like to note that the adopted solution to the
problem of a $d$-dimensional phase space for the heavy quarks is by no
means unique. One could, for example, use the fact that the
$\epsilon$-dimensional components of the heavy quark momentum vectors
are only relevant to the terms singular in $\epsilon$, which are obtained
after one of the massless vectors has been removed (at least one soft
or collinear limit). We then have only two $(d-1)$-dimensional vectors,
and could rotate away the spurious components of
$\vec q_{1,2}$. This approach would only be correct, if the reference frame
for the parameterization of $\vec q_{1,2}$ were defined in relation to
$\vec p_1$ and $\vec k_1 + \vec k_2$. This in turn, would be a
simplification for the massive system, but a complication to the
decomposition of the phase space, which we want to perform in
Section~\ref{sec:decomposition}.


\subsection{Decomposition}
\label{sec:decomposition}

\vspace{.2cm}

\begin{figure}[ht]
\begin{center}
\includegraphics[width=.80\textwidth]{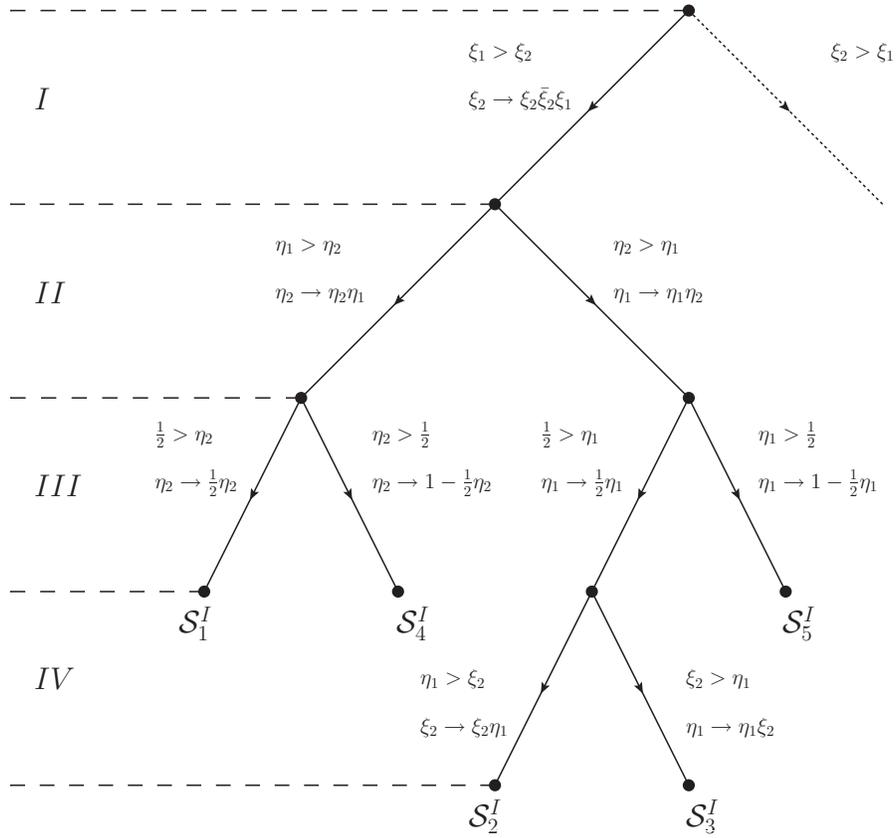}
\end{center}
\caption{Decomposition of the phase space in the triple-collinear
  sector. The variable substitutions, which map the integration range
  onto the unit hypercube are specified. Furthermore, $\hat{\xi}_2 =
  \xi_{max}(\hat{\xi}_1)$ and the second branch starting with the
  dashed line is symmetric to the first.}
\label{fig:dec1}
\end{figure}

\begin{figure}[ht]
\begin{center}
\includegraphics[width=.80\textwidth]{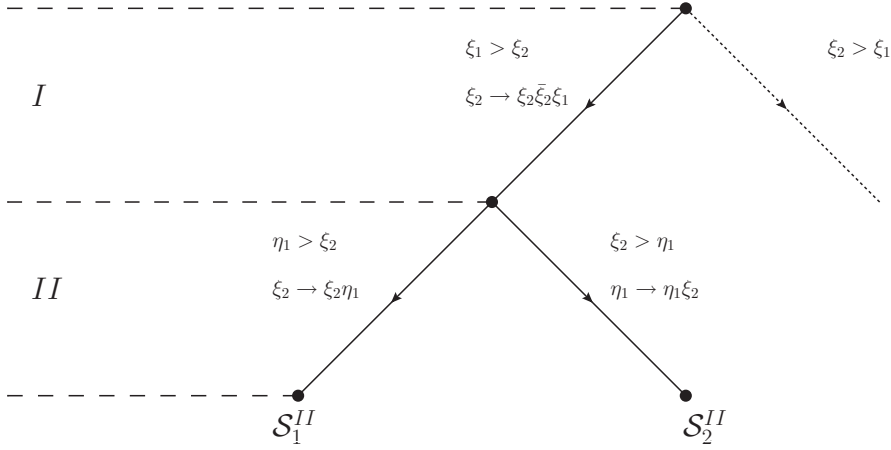}
\end{center}
\caption{Decomposition of the phase space in the double-collinear
  sector. The notation is as in Fig.~\ref{fig:dec1}.}
\label{fig:dec2}
\end{figure}

The last step of our treatment of the phase space is a two-level
decomposition according to singularities. At the first level, we
partition the phase space with suitable selector functions. The latter
are defined on the phase space, add up to unity, and regulate part of
the divergences. In particular, we introduce a selector function for
the triple-collinear sector, in which we allow for collinear
divergences due to partons with momenta $p_1$, $k_1$ and $k_2$, but
not $p_2$. There is also a symmetric function that does just the same
upon replacement of $p_1$ with $p_2$, but we ignore it, as its
contribution can be recovered without additional computation (see
Section~\ref{sec:implementation}). Moreover, we introduce a selector,
which allows for collinear divergences due to $k_1$ being parallel to
$p_1$, or $k_2$ parallel to $p_2$, but no other configuration. This
function defines the double-collinear sector, and has a symmetric
counterpart in $p_1 \leftrightarrow p_2$, which we again do not
discuss any further. The triple- and double-collinear sectors may be
overlapping in the sense that several  selector functions do not
vanish for a given momentum configuration. The only condition is that
the divergences  are properly regulated. In \cite{Czakon:2010td}, we have given two
examples of selector functions, which achieve this goal, one for our
present problem, and one completely general for any number of massless
final states. Apart from numerical efficiency, nothing depends on the
choice. In the present work, we define the sectors implicitly as
follows
\begin{enumerate}
\item $n^z_1 > 0 \; \wedge \; n^z_2 > -\alpha \; n^0_2$ \; in the
  triple-collinear sector;
\item $n^z_1 > 0 \; \wedge \; n^z_2 < -\alpha \; n^0_2$ \; in the
  double-collinear sector,
\end{enumerate}
where $\alpha > 0$ is an arbitrary parameter, which we will take to be
$\alpha = 1/2$ (we checked independence of some results on this
parameter). Notice that the sharp cut $n^z_1 > 0$ in both cases is
necessary for the later use of symmetries. Moreover, we have defined
the conditions through the vectors $n^\mu_{1,2}$, because in the
strict soft limits the actual momentum vectors vanish and it is
impossible to check to which sector they belong.

Having simplified the problem as far as the type of collinear
singularities is concerned, we will perform a second level
decomposition. The purpose is to factorize the divergences of the
propagators in the amplitudes. The set of offending invariants is
\begin{eqnarray}
s_{15} &=& (p_1-k_1)^2 = - s\beta^2 \hat{\xi}_1 \hat{\eta}_1 \; ,
\nonumber \\ s_{16} &=& (p_1-k_2)^2 = - s\beta^2 \hat{\xi}_2
\hat{\eta}_2 \; , \nonumber  \\ s_{26} &=& (p_2-k_2)^2 = - s\beta^2
\hat{\xi}_2 (1-\hat{\eta}_2) \; , \nonumber  \\ s_{56} &=& (k_1+k_2)^2
= s\beta^4 \hat{\xi}_1 \hat{\xi}_2 \eta_3 \; , \nonumber \\ s_{156}
&=& (p_1-k_1-k_2)^2 = - s\beta^2 (\hat{\xi}_1\hat{\eta}_1 +
\hat{\xi}_2\hat{\eta}_2  -\beta^2\hat{\xi}_1\hat{\xi}_2\eta_3) \; ,
\nonumber \\ s_{256} &=& (p_2-k_1-k_2)^2 = - s\beta^2
(\hat{\xi}_1(1-\hat{\eta}_1) +
\hat{\xi}_2(1-\hat{\eta}_2)-\beta^2\hat{\xi}_1\hat{\xi}_2\eta_3) \; .
\label{eq:invariants}
\end{eqnarray}
Assuming that we are only concerned by the collinear singularities,
$s_{15}, s_{16}, s_{56}, s_{156}$ are relevant to the triple-collinear
sector, whereas $s_{15}, s_{26}$ to the double-collinear sector. In
case partons may also become soft, there will be soft-collinear
singularities in the double-collinear sector due to $s_{156}$ and
$s_{256}$. Purely soft singularities may involve other propagators, in
particular the massive, with a general form as follows
\begin{eqnarray}
(p+k_i)^2-p^2 &=& 2\hat{\xi}_i \; p\cdot n_i \; , \;\;\;\; i=1,2 \; ,
  \nonumber \\ (p+k_1+k_2)^2-p^2 &=& 2(\hat{\xi}_1 \; p\cdot n_1 +
  \hat{\xi}_2 \; p\cdot n_2 + \hat{\xi}_1 \hat{\xi}_2 \; n_1\cdot n_2)
  \; .
\end{eqnarray}
Due to the polynomial form of the singular denominators, it is
possible to factorize the divergences with a change of variables in
such a way, that each expression be a product of variables to some
powers, the variables themselves inducing divergences when vanishing,
and a regular function. This is the well known sector decomposition
\cite{Binoth:2000ps}. The difference to the usual treatment is that,
due to the process independent nature of singularities in QCD, we do
not have to consider abstract expressions, but can view the
decomposition as a choice of the order of the soft and collinear
limits. A complete schematic representation of the variable
transformations, which lead to factorization of divergences and
specify the order of singular limits is given in Fig.~\ref{fig:dec1}
for  the triple- and Fig.~\ref{fig:dec2} for the double-collinear
sector.  Each level in these decomposition trees factorizes a certain
type of singularities. For Fig.~\ref{fig:dec1}, we have
\begin{itemize}
\item[I)] factorization of the soft singularities;
\item[II, III)] factorization of the collinear singularities;
\item[IV)] factorization of the soft-collinear singularities.
\end{itemize}
The case of Fig.~\ref{fig:dec2} is even simpler, as some of the levels
disappear. Note that for each tree, we start with a new set of
variables $\eta_{1,2}=\hat{\eta}_{1,2}$ and
$\xi_{1,2}=\hat{\xi}_{1,2}$ at the root, and continue with the
substitutions down to the leaves. The missing right branches
corresponding to $\xi_2 > \xi_1$ can be recovered by changing the
order of the final state partons. For this reason, we will ignore them
as well. Finally, the variable substitutions guarantee that
$\eta_{1,2},\xi_{1,2} \in [0,1]$, as the range Eq.~(\ref{eq:range})
gets expanded.

Proving that this procedure is sufficient for factorization is a
simple matter of substitutions and can be done with pen and paper. We
shall not reproduce the uninteresting transformations here, since we
shall not need them anymore. Nevertheless, we point out that the
factorization of $\eta_3$, present in $s_{56}$, introduces two regular
functions, which we will encounter later. They are defined as
\begin{eqnarray}
\eta_{31}(\eta_1,\eta_2) &=& \frac{1}{\eta_1} \; \eta_3\left(\eta_1,
\frac{1}{2}\eta_1\eta_2\right) \\ &=& \frac{(2-\eta_2)^2}{2
  \left(2+\eta_2(1-2\eta_1)-2 (1-2 \zeta )
  \sqrt{\eta_2(1-\eta_1)(2-\eta_1 \eta_2)}\right)} \; , \nonumber
\\ \nonumber \\ \eta_{32}(\eta_1,\eta_2) &=& \frac{1}{\eta_1\eta_2^2}
\; \eta_3\left( \eta_1, \frac{1}{2}\eta_1(2-\eta_2) \right) \\ &=&
\frac{1}{2\left(
  2+(1-2\eta_1)(2-\eta_2)-2(1-2\zeta)\sqrt{(1-\eta_1)(2-\eta_2)(2-\eta_1(2
    -\eta_2))} \right)} \; , \nonumber 
\end{eqnarray}
with $\eta_3(\hat{\eta}_1,\hat{\eta}_2)$ as in
Eq.~(\ref{eq:eta3second}). The first of these functions is relevant to
the sectors ${\cal S}^I_1,...,{\cal S}^I_3$, whereas the second to
${\cal S}^I_4,{\cal S}^I_5$. The derivation of the subtraction terms
in Section~\ref{sec:subtraction} requires the knowledge of the
behavior of $\eta_{31}(\eta_1,\eta_2)$ for small $\eta_2$
\begin{equation}
\eta_{31}(\eta_1,\eta_2) = 1 + (1-2\zeta)\sqrt{2(1-\eta_1)\eta_2} +
    {\cal O}(\eta_2) \; .
\end{equation}

Tab.~\ref{tab:kinmap} contains the complete set of variable
transformations, {\it i.e.} the expressions of the original kinematic
variables $\hat{\eta}_{1,2},\hat{\xi}_{1,2}$ in terms of the sector
variables $\eta_{1,2},\xi_{1,2}$. For the double-collinear sector, we
have used the transformation $\eta_2 \rightarrow 1-\eta_2$ first, in
order to profit from the phase space measure Eq.~(\ref{eq:mu1}), which
is invariant under this transformation.

\begin{table}[t]
  \begin{center}
    \begin{tabular}{c|c|c|c|c|c|c|c}
      &&&&&&&\\ & ${\cal S}^I_1$ & ${\cal S}^I_2$ & ${\cal S}^I_3$ &
      ${\cal S}^I_4$ & ${\cal S}^I_5$ & ${\cal S}^{II}_1$ & ${\cal
        S}^{II}_2$ \\ &&&&&&&\\ \hline  &&&&&&& \\ $\hat{\eta}_1$ &
      $\eta_1$ & $\frac{1}{2}\eta_1\eta_2$ & $\frac{1}{2}
      \eta_1\eta_2\xi_2$ & $\eta_1$ & $\frac{1}{2}(2-\eta_1)\eta_2$ &
      $\eta_1$ & $\eta_1\xi_2$ \\  &&&&&&&\\ \hline
      &&&&&&&\\ $\hat{\eta}_2$ & $\frac{1}{2}\eta_1\eta_2$ & $\eta_2$
      & $\eta_2$ & $\frac{1}{2}\eta_1(2-\eta_2)$ & $\eta_2$ &
      $1-\eta_2$ & $1-\eta_2$ \\ &&&&&&&\\ \hline
      &&&&&&&\\ $\hat{\xi}_1$ & $\xi_1$ & $\xi_1$ & $\xi_1$ & $\xi_1$
      & $\xi_1$ & $\xi_1$ & $\xi_1$ \\ &&&&&&&\\ \hline
      &&&&&&&\\ $\hat{\xi}_2$ & $\xi_1\xi_2\bar\xi_2$ &
      $\eta_1\xi_1\xi_2\bar\xi_2$ & $\xi_1\xi_2\bar\xi_2$ &
      $\xi_1\xi_2\bar\xi_2$ & $\xi_1\xi_2\bar\xi_2$ &
      $\eta_1\xi_1\xi_2\bar\xi_2$ & $\xi_1\xi_2\bar\xi_2$ \\ &&&&&&&\\
    \end{tabular}
  \end{center}
  \caption{\label{tab:kinmap} Original kinematic variables,
    $\hat{\eta}_1, \hat{\eta}_2, \hat{\xi}_1, \hat{\xi}_2$, expressed
    through the sector variables, $\eta_1, \eta_2, \xi_1, \xi_2$, with
    $\bar\xi_2 = \xi_{max}(\hat{\xi}_1)$.}
\end{table}

\begin{table}[t]
  \begin{center}
    \begin{tabular}{c|c|c}
      && \\ sector & regulator & $\mu^{\mbox{\scriptsize reg}}_{\cal S}$ \\ &&
      \\ \hline  && \\ ${\cal S}^{I}_1$ &
      $\eta_1^{1-2\epsilon}\eta_2^{-\epsilon}\xi_1^{3-4\epsilon}\xi_2^{1-2\epsilon}$
      & $\displaystyle
      \left((1-\eta_1)(2-\eta_1\eta_2)\right)^{-\epsilon}
      \bar{\xi}_2^{\; 2-2\epsilon} \left(
      \frac{\eta_{31}(\eta_1,\eta_2)}{2-\eta_2} \right)^{1-2\epsilon}$
      \\ && \\ \hline  && \\ ${\cal S}^{I}_2$ &
      $\eta_1^{2-3\epsilon}\eta_2^{1-2\epsilon}\xi_1^{3-4\epsilon}\xi_2^{1-2\epsilon}$
      & $\displaystyle
      \left((1-\eta_2)(2-\eta_1\eta_2)\right)^{-\epsilon}
      \bar{\xi}_2^{\; 2-2\epsilon} \left(
      \frac{\eta_{31}(\eta_2,\eta_1)}{2-\eta_1} \right)^{1-2\epsilon}$
      \\ && \\ \hline  && \\ ${\cal S}^{I}_3$ &
      $\eta_1^{-\epsilon}\eta_2^{1-2\epsilon}\xi_1^{3-4\epsilon}\xi_2^{2-3\epsilon}$
      & $\displaystyle
      \left((1-\eta_2)(2-\eta_1\eta_2\xi_2)\right)^{-\epsilon}
      \bar{\xi}_2^{\; 2-2\epsilon} \left(
      \frac{\eta_{31}(\eta_2,\eta_1\xi_2)}{2-\eta_1\xi_2}
      \right)^{1-2\epsilon}$ \\ && \\ \hline  && \\ ${\cal S}^{I}_4$ &
      $\eta_1^{1-2\epsilon}\eta_2^{1-2\epsilon}\xi_1^{3-4\epsilon}\xi_2^{1-2\epsilon}$
      & $\displaystyle
      \left((1-\eta_1)(2-\eta_2)(2-\eta_1(2-\eta_2))\right)^{-\epsilon}
      \bar{\xi}_2^{\; 2-2\epsilon}
      \eta_{32}^{1-2\epsilon}(\eta_1,\eta_2)$ \\ && \\ \hline &&
      \\ ${\cal S}^{I}_5$ &
      $\eta_1^{1-2\epsilon}\eta_2^{1-2\epsilon}\xi_1^{3-4\epsilon}\xi_2^{1-2\epsilon}$
      & $\displaystyle
      \left((1-\eta_2)(2-\eta_1)(2-\eta_2(2-\eta_1))\right)^{-\epsilon}
      \bar{\xi}_2^{\; 2-2\epsilon}
      \eta_{32}^{1-2\epsilon}(\eta_2,\eta_1)$ \\ && \\ \hline &&
      \\ ${\cal S}^{II}_1$ &
      $\eta_1^{2-3\epsilon}\eta_2^{-\epsilon}\xi_1^{3-4\epsilon}\xi_2^{1-2\epsilon}$
      & $\displaystyle \left((1-\eta_1)(1-\eta_2)\right)^{-\epsilon}
      \bar{\xi}_2^{\; 2-2\epsilon} \left(
      \frac{\eta_3}{|1-\eta_1-\eta_2|}\right)^{1-2\epsilon}$ \\ &&
      \\ \hline && \\ ${\cal S}^{II}_2$ &
      $\eta_1^{-\epsilon}\eta_2^{-\epsilon}\xi_1^{3-4\epsilon}\xi_2^{2-3\epsilon}$
      & $\displaystyle
      \left((1-\eta_2)(1-\eta_1\xi_2)\right)^{-\epsilon}
      \bar{\xi}_2^{\; 2-2\epsilon} \left(
      \frac{\eta_3}{|1-\eta_2-\eta_1\xi_2|}\right)^{1-2\epsilon}$
      \\ && \\
    \end{tabular}
  \end{center}
  \caption{\label{tab:meamap} Integration measure, $d\mu_{\eta\xi}$
    expressed through the sector variables, $\eta_{1,2},\xi_{1,2}$,
    decomposed into the product of their powers used to regulate the
    divergences, and a regular function, $\mu^{\mbox{\scriptsize
        reg}}_{\cal S}$, which can be expanded in $\epsilon$.}
\end{table}

The above factorization procedure does not leave the phase space
measure unchanged, even though it is only the factor $d\mu_{\eta\xi}$
that is transformed. The latter assumes, for a given sector ${\cal
  S}$,  the following form
\begin{equation}
\label{eq:asandbs}
d\mu_{\eta\xi} = \eta_1^{a_1+b_1\epsilon}
\eta_2^{a_2+b_2\epsilon} \xi_1^{a_3+b_3\epsilon}
\xi_2^{a_4+b_4\epsilon} \; \mu^{\mbox{\scriptsize
    reg}}_{\cal S} \; d\eta_1 d\eta_2 d\xi_1 d\xi_2 \; ,
\end{equation}
where the first factor regulates the divergences, whereas the second
is a regular function, which we will later on expand in $\epsilon$ in
a straightforward Taylor series. Both factors are given in
Tab.~\ref{tab:meamap}.


\subsection{Normalization of the cross section}

\vspace{.2cm}

Although we have now specified the four-particle phase space
completely, there is still a freedom in defining the divergent cross
section for double-real radiation by including a function, which
equals unity, when $\epsilon = 0$. Usually such a function is
introduced due to the $\epsilon$-dependence of the bare coupling
constant, which, in the $\overline{\mbox{MS}}$ scheme, is chosen to be
\begin{equation}
\label{eq:alphas}
\alpha^0_s = \left( \frac{\mu^2 e^{\gamma_E}}{4\pi} \right)^\epsilon
Z_{\alpha_s}(\alpha_s(\mu^2),\epsilon) \; \alpha_s(\mu^2) \; ,
\end{equation}
where $\mu$ is the renormalization scale and $Z_{\alpha_s}$ the
renormalization constant. Of course, in our case $Z_{\alpha_s} = 1$,
since we are integrating tree-level amplitudes. The problem with this
approach is that our matrix elements are proportional to $\alpha_s^4$,
whereas the dimension of the phase space is the same as that of a
three-loop integral. This would leave unbalanced, large and
dimensionful factors of $\mu^{2\epsilon}$. We can compensate them by
multiplying the cross section with the inverse of the parenthesized
factor in Eq.~(\ref{eq:alphas}) to power $\epsilon$. Equivalently, we
will express all tree level amplitudes by the renormalized strong
coupling and use the following definition of the partonic cross
section
\begin{equation}
\label{eq:sigma}
\sigma_{\cal O} = \frac{1}{2s} \left( \frac{\mu^2 e^{\gamma_E}}{4\pi}
\right)^{3\epsilon} \int d\Phi_4 \; F_J \; \overline{|{\cal M}_4|^2}  \; ,
\end{equation}
where the overline over the matrix element squared signifies the sums
and averages in color and spin,  as well as statistical factors for
identical final states. $F_J$ is a jet function defining the
observable ${\cal O}$. In what follows, we will mostly use a trivial
jet function $F_J = 1$. Notice, however, that a test in
Section~\ref{sec:implementation} will be performed with a non-trivial
$F_J$.

Let us stress, that we could have considered a different factor in
Eq.~(\ref{eq:sigma}), as long as we would include it in all the other
contributions that enter the finite physical cross section. From the
point of view of physics, this factor is irrelevant, and yet it can
play a substantial r\^ole in obtaining precise numerical values. For
example, multiplying by $\beta^{10\epsilon}$, we would remove dominant
logarithms of $\beta$ close to threshold, which would substantially
lower the double-real cross section, while enhancing the others. We
can also lower the contribution of the finite part of the cross
section by multiplying with a constant, but $\epsilon$ dependent
factor. The decision on what to enhance and what to diminish can only
be taken once all terms are implemented in a numerical program,
because only then can we check, what contributes the largest absolute
errors. We will leave this problem to future studies.


\section{Subtraction and integrated subtraction terms}


\subsection{Derivation}
\label{sec:subtraction}

\vspace{.2cm}

The decomposition of the phase space introduced in the previous
sections is sufficient to derive Laurent expansions of arbitrary
infrared safe observables. In order to obtain explicit expressions for
a given sector ${\cal S}$, we define
\begin{equation}
\mathfrak{M}_\mathcal{S} = \eta_1^{1+a_1} \eta_2^{1+a_2} \xi_1^{1+a_3}
\xi_2^{1+a_4} \; \overline{|{\cal M}_4|^2} \; ,
\end{equation}
where the $a_i$ constants have been defined in Eq.~(\ref{eq:asandbs}),
and are given for each sector in Tab.~\ref{tab:meamap}. The averaged
matrix element, $\overline{|{\cal M}_4|^2}$, has been introduced in
Eq.~(\ref{eq:sigma}). $\mathfrak{M}_\mathcal{S}$ must be regular, by
infrared power counting in QCD, in limits of any of
$\eta_{1,2},\xi_{1,2}$ vanishing. This can be checked
explicitly, with the formulae introduced later in this section.

The cross section is now
\begin{equation}
\sigma_{\cal O} = \sum_{\cal S} \sigma_{\cal O}^{(\cal S)} \; ,
\end{equation}
where
\begin{eqnarray}
\label{eq:maincross}
\sigma_{\cal O}^{(\cal S)} &=& \frac{1}{2s} \left( \frac{\mu^2 e^{\gamma_E}}{4\pi}
\right)^{3\epsilon} \int d\mu_\zeta \; d\eta_1 \; d\eta_2 \;
d\xi_1 \; d\xi_2 \; d\Phi_2 \; \mu_{\cal S}^{\mbox{\scriptsize reg}} \;
\theta_{\cal S} \; F_J \; \frac{1}{\eta_1^{1-b_1\epsilon}}
\frac{1}{\eta_2^{1-b_2\epsilon}} \frac{1}{\xi_1^{1-b_3\epsilon}}
\frac{1}{\xi_2^{1-b_4\epsilon}} \; \mathfrak{M}_\mathcal{S} \nonumber
\\ &=&
\int d\zeta \; d\eta_1 \; d\eta_2 \; d\xi_1 \; d\xi_2 \; d\cos\theta_Q
\; d\phi_Q \; d\cos\rho_Q \; \Sigma_{\cal O}^{(\cal S)} \; ,
\end{eqnarray}
with the integrand
\begin{equation}
\Sigma_{\cal O}^{(\cal S)} = \frac{1}{2s} \left( \frac{\mu^2
  e^{\gamma_E}}{4\pi} \right)^{3\epsilon} \mu_\zeta \; \mu_{\cal
  S}^{\mbox{\scriptsize reg}} \; \mu_2 \; \theta_{\cal S} \; F_J
\frac{1}{\eta_1^{1-b_1\epsilon}} \frac{1}{\eta_2^{1-b_2\epsilon}}
\frac{1}{\xi_1^{1-b_3\epsilon}} \frac{1}{\xi_2^{1-b_4\epsilon}} \;
\mathfrak{M}_\mathcal{S} \; .
\end{equation}
The $b_i$ constants have been defined in Eq.~(\ref{eq:asandbs}), and
are given for each sector in Tab.~\ref{tab:meamap}. The jet function
$F_J$ has been introduced in Eq.~(\ref{eq:sigma}). Finally,
$\theta_{\cal S}$ is the selector function described at the beginning
of Section~\ref{sec:decomposition}. We remind the reader that the
full phase space is covered by changing the order of the final state
massless partons, and swapping $p_1$ and $p_2$, which can also be
thought of as changing the order of the initial state massless
partons.

The Laurent expansion of the cross section contribution, $\sigma_{\cal
  O}^{\cal S}$, is obtained by using
\begin{equation}
\label{eq:sub1}
\frac{1}{\lambda^{1-b \epsilon}} = \frac{1}{b}\frac{\delta(\lambda)}{\epsilon}
+ \sum_{n=0}^\infty \frac{(b \epsilon)^n}{n!} \left[ \frac{\ln^n
    (\lambda)}{\lambda} \right]_+ \; ,
\end{equation}
where $\lambda=\eta_{1,2},\xi_{1,2}$, and the ``+''-distribution is
\begin{equation}
\label{eq:sub2}
\int_0^1 d\lambda \; \left[ \frac{\ln^n (\lambda)}{\lambda} \right]_+
f(\lambda) = \int_0^1 \frac{\ln^n (\lambda)}{\lambda} (f(\lambda) -
f(0)) \; .
\end{equation}
A more practical, albeit equivalent, application of these formulae is
\begin{equation}
\label{eq:maintool}
\int_0^1 \frac{d\lambda}{\lambda^{1-b \epsilon}} f(\lambda)
\longrightarrow \int_0^1 d\lambda \left[ \frac{f(0)}{b \epsilon} +
  \frac{f(\lambda) - f(0)}{\lambda^{1-b \epsilon}} \right] \; .
\end{equation}
In any case, Eq.~(\ref{eq:maincross}) involves four singular
integrations and the above formula has to be applied iteratively. The
result contains the integrand at sixteen different points obtained by
setting the variables in all possible subsets of $\{\eta_1, \eta_2,
\xi_1, \xi_2\}$ to zero. Using Eq.~(\ref{eq:maintool}) gives a
convergent integrand
\begin{equation}
\Sigma_{\cal O}^{(\cal S)} \longrightarrow \left[ \Sigma_{\cal
    O}^{(\cal S)} \right] \; .
\end{equation}
Considered differently, the terms in $\left[ \Sigma_{\cal O}^{(\cal
    S)} \right]$ proportional to negative powers of $\epsilon$ are
called integrated subtraction terms, those free of singularities are
simply called subtraction terms if any of the variables vanishes, just
as in the classic approach to subtraction schemes. Notice that the
only analytic integration that is needed here is the rather trivial
integral of $1/\lambda^{1-b \epsilon}$. This is the main difference to
the traditional approach. The limits of $d\Phi_2$, $\mu_{\cal
  S}^{\mbox{\scriptsize reg}}$, selector and jet functions are
obtained by directly setting variables to zero, and are process
independent. The only process dependent information is in
$\mathfrak{M}_\mathcal{S}$. The vanishing of the sector variables
corresponds, however, to singular limits of QCD amplitudes. Thus, we
can obtain the subtraction and integrated subtraction terms from the
splitting functions and soft currents exactly as it is done at
NLO. The sector decomposition of Section~\ref{sec:decomposition}
guarantees the independence of the result from the order, in which the
limits are taken. The process dependent information will now be
shifted to reduced $d$-dimensional matrix elements.

In order to derive the relevant formulae, let $\mathbf{X} \subseteq \{
  \eta_1, \eta_2, \xi_1, \xi_2\}$ be the subset of vanishing variables
in a given limit, and define 
\begin{equation}
\lim_{\mathbf{X} \to \mathbf{0}}
\mathfrak{M}_\mathcal{S} = g^2 \; \langle {\cal M}_3 | \mathbf{V} |
{\cal M}_3 \rangle \;\;\;\; \mbox{or} \;\;\;\; 
\lim_{\mathbf{X} \to \mathbf{0}}
\mathfrak{M}_\mathcal{S} = g^4 \; \langle {\cal M}_2 | \mathbf{V} |
{\cal M}_2 \rangle \; ,
\end{equation}
depending on whether the Born matrix elements above correspond to
reduced processes with one, $| {\cal M}_3 \rangle$, or two partons, $|
{\cal M}_2 \rangle$, less. $\mathbf{V}$ is an operator in spin and
color space, and depends on the flavors of the partons involved. Let
us also define a shorthand notation for the divergence regulating
product of integration variables
\begin{equation}
\mathfrak{R}_\mathcal{S} = \eta_1^{1+a_1} \eta_2^{1+a_2} \xi_1^{1+a_3}
\xi_2^{1+a_4} \; .
\end{equation}
We now consider the various limiting cases starting from those
relevant to the triple-collinear sector. With the flavor assignment
\begin{equation}
a_1(p_1) + a_2(p_2) \rightarrow t(q_1)  + \bar t(q_2) + a_5(k_1) +
a_6(k_2) \; ,
\end{equation}
there are nine cases, which are identified from $\mathbf{X}$ with the
help of Tab.~\ref{tab:kinmap}
\begin{enumerate}

\item \framebox{$\hat\eta_1 = \hat\eta_2 = 0$}
\begin{equation}
\mathbf{V}^{ss'}_{a_1a_5a_6} = \lim_{\mathbf{X} \to \mathbf{0}}
\mathfrak{R}_\mathcal{S} \frac{4\hat P^{ss'}_{a_1a_5a_6}}{s^2_{156}} \; .
\end{equation}
The splitting functions $\hat P^{ss'}_{a_1a_5a_6}$ are given in
\ref{sec:splitting}. They depend on the following variables
\begin{equation}
x_1 = -1 \; , \;\;\;\;
x_5 = \beta^2\hat\xi_1 \; , \;\;\;\;
x_6 = \beta^2\hat\xi_2 \; ,
\end{equation}
and
\begin{equation}
k^\mu_{\perp 1} = 0\; , \;\;\;\;
k^\mu_{\perp 5} = \beta^2 \hat\xi_1 \sqrt{\hat\eta_1} \;
\bar k^\mu_{\perp 5} \; , \;\;\;\;
k^\mu_{\perp 6} = \beta^2 \hat\xi_2 \sqrt{\hat\eta_2} \;
\bar k^\mu_{\perp 6}(\hat\eta_1, \hat\eta_2) \; ,
\end{equation}
with
\begin{eqnarray}
\bar k^\mu_{\perp 5} &=& \big( 0,0,1,0 \big) \; , \\ 
\bar k^\mu_{\perp 6}(\hat\eta_1, \hat\eta_2) &=&
\frac{1}{\hat\eta_1 + \hat\eta_2 - 2(1-2\zeta) \sqrt{\hat\eta_1
    \hat\eta_2}} \nonumber \\ &\times& \Big( 0, \; 2 |\hat\eta_1 - \hat\eta_2|
\sqrt{\zeta(1-\zeta)}, \; 2\sqrt{\hat\eta_1 \hat\eta_2} -(\hat\eta_1 +
\hat\eta_2) (1-2\zeta), \; 0 \Big) \; .
\end{eqnarray}
The last vector is symmetric and homogeneous in $\hat\eta_1$ and
$\hat\eta_2$, $\bar k^\mu_{\perp 6}(\hat\eta_1, \hat\eta_2) = \bar
k^\mu_{\perp 6}(\hat\eta_2, \hat\eta_1) = \bar k^\mu_{\perp
  6}(1, \hat\eta_1/\hat\eta_2) = \bar k^\mu_{\perp 6}(1, \hat\eta_2/
\hat\eta_1)$. Moreover
\begin{equation}
\bar k^\mu_{\perp 6}(1, \rho) = \bar k^\mu_{\perp 5} +
\frac{1}{2} \sqrt{\frac{1-\zeta}{\zeta}} \; |1-\rho| \; \bar k^\mu_{\perp 0}
+ {\cal O} \left( (1-\rho)^2 \right) \; ,
\end{equation}
with
\begin{equation}
\bar k^\mu_{\perp 0} = (0,1,0,0) \; .
\end{equation}
This asymptotic behavior is necessary for $\mathfrak{M}_\mathcal{S}$
not to be singular. This limit is usually responsible for eight of the
fifteen subtraction terms.

\vspace{.2cm}
\item \framebox{$\hat\eta_1 = 0$}
\begin{equation}
\mathbf{V}^{ss'}_{a_1a_5} = \lim_{\mathbf{X} \to \mathbf{0}}
\mathfrak{R}_\mathcal{S} \frac{2\hat P^{ss'}_{a_1a_5}}{s_{15}} \; ,
\end{equation}
with
\begin{equation}
\label{eq:eta1}
z = \frac{1}{1-\beta^2\hat\xi_1} \; , \;\;\;\; k^\mu_{\perp} =
(0,0,1,0) \; .
\end{equation}

\vspace{.2cm}
\item \framebox{$\hat\eta_2 = 0$}
\begin{equation}
\mathbf{V}^{ss'}_{a_1a_6} = \lim_{\mathbf{X} \to \mathbf{0}}
\mathfrak{R}_\mathcal{S} \frac{2\hat P^{ss'}_{a_1a_6}}{s_{16}} \; ,
\end{equation}
with
\begin{equation}
\label{eq:eta2}
z = \frac{1}{1-\beta^2\hat\xi_2} \; , \;\;\;\; k^\mu_{\perp} =
\Big( 0, \; 2\sqrt{\zeta(1-\zeta)}, \; 2\zeta - 1, \; 0 \Big) \; .
\end{equation}

\vspace{.2cm}
\item \framebox{$\hat\eta_1 = \hat\eta_2$}
\begin{equation}
\mathbf{V}^{ss'}_{a_5a_6} = \lim_{\mathbf{X} \to \mathbf{0}}
\mathfrak{R}_\mathcal{S} \frac{2\hat P^{ss'}_{a_5a_6}}{s_{56}} \; ,
\end{equation}
with
\begin{equation}
\label{eq:eta3}
z = \frac{\hat\xi_1}{\hat\xi_1 + \hat\xi_2} \; , \;\;\;\; k^\mu_{\perp} =
\Big( 0, \; \sqrt{1-\zeta}, \; \mbox{sgn}(\hat\eta_2 - \hat\eta_1)
(1-2\hat\eta_1) \sqrt{\zeta}, \; -2 \; \mbox{sgn}(\hat\eta_2 - \hat\eta_1)
\sqrt{\hat\eta_1(1 - \hat\eta_1)\zeta} \Big) \; .
\end{equation}

\vspace{.2cm}
\item \framebox{$\hat\xi_2 = 0 \wedge \hat\eta_2 \neq
  0 \wedge \hat\eta_2 \neq 1 \wedge \hat\eta_2 \neq \hat\eta_1
  \wedge a_6 = g$}
\begin{equation}
\mathbf{V} = -\lim_{\mathbf{X} \to \mathbf{0}}
\mathfrak{R}_\mathcal{S} \; \frac{1}{\hat\xi_2^2} \sum_{ij=1}^5 {\cal
 S}_{ij}(n_2) \; \mathbf{T}_i \cdot \mathbf{T}_j
\; ,
\end{equation}
where
\begin{equation}
\label{eq:sij1}
{\cal S}_{ij}(k) = \frac{p_i \cdot p_j}{(p_i \cdot k)(p_j \cdot
  k)} \; ,
\end{equation}
and $p_{i,j}$ is one of $p_1, p_2, q_1, q_2, k_1$. $\mathbf{T}_i$  are
the standard \cite{Catani:1996vz} color operators.

\vspace{.2cm}
\item \framebox{$\hat\xi_1 = \hat\xi_2 = 0 \wedge \hat\eta_{1,2} \neq
  0 \wedge \hat\eta_{1,2} \neq 1 \wedge \hat\eta_1 \neq \hat\eta_2$}
\vspace{.2cm}

In the case of a gluon pair the limit is given by
\begin{eqnarray}
\mathbf{V} &=& \lim_{\mathbf{X} \to \mathbf{0}}
\mathfrak{R}_\mathcal{S} \\ \nonumber &\times& \left( \sum_{ijkl=1}^4
\frac{1}{2} {\cal S}_{ij}(\hat\xi_1 n_1) {\cal S}_{kl}(\hat\xi_2 n_2)
\Big\{ \mathbf{T}_i \cdot \mathbf{T}_j, \; \mathbf{T}_k \cdot
\mathbf{T}_l \Big\} - C_A \sum_{ij=1}^4 {\cal S}_{ij}(\hat\xi_1 n_1,
\; \hat\xi_2 n_2) \; \mathbf{T}_i \cdot \mathbf{T}_j \right) \; .
\end{eqnarray}
This approximation is discussed in \ref{sec:soft}, where we also
define ${\cal S}_{ij}(k_1,k_2)$. In the case of a quark pair we have 
\begin{equation}
\mathbf{V} = \lim_{\mathbf{X} \to \mathbf{0}}
\mathfrak{R}_\mathcal{S} \; T_F \sum_{ij=1}^4 {\cal I}_{ij}(\hat\xi_1 n_1,
\; \hat\xi_2 n_2) \; \mathbf{T}_i \cdot \mathbf{T}_j \; ,
\end{equation}
with \cite{Catani:1999ss}
\begin{equation}
  {\cal I}_{ij}(k_1, k_2) =  \f{(p_i \cdot k_1)\, (p_j \cdot k_2)
+ (p_j \cdot k_1)\, (p_i \cdot k_2) - (p_i \cdot p_j) 
\,(k_1 \cdot k_2)}{(k_1 \cdot k_2)^2 
\,[p_i\cdot (k_1+k_2)]\, [p_j \cdot (k_1+k_2)]} \; ,
\end{equation}
and $p_{i,j}$ in the expressions above is one of $p_1, p_2, q_1,
q_2$.

\vspace{.2cm}
\item \framebox{$\hat\eta_1 = \hat\xi_2 = 0 \wedge \hat\eta_2 \neq 0
  \wedge \hat\eta_2 \neq 1 \wedge a_6 = g$}
\begin{equation}
\mathbf{V}^{ss'}_{a_1a_5} = -\lim_{\mathbf{X} \to \mathbf{0}}
\mathfrak{R}_\mathcal{S} \frac{2\hat P^{ss'}_{a_1a_5}}{s_{15}} \sum_{ij=1}^4
\frac{1}{\hat\xi_2^2} {\cal S}_{ij}(n_2) \; \mathbf{T}_i \cdot
\mathbf{T}_j \; ,
\end{equation}
with the collinear parameters specified in Eq.~(\ref{eq:eta1}),
whereas $p_{i,j}$ in Eq.~(\ref{eq:sij1}) is one of $p_1-k_1, p_2,
q_1, q_2$.

\vspace{.2cm}
\item \framebox{$\hat\eta_2 = \hat\xi_1 = 0 \wedge \hat\eta_1 \neq 0
  \wedge \hat\eta_1 \neq 1 \wedge a_5 = g$}
\begin{equation}
\mathbf{V}^{ss'}_{a_1a_6} = -\lim_{\mathbf{X} \to \mathbf{0}}
\mathfrak{R}_\mathcal{S} \frac{2\hat P^{ss'}_{a_1a_6}}{s_{16}} \sum_{ij=1}^4
\frac{1}{\hat\xi_1^2} {\cal S}_{ij}(n_1) \; \mathbf{T}_i \cdot
\mathbf{T}_j \; ,
\end{equation}
with the collinear parameters specified in Eq.~(\ref{eq:eta2}),
whereas $p_{i,j}$ in Eq.~(\ref{eq:sij1}) is one of $p_1-k_2, p_2,
q_1, q_2$.

\vspace{.2cm}
\item \framebox{$\hat\eta_1 = \hat\eta_2 \wedge \hat\xi_1 = \hat\xi_2
  = 0 \wedge \hat\eta_{1,2} \neq 0 \wedge \hat\eta_{1,2} \neq 1$}
\begin{equation}
\mathbf{V}_{a_5a_6} = \lim_{\mathbf{X} \to \mathbf{0}}
\mathfrak{R}_\mathcal{S} \;{\bf J}^\dagger_\mu(\hat\xi_1 n_1 +
\hat\xi_2 n_2) \frac{2 \hat{P}_{a_5a_6}^{\mu\nu}}{s_{56}} {\bf
  J}_\nu(\hat\xi_1 n_1 + \hat\xi_2 n_2) \; .
\end{equation}
This approximation is discussed in \ref{sec:coll2soft}, and
the collinear parameters are given in Eq.~(\ref{eq:eta3})

\end{enumerate}
The conditions for the various limits have been chosen such that only
one of the above expressions generates a non-vanishing $\mathbf{V}$
for a given $\mathbf{X}$. We have also minimized the use of soft
limits. For example, the double-soft limit contains the
double-collinear and double-soft limit, nevertheless the expressions
are much lengthier than the double-soft limit of the double-collinear
limit, since the latter contains no color correlators. This is only a
practical choice, since all expressions would give the same after
taking into account color conservation.

The double-collinear sector requires the cases 2, 5, 6 and 7 from the
triple-collinear sector, as well as the following
\begin{enumerate}

\vspace{.2cm}
\item \framebox{$\hat\eta_1 = 0 \wedge \hat\eta_2 = 1$}
\begin{equation}
\mathbf{V}^{ss's''s'''}_{a_1a_5a_2a_6} = \lim_{\mathbf{X} \to \mathbf{0}}
\mathfrak{R}_\mathcal{S} \frac{2\hat P^{ss'}_{a_1a_5}}{s_{15}}
\frac{2\hat P^{s''s'''}_{a_2a_6}}{s_{26}} \; , 
\end{equation}
with the collinear variables defined in Eq.~(\ref{eq:eta1}) for $\hat
P^{ss'}_{a_1a_5}$ and in Eq.~(\ref{eq:eta2}) for $\hat
P^{s''s'''}_{a_2a_6}$.

\vspace{.2cm}
\item \framebox{$\hat\eta_2 = 1$}
\begin{equation}
\mathbf{V}^{ss'}_{a_2a_6} = \lim_{\mathbf{X} \to \mathbf{0}}
\mathfrak{R}_\mathcal{S} \frac{2\hat P^{ss'}_{a_2a_6}}{s_{26}} \; ,
\end{equation}
with the collinear variables defined in Eq.~(\ref{eq:eta2}).

\vspace{.2cm}
\item \framebox{$\hat\eta_2 = 1 \wedge \hat\xi_1 = 0 \wedge \hat\eta_1
  \neq 0 \wedge \hat\eta_1 \neq 1 \wedge a_5 = g$}
\begin{equation}
\mathbf{V}^{ss'}_{a_2a_6} = -\lim_{\mathbf{X} \to \mathbf{0}}
\mathfrak{R}_\mathcal{S} \frac{2\hat P^{ss'}_{a_2a_6}}{s_{26}} \sum_{ij=1}^4
\frac{1}{\hat\xi_1^2} {\cal S}_{ij}(n_1) \; \mathbf{T}_i \cdot
\mathbf{T}_j \; ,
\end{equation}
with the collinear parameters specified in Eq.~(\ref{eq:eta2}),
whereas $p_{i,j}$ in Eq.~(\ref{eq:sij1}) is one of $p_1, p_2-k_2,
q_1, q_2$.

\end{enumerate}

Since the $\mathbf{V}$ operators are used both for subtraction and
integrated subtraction terms, there is one more difference of the
present approach to the traditional one. There will be transverse
vectors $k_\perp$ in the integrated subtraction terms. These are
usually removed (averaged over) using the fact that once in the
collinear limit, one can integrate over the azimuthal angle. Keeping them
makes our approach less sensitive to simple errors, while not
hampering efficiency.

We do not provide the explicit expressions for all the limits derived
according to the rules above. On the one hand, it is easy to obtain
them, on the other, the formulae are extremely lengthy. We believe
that it only makes sense to provide a complete, general, working
implementation of our subtraction scheme. We will return to this in
the future.

There is one more aspect that we can discuss now, namely
convergence. Due to the well known pointwise nature of the listed
limits, when using polarized splitting kernels, the convergence of the
cross section integrands will be pointwise. We can also assess the
rate of convergence. Assume that one of the variables
$x \in \{ \eta_1, \eta_2, \xi_1, \xi_2 \}$ is rescaled as $x \to
\kappa \; x$, $\kappa \to 0$, while the others remain fixed. If
$\kappa = 0$ implies $\hat\xi_1 = 0$ or $\hat\xi_2 = 0$ or $\hat\eta_1 =
\hat\eta_2$, but neither $\hat\eta_{1,2} = 0$ nor $\hat\eta_{1,2} =
1$, then, ignoring logarithmic enhancements,
\begin{equation}
\left[ \Sigma_{\cal O}^{(\cal S)} \right](\kappa \; x)  \approx \left[
  \Sigma_{\cal O}^{(\cal S)} \right](x) \; .
\end{equation}
On the other hand, if $\kappa = 0$ implies any of $\hat\eta_1 = 0$,
$\hat\eta_2 = 0$, $\hat\eta_1 = 1$, $\hat\eta_2 = 1$, then again up to
logarithmic enhancements
\begin{equation}
\label{eq:invsqrt}
\left[ \Sigma_{\cal O}^{(\cal S)} \right](\kappa \; x)  \approx
\frac{1}{\sqrt{\kappa}}\left[ \Sigma_{\cal O}^{(\cal S)} \right](x) \;. 
\end{equation}
This is the well known inverse square root behavior of collinear
limits. The lack of such a behavior in the case $\hat\eta_1 =
\hat\eta_2$ is due to the dependence of the relative angle parameter
$\eta_3$ on the difference $\hat\eta_1-\hat\eta_2$, which is quadratic
as seen in Eq.~(\ref{eq:eta3second}). Due to the iterative derivation
of the integrand, rescaling several variables leads to a scaling,
which can also be obtained iteratively from the above formulae. Let us
stress, that unless the given limit is a single-soft limit for a
final state quark, the unsubtracted integrand behaves as
\begin{equation}
\Sigma_{\cal O}^{(\cal S)}(\kappa \; x)  \approx
\frac{1}{\kappa}\Sigma_{\cal O}^{(\cal S)}(x) \;. 
\end{equation}
%


\subsection{Leading divergences and leading logs}
\label{sec:leading}

\vspace{.2cm}

We have stressed in the Introduction that one of the main ideas behind
our subtraction scheme is to avoid any non-trivial analytic
integration and perform the entire calculation purely numerically. It
is, however, advantageous to have at least some analytic formulae to
perform tests of the implementation. This was already our motivation
in deriving the volume of the phase space. It turns out that we can
also obtain the leading singularity in $\epsilon$ directly from the
subtraction terms. After all, the $1/\epsilon^4$ term corresponds to
$\eta_1 = \eta_2 = \xi_1 = \xi_2 = 0$, which means that the reduced
matrix element is that of the leading order process. The integral in
$\zeta$ is then trivial
\begin{equation}
\int_0^1 d\zeta \frac{1}{\sqrt{\zeta(1-\zeta)}} = \pi \; .
\end{equation}
What remains is the two-particle phase space of the leading order
cross section. In consequence, we obtain
\begin{eqnarray}
\label{eq:divgg}
\sigma^{RR}_{gg\rightarrow t\bar tgg} &=& 10C_A^2\frac{1}{\epsilon^4}
\left(\frac{\alpha_s}{4\pi}\right)^2 \sigma^{B}_{gg\rightarrow t\bar
  t} + {\cal O}\left(\frac{1}{\epsilon^3}\right) \; , \\ \label{eq:divqq}
\sigma^{RR}_{q\bar q\rightarrow t\bar tgg} &=& 2C_F(C_A+4C_F)\frac{1}{\epsilon^4}
\left(\frac{\alpha_s}{4\pi}\right)^2 \sigma^{B}_{q\bar q\rightarrow t\bar
  t} + {\cal O}\left(\frac{1}{\epsilon^3}\right) \; , \\
\sigma^{RR}_{gg\rightarrow t\bar tq\bar q} &=& {\cal
  O}\left(\frac{1}{\epsilon^3}\right) \; , \\ 
\sigma^{RR}_{q\bar q\rightarrow t\bar tq'\bar q'} &=& {\cal
  O}\left(\frac{1}{\epsilon^3}\right) \; ,
\end{eqnarray}
where $\sigma^{B}$ is the Born cross section for the two channels and
can be found in \ref{sec:born}. Later on, we will use these formulae
to test the normalization and precision of our numerical
calculation. At this point, we can, however, verify the cancellation
of the $1/\epsilon^4$ singularities in the inclusive top quark pair
production cross section, since the divergences of the other
contributions can be found in the literature. Indeed, we have
\begin{eqnarray}
\sigma^{VV}_{gg\rightarrow t\bar t} &=& (4C_A^2+4C_A^2)\frac{1}{\epsilon^4}
\left(\frac{\alpha_s}{4\pi}\right)^2 \sigma^{B}_{gg\rightarrow t\bar
  t} + {\cal O}\left(\frac{1}{\epsilon^3}\right) \; , \\
\sigma^{VV}_{q\bar q\rightarrow t\bar t} &=& (4C_F^2+4C_F^2)\frac{1}{\epsilon^4}
\left(\frac{\alpha_s}{4\pi}\right)^2 \sigma^{B}_{q\bar q\rightarrow t\bar
  t} + {\cal O}\left(\frac{1}{\epsilon^3}\right) \; .
\end{eqnarray}
The two color factors in the parentheses in both equations have a
different origin. One is given by the two-loop virtual corrections and
can be read off from the explicit results of \cite{Czakon:2007ej,
  Czakon:2007wk} in the high
energy limit (under the assumption that the leading divergence is
proportional to the exact Born matrix element). Otherwise, it can be
obtained from the complete divergence structure presented in
\cite{Ferroglia:2009ii}. The second color factor is given by the square of the
one-loop matrix element, and is easily obtained with the help of the
$\mathbf{I}$ operator from \cite{Catani:2000ef}.

We still need to derive the divergences of the real-virtual
corrections. This is much more difficult, because the structure of the
singular limits of one-loop amplitudes is not the same as that of
tree-level amplitudes. In other words, it is not enough to just take
the $1/\epsilon^2$ term from the one-loop amplitude using the
$\mathbf{I}$ operator and then use the same operator to obtain the
$1/\epsilon^2$ divergences from the phase space integration. The
situation seems to be even more involved, because this problem has never been
studied for hadronic heavy quark production, where the initial states
are also partons. Fortunately, the leading singularity is only due to
the purely massless states and has to factorize, and since it is due
to the soft and collinear limit, it is the same irrespective of
whether we consider initial or final states. In view of these
considerations, we can use the results obtained for jet cross sections
in $e^+e^-$ annihilation \cite{Somogyi:2006db}. Applying the general
formulae obtained there, we have
\begin{eqnarray}
\sigma^{RV}_{gg\rightarrow t\bar tg} &=& -18C_A^2\frac{1}{\epsilon^4}
\left(\frac{\alpha_s}{4\pi}\right)^2 \sigma^{B}_{gg\rightarrow t\bar
  t} + {\cal O}\left(\frac{1}{\epsilon^3}\right) \; , \\
\sigma^{RV}_{q\bar q\rightarrow t\bar tg} &=& -2C_F(C_A+8C_F)\frac{1}{\epsilon^4}
\left(\frac{\alpha_s}{4\pi}\right)^2 \sigma^{B}_{q\bar q\rightarrow t\bar
  t} + {\cal O}\left(\frac{1}{\epsilon^3}\right) \; .
\end{eqnarray}

Combining the three contributions, we have indeed
\begin{eqnarray}
\sigma^{\mbox{\scriptsize NNLO}}_{gg\rightarrow t\bar t+X} &=& {\cal
  O}\left(\frac{1}{\epsilon^3}\right) \; , \\ 
\sigma^{\mbox{\scriptsize NNLO}}_{q\bar q\rightarrow t\bar t+X} &=&
      {\cal O}\left(\frac{1}{\epsilon^3}\right) \; . 
\end{eqnarray}
Proving the cancellation of all divergences is much more difficult,
even though we known that the cross section is finite by the
Kinoshita-Lee-Nauenberg and the factorization theorems (the latter to
remove initial state collinear divergences). Let us only note, that it
is possible to write the coefficients of the $1/\epsilon^3$
singularities through the Born cross sections using convolutions in the worst
case. Nevertheless, we will refrain from this exercise.

The knowledge of the leading singularities can also be exploited in
another way. Indeed, it allows to predict the leading logarithms of
$\beta$. It is enough to know the $d$-dimensional behavior of the
phase space near threshold, which we have derived in
Eq.~(\ref{eq:betaphase}) as being $\beta^{-10\epsilon}$. Performing the
expansion down to the finite part, we obtain
\begin{eqnarray}
\label{eq:loggg}
\sigma^{RR}_{gg\rightarrow t\bar tgg} &=& \frac{12500}{3}C_A^2\log^4\beta
\left(\frac{\alpha_s}{4\pi}\right)^2 \sigma^{B}_{g\bar g\rightarrow t\bar
  t} + {\cal O}\left(\log^3\beta\right) \; , \\ \label{eq:logqq}
\sigma^{RR}_{q\bar q\rightarrow t\bar tgg} &=& \frac{2500}{3}C_F(C_A+4C_F)\log^4\beta
\left(\frac{\alpha_s}{4\pi}\right)^2 \sigma^{B}_{q\bar q\rightarrow t\bar
  t} + {\cal O}\left(\log^3\beta\right) \; , \\
\sigma^{RR}_{gg\rightarrow t\bar tq\bar q} &=& {\cal
  O}\left(\log^3\beta\right) \; , \\ 
\sigma^{RR}_{q\bar q\rightarrow t\bar tq'\bar q'} &=& {\cal
  O}\left(\log^3\beta\right) \; .
\end{eqnarray}
Notice that the coefficients are very large. In fact they are about an
order of magnitude larger than the coefficients of $\log^4\beta$ in
the total cross section \cite{Beneke:2009ye}. This will be reflected in the very
large values of the cross section near threshold.


\section{Implementation}
\label{sec:implementation}

\begin{figure}[ht]
\begin{center}
\includegraphics[width=1.0\textwidth]{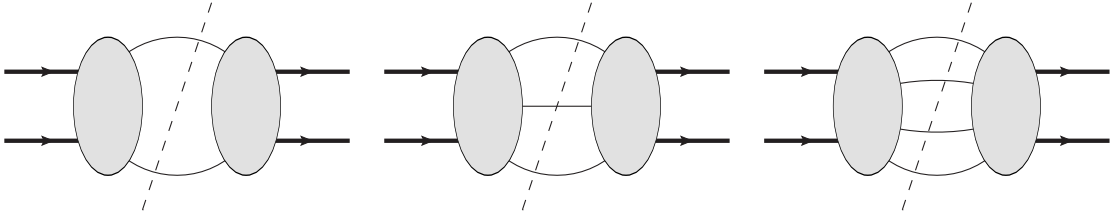}
\end{center}
\caption{Cut graph classes used for generation of tree level
  amplitudes squared. The external thick lines represent top quarks,
  the cut thin lines are gluons, ghosts and light quarks, whereas
  blobs contain trees. Color operators may be assigned to each visible
  line on the left of the cut. Moreover, cut gluon lines may contain
  spin correlators.}
\label{fig:cuts}
\end{figure}

The implementation of the subtraction scheme described in the previous
sections for the particular case of top quarks involves a large set of
tree level matrix elements, which can, moreover, be spin and color
correlated. There are many methods to evaluate these in four
dimensions. Nevertheless, we decided to work in conventional
dimensional regularization (CDR), thus keeping the full
$d$-dimensional dependence of the matrix elements. A simple way to
derive explicit expressions in such a case is to generate cut graphs
for forward scattering amplitudes with up to three loops. This
procedure based on Cutkosky rules leaves the freedom of the choice of
the external states, as any configuration can be obtained by
crossing. We decided to take the top quarks on the external lines of
the cut graphs. The tree-level amplitudes may have up to four external
gluons, which means that there will be a lot of graphs related by
symmetry (exchange of the gluon momenta). Taking the massless partons
as virtual in the cut graphs takes care of the symmetry, in the sense
that only one configuration is generated and one can symmetrize the
complete amplitudes at the end. Thus, our expressions are
substantially shorter than they would be, if we just squared
tree-level amplitudes. Our procedure is a simpler version of the old
approach of \cite{Ellis:1985er}, but we would prefer the latter if we only had
massless partons. The cut graph classes are shown in
Fig.~\ref{fig:cuts}. We insert color operators $\mathbf{T}_i$ to all
external lines on the left of the cut, with up to four operators in
total. We take the cut gluon propagators to be in the Feynman gauge
and compensate the gauge invariance violation by cut ghost lines. We
also insert spin correlators according to
\begin{equation}
\sum_\lambda \epsilon^\mu_\lambda \epsilon^{\nu \; *}_\lambda
\longrightarrow \frac{1}{2} \left( k^\mu_{\perp \; 1} k^\nu_{\perp \;
  2} + k^\nu_{\perp \; 1} k^\mu_{\perp \; 2} \right) \; ,
\end{equation}
where the symmetrized version is all that is required in practice, but
one may need two different transverse vectors as demonstrated in
\ref{sec:splitting}. As far as the color correlators are concerned, we
have exploited color conservation to reduce the number of needed
matrix elements. For an amplitudes with five partons, we have
\begin{equation}
\mathbf{T}_5 | {\cal M}_3 \rangle = -(\mathbf{T}_1 + \mathbf{T}_2 +
\mathbf{T}_3 + \mathbf{T}_4) | {\cal M}_3 \rangle \; .
\end{equation}
Moreover, since $\mathbf{T}_5^2 = C_A$ (for all the channels
considered) or $\mathbf{T}_5^2 = C_F$ (in general), we have only five
correlators to consider
\begin{equation}
\mathbf{T}_1 \cdot \mathbf{T}_2 , \; 
\mathbf{T}_1 \cdot \mathbf{T}_3 , \; 
\mathbf{T}_1 \cdot \mathbf{T}_4 , \; 
\mathbf{T}_2 \cdot \mathbf{T}_3 , \; 
\mathbf{T}_2 \cdot \mathbf{T}_4 \; .
\end{equation}
Similarly, in the case of four-parton amplitudes, we have the
following two double-correlators
\begin{equation}
\mathbf{T}_1 \cdot \mathbf{T}_2 , \; 
\mathbf{T}_1 \cdot \mathbf{T}_3 , \; 
\end{equation}
which may be additionally spin correlated. Amplitudes with four
partons require, however, also quadruple-correlators. At the level of
the amplitude
\begin{equation}
(\mathbf{T}_1 \cdot \mathbf{T}_2)(\mathbf{T}_1 \cdot \mathbf{T}_3) \; |
{\cal M}_2 \rangle \neq (\mathbf{T}_1 \cdot \mathbf{T}_3)(\mathbf{T}_1
\cdot \mathbf{T}_2) \; | {\cal M}_2 \rangle \; ,
\end{equation}
nevertheless, since the matrix elements squared are real, we have
\begin{eqnarray}
\langle {\cal M}_2 | (\mathbf{T}_1 \cdot \mathbf{T}_2)(\mathbf{T}_1
\cdot \mathbf{T}_3) \; | {\cal M}_2 \rangle &=& \langle {\cal M}_2 |
(\mathbf{T}_1 \cdot \mathbf{T}_2)(\mathbf{T}_1 \cdot \mathbf{T}_3) \;
| {\cal M}_2 \rangle^* \nonumber \\ &=&
\langle {\cal M}_2 | (\mathbf{T}_1 \cdot \mathbf{T}_3)^\dagger
(\mathbf{T}_1 \cdot \mathbf{T}_2)^\dagger \; | {\cal M}_2 \rangle
\nonumber \\ &=&
\langle {\cal M}_2 | (\mathbf{T}_1 \cdot \mathbf{T}_3)
(\mathbf{T}_1 \cdot \mathbf{T}_2) \; | {\cal M}_2 \rangle \; .
\end{eqnarray}
We thus only evaluate the following correlators
\begin{equation}
(\mathbf{T}_1 \cdot \mathbf{T}_2)(\mathbf{T}_1 \cdot \mathbf{T}_2) , \; 
(\mathbf{T}_1 \cdot \mathbf{T}_2)(\mathbf{T}_1 \cdot \mathbf{T}_3) , \; 
(\mathbf{T}_1 \cdot \mathbf{T}_3)(\mathbf{T}_1 \cdot \mathbf{T}_3) \; .
\end{equation}
Notice, that we have not exploited color conservation at the level
of unsymmetrized amplitudes. This could provide another minor speedup.

Due to the selector functions and sectors chosen, we are missing phase
space. Indeed, following Section~\ref{sec:decomposition}, we have
$k_1^z > 0$ and $k_1^0 > k_2^0$. For a given channel specified by
flavor assignments to the initial and final partons, the additional
contributions can be recovered by permuting initial and final states
independently. Nevertheless, we can use charge conjugation invariance
of QCD amplitudes together with rotation invariance, which allows to
swap $p_1 \leftrightarrow p_2$, to reduce the number of
configurations, which actually need to be evaluated. At the level of
six-parton amplitudes, we only need the five cases
\begin{equation}
gg \to t\bar tgg , \; 
gg \to t\bar tq\bar q , \; 
q\bar q \to t\bar tgg , \; 
q\bar q \to t\bar tq'\bar q' , \; 
q\bar q \to t\bar t\bar q'q' \; .
\end{equation}
We will not explicitly mention the last amplitude
anymore. Interestingly, however, its contribution is, for most cases,
very close to that with swapped quark and anti-quark. The complete
list of matrix elements is given in Tab.~\ref{tab:amplitudes}.

\begin{table}[t]
  \begin{center}
    \begin{tabular}{l|c|c||l|c|c}
      &&&&&\\
      & spin & color && spin & color \\
      amplitude & correlated & correlated & amplitude & correlated & correlated \\
      & lines & lines && lines & lines \\
      &&&&&\\
      \hline 
      $gg\to t\bar tgg$ &&& $q\bar q\to t\bar tgg$ && \\
      $gg\to t\bar tq\bar q$ &&& $q\bar q\to t\bar tq'\bar q'$ && \\
      $gg\to t\bar tg$ &&& $q\bar q\to t\bar t\bar q'q'$ && \\
      $gg\to t\bar tg$ & 1 && $q\bar q\to t\bar tg$ && \\
      $gg\to t\bar tg$ & 2 && $q\bar q\to t\bar tg$ & 5 & \\
      $gg\to t\bar tg$ & 5 && $q\bar q\to t\bar tg$ && (1,2) \\
      $gg\to t\bar tg$ && (1,2) & $q\bar q\to t\bar tg$ && (1,3) \\
      $gg\to t\bar tg$ && (1,3) & $q\bar q\to t\bar tg$ && (1,4) \\
      $gg\to t\bar tg$ && (1,4) & $q\bar q\to t\bar tg$ && (2,3) \\
      $gg\to t\bar tg$ && (2,3) & $q\bar q\to t\bar tg$ && (2,4) \\
      $gg\to t\bar tg$ && (2,4) & $q\bar q\to t\bar t$ && \\
      $gg\to t\bar t$ &&& $q\bar q\to t\bar t$ && (1,2) \\
      $gg\to t\bar t$ & 1 && $q\bar q\to t\bar t$ && (1,3) \\
      $gg\to t\bar t$ & 2 && $q\bar q\to t\bar t$ && (1,2)(1,2) \\
      $gg\to t\bar t$ & 1,2 && $q\bar q\to t\bar t$ && (1,2)(1,3) \\
      $gg\to t\bar t$ && (1,2) & $q\bar q\to t\bar t$ && (1,3)(1,3) \\
      $gg\to t\bar t$ && (1,3) & && \\
      $gg\to t\bar t$ && (1,2)(1,2) & $\bar qq\to t\bar t$ && \\
      $gg\to t\bar t$ && (1,2)(1,3) & && \\
      $gg\to t\bar t$ && (1,3)(1,3) & $gq\to t\bar tq$ && \\
      $gg\to t\bar t$ & 1 & (1,2) & $qg\to t\bar tq$ && \\
      $gg\to t\bar t$ & 1 & (1,3) & $\bar qg\to t\bar t\bar q$ && \\
    \end{tabular}
  \end{center}
  \caption{\label{tab:amplitudes} The 42 spin and color correlated
    amplitudes with four, five and six partons, which are needed for
    the four main channels of top quark pair production.}
\end{table}

\begin{table}[t]
  \begin{center}
    \begin{tabular}{l|c|c|c|c}
              &&&&\\
              & matrix  & using       & including   & in quadruple \\
      process & element & {\sc Helac} & subtraction & precision \\
              & [msec]  & [msec]      & [msec]      & [msec] \\
              &&&&\\
      \hline 
              &&&&\\
      $gg \rightarrow t\bar tgg$ & 13 & 53 & 18 & 450 \\
      &&&&\\
      \hline
      &&&&\\
      $q\bar q \rightarrow t\bar tgg$ & 0.71 & 6.5 & 0.81 & 27 \\
      &&&&\\
      \hline
      &&&&\\
      $gg \rightarrow t\bar tq\bar q$ & 0.71 & 6.3 & 0.97 & 32 \\
      &&&&\\
      \hline
      &&&&\\
      $q\bar q \rightarrow t\bar tq'\bar q'$ & 0.015 & 0.52 & 0.041 & 1.5 \\
      &&&&\\
    \end{tabular}
  \end{center}
  \caption{\label{tab:timings} Single phase space point timings for
    the evaluation of the matrix element with or without
    subtraction. The values in the fourth and fifth columns are the
    worst timings of all the seven sectors. The matrix elements of the
    present implementation have been compiled without
    optimization, because of the size of the expressions. Quadruple
    precision is obtained using the native implementation of the Intel
    Fortran compiler.}
\end{table}

It is interesting to measure the evaluation time of the integrands for
a single phase space point. The results of such a measurement are
shown in Tab.~\ref{tab:timings}. The main point of this table is to
show that the subtraction and integrated subtraction terms are faster
in evaluation than the matrix element of the double-real radiation
process. This is true in all the cases, but the simplest involving six
quarks. The latter has a very short expression for the amplitude, and
it is simply impossible to have even simpler subtraction terms. One
could, of course, suppose that the relative efficiency of the
subtraction scheme is simply due to the very inefficient
implementation of the six parton matrix elements themselves. That this
is not the case is shown by comparing our implementation with that of
\textsc{Helac} \cite{Kanaki:2000ey, Cafarella:2007pc}. Reversing the
argument, one could suppose that 
\textsc{Helac} is inefficient, since our matrix elements have been
obtained in the most naive way, and have, moreover, been compiled
without optimization. This is not true in practice, as \textsc{Helac}
uses helicity sampling (or rather random polarization vectors), and
has not been optimized for spin summed amplitudes, which we need
here. It would certainly be advantageous to have an implementation of
our scheme allowing for helicity sampling, as it was done in \cite{Czakon:2009ss}
at NLO. This requires, however, a tremendous effort, which we leave
for the future. In Tab.~\ref{tab:timings}, we have also quoted timings
for computations in quadruple precision. In fact, we have used quadruple
precision for all the values presented in the next section. Due to the
cancellations inherent in subtraction schemes, there is always a risk
of numerical instabilities. In this first study, we have not made any
analysis in this direction, and decided to avoid the problem
altogether using higher precision. This is certainly an issue, which
requires improvements. We would also like to point out that the
implementation of quadruple precision of the Intel Fortran
compiler, which we used, is rather inefficient and we could have
gained a speedup factor of at least tree by switching to an external
library.

Let us now discuss our implementation of the phase space
integration. The evaluation of the cross sections requires seven- and
eight-dimensional integrals, corresponding to the two contributions
from the two-particle phase space Eq.~(\ref{eq:phi2splitting}). The
integrands are indeed integrable, but not free of singularities. Besides
logarithmic singularities due to the $\epsilon$-expansion, there are
inverse square root singularities as shown in
Eq.~(\ref{eq:invsqrt}). Our strategy is to use adaptive Monte Carlo
integration techniques to improve convergence. We do not even bother
with remappings for the square roots, which could perhaps help, but
would be against our approach of avoiding any complications, unless,
well, unavoidable. Thus, we use \textsc{Parni}
\cite{vanHameren:2007pt} to take care of all 
singularities. The seven sectors are evaluated with the same program
in one run, and we even have implemented variance optimization {\it \`a
  la} stratified sampling, but we have not used this feature for
the results presented here. The main computing time goes into the
seven-dimensional integrals, as they involve the six-parton
amplitudes. The eight-dimensional integrals have smaller contributions
and are faster per phase space point. They are thus negligible as far
as resource requirements are concerned.

There is one more issue connected to numerics that we need to address,
and that cannot be solved
with higher precision. Inherent to subtraction is the fact that the
integrands involve a cancellation of many digits close to
singularities between the matrix elements and their approximations. If
there were no inverse square roots, this might not have been a
problem, but high precision of the results requires evaluation close
to singularities. We thus need to cutoff the phase space to avoid
instabilities. The form of the cutoff condition can be chosen at will,
but since the matrix elements grow as $1/\eta_1\eta_2\xi_1\xi_2$, when
the sector variables become small, we have decided to require the
following
\begin{equation}
\label{eq:cutoff}
\eta_1 \; \eta_2 \; \xi_1 \; \xi_2 > \Delta \; .
\end{equation}
The values we chose for $\Delta$ are given in the next section. Here,
we evaluate the size the missing phase space due to this condition. In
fact, one can show that
\begin{equation}
\delta \Phi_n(\Delta) = \int_0^1 \prod_{i=1}^n \; d \alpha_i \;
\theta\left(\prod_{j=1}^n \alpha_j < \Delta\right) = 
\Delta \sum_{i=0}^{n-1} \frac{1}{i!}
\log^i\left(\frac{1}{\Delta}\right) \; .
\end{equation}
In our case $n=4$. The size of the missing phase space cannot be
used to estimate the implied error on the total cross section, due to
the presence of square roots and logarithms. In practice, one
evaluates the cross section at two values of the cutoff, say $\Delta$
and $\Delta'$, with $\Delta' < \Delta$. It is expected that the
improvement in the error will be of the order of
$\sqrt{\Delta/\Delta'}$.

The final issue we would like to discuss is that of testing. In a
project this size, it is important to check correctness at all
stages. We have performed the following checks
\begin{enumerate}

\item \framebox{Phase space volume} We have evaluated the volume
  analytically in Section~\ref{sec:volume}. By setting the matrix
  elements to unity, we can obtain a numeric result with our
  implementation. For example, with
  \begin{equation}
    s = 2 \; , \;\;\;\; m = \frac{2}{3} \; , \;\;\;\; \mu =
    \frac{1}{3} \; ,
  \end{equation}
  the exact value reads
  \begin{eqnarray}
    \left( \frac{\mu^2 e^{\gamma_E}}{4\pi}
    \right)^{3\epsilon} \int d\Phi_4 &=& \Big( 
    5.9719120
    +87.151375 \; \epsilon
    +630.84755 \; \epsilon^2 \nonumber \\ &&
    +3019.2212 \; \epsilon^3
    +10746.200 \; \epsilon^4
    \Big) \times 10^{-12} \; ,
  \end{eqnarray}
  whereas 10 000 000 generated points suffice to obtain
  \begin{eqnarray}
    \left( \frac{\mu^2 e^{\gamma_E}}{4\pi}
    \right)^{3\epsilon} \int d\Phi_4 &\approx& \Big(
    (5.9697 \pm 0.0028)
    +(87.131 \pm 0.033) \; \epsilon
    +(630.77 \pm 0.18) \; \epsilon^2 \nonumber \\ &&
    +(3019.2 \pm 0.64) \; \epsilon^3
    +(10747 \pm 2.1) \; \epsilon^4
    \Big) \times 10^{-12} \; .
  \end{eqnarray}

\vspace{.2cm}
\item \framebox{$\epsilon$-contribution to the two-particle phase
  space} Section~\ref{sec:volume} contains a result for the integral
  of $(k_1 \cdot q_1)^2$. This integral is the simplest object
  sensitive to $d\Phi_2^{(\epsilon)}$. Unfortunately, the contribution
  is tiny at best. The analytic result for $s=1, m^2=\mu^2=2/9$ is
  \begin{eqnarray}
    \label{eq:value1}
    \frac{\mu^{6\epsilon}}{P_4(s,\epsilon)} \int d\Phi_4 \left(
    \frac{k_1 \cdot q_1}{s} \right)^2 &=& \Big( 
    1.4553533
    +10.106976 \; \epsilon
    +34.956382 \; \epsilon^2 \nonumber \\ &&
    +80.126733 \; \epsilon^3
    +136.49975 \; \epsilon^4
    \Big) \times 10^{-9} \; ,
  \end{eqnarray}
  whereas with very high statistics, we have
  \begin{eqnarray}
    \frac{\mu^{6\epsilon}}{P_4(s,\epsilon)} \int
    d\Phi_4^{(d|\epsilon)} \left( \frac{k_1 \cdot q_1}{s} \right)^2
    &=& \Big(
    (1.45536 \pm 0.000014)
    +(10.1056 \pm 0.000085) \; \epsilon \nonumber \\ &&
    +(34.9482 \pm 0.00035) \; \epsilon^2
    +(80.1058 \pm 0.0017) \; \epsilon^3 \nonumber \\ &&
    +(136.473 \pm 0.0084) \; \epsilon^4
    \Big) \times 10^{-9} \; , \\
    \frac{\mu^{6\epsilon}}{P_4(s,\epsilon)} \int
    d\Phi_4^{(\epsilon)} \left( \frac{k_1 \cdot q_1}{s} \right)^2
    &=& \Big(
    (0.001413 \pm 0.000011) \; \epsilon
    +(0.008145 \pm 0.000056) \; \epsilon^2 \nonumber \\ &&
    +(0.02064 \pm 0.00013) \; \epsilon^3
    +(0.02469 \pm 0.00020) \; \epsilon^4
    \Big) \times 10^{-9} \; . \nonumber \\
  \end{eqnarray}
  Clearly, there is disagreement between the $d\Phi_4^{(d|\epsilon)}$
  values and Eq.~(\ref{eq:value1}) starting from order
  $\epsilon$. Besides the highest order, $\epsilon^4$, the 
  difference is at least at the level of $10\sigma$. Together with the
  $\epsilon$-contribution we obtain, however,
  \begin{eqnarray}
    \frac{\mu^{6\epsilon}}{P_4(s,\epsilon)} \int
    d\Phi_4 \left( \frac{k_1 \cdot q_1}{s} \right)^2
    &=& \Big(
    (1.45536 \pm 0.000014)
    +(10.1070 \pm 0.000086) \; \epsilon \nonumber \\ &&
    +(34.9564 \pm 0.00035) \; \epsilon^2
    +(80.1265 \pm 0.0017) \; \epsilon^3 \nonumber \\ &&
    +(136.498 \pm 0.0084) \; \epsilon^4
    \Big) \times 10^{-9} \; .
  \end{eqnarray}

\vspace{.2cm}
\item \framebox{Leading order cross sections} With a jet function that
  forces all massless partons to be separated, we obtain the leading
  order cross section for top quark pair production in association
  with two jets. We use the following simple setup
  \begin{equation}
    k_{1,2} \cdot p_{1,2} > 10^{-4} \; s \; , \; k_1 \cdot k_2 >
    10^{-6} \; s \; ,
  \end{equation}
  together with
  \begin{equation}
    E_{\text{CM}} = 400 \; \mbox{GeV} \; , \; m_t = 172.6 \; \mbox{GeV} \; ,
    \; \alpha_s(m_t) = 0.107639510785815 \; .
  \end{equation}
  We obtain
  \begin{eqnarray}
    \sigma_{gg \to t\bar tgg} &=& (7.75 \pm 0.018) \times 10^{-2} \; \mbox{nb} \; , \\
    \sigma_{gg \to t\bar tq\bar q} &=& (4.81 \pm 0.019) \times 10^{-4}
    \; \mbox{nb} \; , \\
    \sigma_{q\bar q \to t\bar tgg} &=& (2.86 \pm 0.0065) \times
    10^{-2} \; \mbox{nb} \; , \\
    \sigma_{q\bar q \to t\bar tq'\bar q'} &=& (3.55 \pm 0.010) \times
    10^{-4} \; \mbox{nb} \; ,
  \end{eqnarray}
  whereas \textsc{Helac} gives
  \begin{eqnarray}
    \sigma_{gg \to t\bar tgg} &=& (7.76 \pm 0.039) \times 10^{-2} \; \mbox{nb} \; , \\
    \sigma_{gg \to t\bar tq\bar q} &=& (4.77 \pm 0.025) \times 10^{-4}
    \; \mbox{nb} \; , \\
    \sigma_{q\bar q \to t\bar tgg} &=& (2.88 \pm 0.0078) \times
    10^{-2} \; \mbox{nb} \; , \\
    \sigma_{q\bar q \to t\bar tq'\bar q'} &=& (3.54 \pm 0.019) \times
    10^{-4} \; \mbox{nb} \; .
  \end{eqnarray}
  We point out that the results have been obtained with 1 000 000
  Monte Carlo events, and our phase space parameterization seems to be
  more efficient than that of \textsc{Helac}.

  We have also checked our matrix elements
  for the six-parton processes at chosen phase space points against
  those obtained with \textsc{Helac}. We have reached agreement within
  expected numerical accuracy.

\vspace{.2cm}
\item \framebox{Pointwise convergence in $d$-dimensions} We can test
  the scaling given in Section~\ref{sec:subtraction} in all the
  limits, while expanding the results up to $\epsilon^4$. This has the
  advantage of showing that the subtraction terms are correct to all
  orders in $\epsilon$, since the highest order of expansion of the
  matrix elements is $\epsilon^4$. The further terms of the expansion
  of $\left[ \Sigma_{\cal O}^{(\cal S)} \right]$ follow from the
  $\epsilon$-dependence of the measure. We have evaluated all the
  fifteen limits of the seven sectors of the four main processes,
  which amounts to 420 cases. Needless to say, we obtained the correct
  behavior. To illustrate the tests, we only give one case, which
  sector ${\cal S}^I_4$ of the process $gg \to t\bar tgg$ for the
  following configuration
  \begin{equation}
    s = 1 \; , \;
    \beta = \frac{1}{2} \; , \;
    \zeta = \frac{1}{3} \; , \;
    \eta_1 = \frac{1}{4} \; , \;
    \eta_2 = \frac{1}{5} \times 10^{-40} \; , \;
    \xi_1 = \frac{1}{6} \; , \;
    \xi_2 = \frac{1}{7} \times 10^{-40} \; , \; \nonumber
  \end{equation}
  \begin{equation}
    \cos\theta_Q = \frac{1}{8} \; , \;
    \phi_Q = \frac{1}{9} \; , \;
    \cos\rho_Q = 1 \; .
  \end{equation}
  The results quoted correspond to $g_s = 1, \mu = m$, and we did not
  include the factors for spin and color averages, identical final
  states and flux. Without subtraction the integrand is
  \begin{eqnarray}
    \Sigma^{({\cal S}^I_4)} &=& 
      2.92\times 10^{79}
      +1.14003\times 10^{82} \; \epsilon
      +2.22528\times 10^{84} \; \epsilon^2 \nonumber \\ &&
      +2.89548\times 10^{86} \; \epsilon^3
      +2.82539\times 10^{88} \; \epsilon^4 \; .
  \end{eqnarray}
  The size of the coefficients is due to the expected growth of the
  integrand proportional to the inverse of the product of the four
  sector variables, which in this case is of the order $10^{80}$. Once
  subtraction and integrated subtraction terms are included, the
  integrand becomes
  \begin{eqnarray}
    \left[ \Sigma^{({\cal S}^I_4)} \right] &=& 
    9.98769\times 10^{-6} \; \frac{1}{\epsilon^4}
    +0.0000869195 \; \frac{1}{\epsilon^3}
    +0.00676374 \; \frac{1}{\epsilon^2}
    +0.641465 \; \frac{1}{\epsilon} \nonumber \\ &&
    +40.9821
    +1961.23 \; \epsilon
    +75083.4 \; \epsilon^2
    +2.39952\times 10^6 \; \epsilon^3
    +6.61228\times 10^7 \; \epsilon^4 \; . \nonumber \\
  \end{eqnarray}
  We observe no growth of the coefficients, since the rescaling of
  $\eta_2$ and $\xi_2$ corresponds to the limit $\hat\eta_1 =
  \hat\eta_2$ and $\hat\xi_2 = 0$. Amongst all terms in $\left[
    \Sigma^{({\cal S}^I_4)} \right]$, there is one that cancels the
  divergence of $\Sigma^{({\cal S}^I_4)}$ alone, and contains the
  limit of $\mathfrak{M}_\mathcal{S}$ at $\hat\eta_1 = \hat\eta_2$ and
  $\hat\xi_2 = 0$. We can compare it order by order in $\epsilon$ to $
  \Sigma^{({\cal S}^I_4)}$. The result is
  \begin{eqnarray}
    \sum_{i=0}^4 \left( 1-\frac{ \Sigma^{({\cal S}^I_4)}_\text{approx} |_{\epsilon^i}
    }{ \Sigma^{({\cal S}^I_4)} |_{\epsilon^i} }\right) \epsilon^i &=& 
    -2.68 \times 10^{-41}
    -2.68 \times 10^{-41} \; \epsilon
    -2.68 \times 10^{-41} \; \epsilon^2 \nonumber \\ &&
    -2.67 \times 10^{-41} \; \epsilon^3
    -2.67 \times 10^{-41} \; \epsilon^4 \; .
  \end{eqnarray}
  The very high numerical precision of these tests is important,
  because minor mistakes can sometimes show, for example, at the eight
  digit and would be impossible to find with a double precision
  implementation in Fortran. The above numbers have been obtained with
  an arbitrary precision implementation in \textsc{Mathematica}. We
  have also compared quadruple precision results from Fortran with
  those of \textsc{Mathematica} at ordinary points making sure that
  both implementations agree.

\end{enumerate}
Further tests, most notably the agreement of the numerical values for
the leading divergence with the analytic formulae from
Section~\ref{sec:leading}, as well as cutoff independence are described in
the next section.

Finally, let us point out that this project has only been possible
thanks to the use of complicated external software systems. These are
listed in \ref{sec:software}.


\section{Results}

\begin{table}[ht]
  \begin{center}
    \begin{tabular}{c|ll|ll}
&&&& \\ $\beta$ && $\epsilon^{-3}$ && $\epsilon^{-2}$  \\ &&&& \\ \hline
0.001 & $+3.501\times 10^{\text{-3}}$ & $\pm \;\;\;\;1.7\times 10^{\text{-6}}$ & $+1.245\times 10^{\text{-1}}$ & $\pm \;\;\;\;6.1\times 10^{\text{-5}}$\\
0.025 & $+4.796\times 10^{\text{-2}}$ & $\pm \;\;\;\;1.7\times 10^{\text{-5}}$ & $+9.312\times 10^{\text{-1}}$ & $\pm \;\;\;\;3.3\times 10^{\text{-4}}$\\
0.075 & $+1.026\times 10^{\text{-1}}$ & $\pm \;\;\;\;3.9\times 10^{\text{-5}}$ & $+1.422\times 10^0$ & $\pm \;\;\;\;5.2\times 10^{\text{-4}}$\\
0.125 & $+1.377\times 10^{\text{-1}}$ & $\pm \;\;\;\;5.7\times 10^{\text{-5}}$ & $+1.544\times 10^0$ & $\pm \;\;\;\;6.0\times 10^{\text{-4}}$\\
0.175 & $+1.603\times 10^{\text{-1}}$ & $\pm \;\;\;\;6.8\times 10^{\text{-5}}$ & $+1.512\times 10^0$ & $\pm \;\;\;\;6.2\times 10^{\text{-4}}$\\
0.225 & $+1.731\times 10^{\text{-1}}$ & $\pm \;\;\;\;7.5\times 10^{\text{-5}}$ & $+1.393\times 10^0$ & $\pm \;\;\;\;6.1\times 10^{\text{-4}}$\\
0.275 & $+1.777\times 10^{\text{-1}}$ & $\pm \;\;\;\;8.0\times 10^{\text{-5}}$ & $+1.225\times 10^0$ & $\pm \;\;\;\;5.6\times 10^{\text{-4}}$\\
0.325 & $+1.750\times 10^{\text{-1}}$ & $\pm \;\;\;\;8.2\times 10^{\text{-5}}$ & $+1.025\times 10^0$ & $\pm \;\;\;\;5.1\times 10^{\text{-4}}$\\
0.375 & $+1.662\times 10^{\text{-1}}$ & $\pm \;\;\;\;8.0\times 10^{\text{-5}}$ & $+8.134\times 10^{\text{-1}}$ & $\pm \;\;\;\;4.5\times 10^{\text{-4}}$\\
0.425 & $+1.520\times 10^{\text{-1}}$ & $\pm \;\;\;\;7.6\times 10^{\text{-5}}$ & $+6.006\times 10^{\text{-1}}$ & $\pm \;\;\;\;3.9\times 10^{\text{-4}}$\\
0.475 & $+1.332\times 10^{\text{-1}}$ & $\pm \;\;\;\;7.9\times 10^{\text{-5}}$ & $+3.978\times 10^{\text{-1}}$ & $\pm \;\;\;\;3.9\times 10^{\text{-4}}$\\
0.525 & $+1.108\times 10^{\text{-1}}$ & $\pm \;\;\;\;6.4\times 10^{\text{-5}}$ & $+2.132\times 10^{\text{-1}}$ & $\pm \;\;\;\;6.5\times 10^{\text{-4}}$\\
0.575 & $+8.609\times 10^{\text{-2}}$ & $\pm \;\;\;\;5.6\times 10^{\text{-5}}$ & $+5.670\times 10^{\text{-2}}$ & $\pm \;\;\;\;2.9\times 10^{\text{-4}}$\\
0.625 & $+5.967\times 10^{\text{-2}}$ & $\pm \;\;\;\;4.8\times 10^{\text{-5}}$ & $-6.833\times 10^{\text{-2}}$ & $\pm \;\;\;\;2.7\times 10^{\text{-4}}$\\
0.675 & $+3.315\times 10^{\text{-2}}$ & $\pm \;\;\;\;4.1\times 10^{\text{-5}}$ & $-1.548\times 10^{\text{-1}}$ & $\pm \;\;\;\;3.2\times 10^{\text{-4}}$\\
0.725 & $+7.890\times 10^{\text{-3}}$ & $\pm \;\;\;\;3.6\times 10^{\text{-5}}$ & $-1.983\times 10^{\text{-1}}$ & $\pm \;\;\;\;2.5\times 10^{\text{-4}}$\\
0.775 & $-1.472\times 10^{\text{-2}}$ & $\pm \;\;\;\;4.0\times 10^{\text{-5}}$ & $-1.974\times 10^{\text{-1}}$ & $\pm \;\;\;\;2.2\times 10^{\text{-4}}$\\
0.825 & $-3.278\times 10^{\text{-2}}$ & $\pm \;\;\;\;3.9\times 10^{\text{-5}}$ & $-1.540\times 10^{\text{-1}}$ & $\pm \;\;\;\;1.8\times 10^{\text{-4}}$\\
0.875 & $-4.402\times 10^{\text{-2}}$ & $\pm \;\;\;\;4.0\times 10^{\text{-5}}$ & $-7.173\times 10^{\text{-2}}$ & $\pm \;\;\;\;1.5\times 10^{\text{-4}}$\\
0.925 & $-4.497\times 10^{\text{-2}}$ & $\pm \;\;\;\;3.2\times 10^{\text{-5}}$ & $+3.532\times 10^{\text{-2}}$ & $\pm \;\;\;\;1.3\times 10^{\text{-4}}$\\
0.975 & $-2.793\times 10^{\text{-2}}$ & $\pm \;\;\;\;1.5\times 10^{\text{-5}}$ & $+1.181\times 10^{\text{-1}}$ & $\pm \;\;\;\;8.3\times 10^{\text{-5}}$\\
0.999 & $-2.596\times 10^{\text{-3}}$ & $\pm \;\;\;\;2.0\times 10^{\text{-6}}$ & $+3.140\times 10^{\text{-2}}$ & $\pm \;\;\;\;2.2\times 10^{\text{-5}}$\\
    \end{tabular}
  \end{center}
  \caption{\label{tab:qqttgg1} Coefficients of the Laurent expansion of the
    $f_{q\bar q \rightarrow t\bar tgg}$ function.}
\end{table}
\begin{table}[ht]
  \begin{center}
    \begin{tabular}{c|ll|ll}
&&&& \\ $\beta$ && $\epsilon^{-1}$ && $\epsilon^{0}$  \\ &&&& \\ \hline
0.001 & $+2.947\times 10^0$ & $\pm \;\;\;\;1.4\times 10^{\text{-3}}$ & $+5.218\times 10^1$ & $\pm \;\;\;\;2.9\times 10^{\text{-2}}$\\
0.025 & $+1.196\times 10^1$ & $\pm \;\;\;\;4.4\times 10^{\text{-3}}$ & $+1.143\times 10^2$ & $\pm \;\;\;\;4.6\times 10^{\text{-2}}$\\
0.075 & $+1.293\times 10^1$ & $\pm \;\;\;\;4.8\times 10^{\text{-3}}$ & $+8.660\times 10^1$ & $\pm \;\;\;\;3.6\times 10^{\text{-2}}$\\
0.125 & $+1.127\times 10^1$ & $\pm \;\;\;\;4.5\times 10^{\text{-3}}$ & $+5.975\times 10^1$ & $\pm \;\;\;\;3.0\times 10^{\text{-2}}$\\
0.175 & $+9.161\times 10^0$ & $\pm \;\;\;\;4.1\times 10^{\text{-3}}$ & $+3.949\times 10^1$ & $\pm \;\;\;\;2.9\times 10^{\text{-2}}$\\
0.225 & $+7.062\times 10^0$ & $\pm \;\;\;\;4.3\times 10^{\text{-3}}$ & $+2.454\times 10^1$ & $\pm \;\;\;\;2.6\times 10^{\text{-2}}$\\
0.275 & $+5.149\times 10^0$ & $\pm \;\;\;\;3.1\times 10^{\text{-3}}$ & $+1.375\times 10^1$ & $\pm \;\;\;\;1.9\times 10^{\text{-2}}$\\
0.325 & $+3.469\times 10^0$ & $\pm \;\;\;\;2.7\times 10^{\text{-3}}$ & $+6.118\times 10^0$ & $\pm \;\;\;\;1.6\times 10^{\text{-2}}$\\
0.375 & $+2.063\times 10^0$ & $\pm \;\;\;\;2.4\times 10^{\text{-3}}$ & $+9.998\times 10^{\text{-1}}$ & $\pm \;\;\;\;1.5\times 10^{\text{-2}}$\\
0.425 & $+9.289\times 10^{\text{-1}}$ & $\pm \;\;\;\;2.1\times 10^{\text{-3}}$ & $-2.245\times 10^0$ & $\pm \;\;\;\;1.3\times 10^{\text{-2}}$\\
0.475 & $+7.292\times 10^{\text{-2}}$ & $\pm \;\;\;\;2.2\times 10^{\text{-3}}$ & $-3.973\times 10^0$ & $\pm \;\;\;\;1.3\times 10^{\text{-2}}$\\
0.525 & $-5.197\times 10^{\text{-1}}$ & $\pm \;\;\;\;2.7\times 10^{\text{-3}}$ & $-4.575\times 10^0$ & $\pm \;\;\;\;1.1\times 10^{\text{-2}}$\\
0.575 & $-8.661\times 10^{\text{-1}}$ & $\pm \;\;\;\;1.7\times 10^{\text{-3}}$ & $-4.393\times 10^0$ & $\pm \;\;\;\;1.1\times 10^{\text{-2}}$\\
0.625 & $-9.995\times 10^{\text{-1}}$ & $\pm \;\;\;\;1.6\times 10^{\text{-3}}$ & $-3.711\times 10^0$ & $\pm \;\;\;\;8.9\times 10^{\text{-3}}$\\
0.675 & $-9.525\times 10^{\text{-1}}$ & $\pm \;\;\;\;1.5\times 10^{\text{-3}}$ & $-2.772\times 10^0$ & $\pm \;\;\;\;8.9\times 10^{\text{-3}}$\\
0.725 & $-7.695\times 10^{\text{-1}}$ & $\pm \;\;\;\;1.9\times 10^{\text{-3}}$ & $-1.823\times 10^0$ & $\pm \;\;\;\;8.5\times 10^{\text{-3}}$\\
0.775 & $-5.075\times 10^{\text{-1}}$ & $\pm \;\;\;\;1.1\times 10^{\text{-3}}$ & $-1.023\times 10^0$ & $\pm \;\;\;\;5.6\times 10^{\text{-3}}$\\
0.825 & $-2.310\times 10^{\text{-1}}$ & $\pm \;\;\;\;8.5\times 10^{\text{-4}}$ & $-5.028\times 10^{\text{-1}}$ & $\pm \;\;\;\;4.5\times 10^{\text{-3}}$\\
0.875 & $-2.577\times 10^{\text{-2}}$ & $\pm \;\;\;\;6.8\times 10^{\text{-4}}$ & $-3.085\times 10^{\text{-1}}$ & $\pm \;\;\;\;3.5\times 10^{\text{-3}}$\\
0.925 & $+1.082\times 10^{\text{-3}}$ & $\pm \;\;\;\;5.2\times 10^{\text{-4}}$ & $-2.996\times 10^{\text{-1}}$ & $\pm \;\;\;\;2.6\times 10^{\text{-3}}$\\
0.975 & $-2.722\times 10^{\text{-1}}$ & $\pm \;\;\;\;3.0\times 10^{\text{-4}}$ & $+2.370\times 10^{\text{-1}}$ & $\pm \;\;\;\;1.5\times 10^{\text{-3}}$\\
0.999 & $-2.158\times 10^{\text{-1}}$ & $\pm \;\;\;\;1.9\times 10^{\text{-4}}$ & $+9.970\times 10^{\text{-1}}$ & $\pm \;\;\;\;1.5\times 10^{\text{-3}}$\\
    \end{tabular}
  \end{center}
  \caption{\label{tab:qqttgg2} Coefficients of the Laurent expansion of the
    $f_{q\bar q \rightarrow t\bar tgg}$ function.}
\end{table}

We are now ready to present our numerical results for the Laurent
expansions of the cross sections. We turn to dimensionless functions
of the velocity $\beta$ with the help of the following definition
\begin{equation}
\sigma^{RR}_{ab\rightarrow t\bar tcd}(s,m^2,\mu^2 = m^2,\alpha_s,\epsilon) =
\frac{\alpha_s^4}{m^2} f_{ab\rightarrow t\bar tcd}(\beta,\epsilon) \; , 
\end{equation}
where $a,b$ are initial, and $c,d$ final state massless
partons. As shown on the left hand side above, we only give
values for the case $\mu = m$, since the dependence on the scale can
be obtained from renormalization group equations. The $f$ functions
admit the following expansions 
\begin{equation}
f_{ab\rightarrow t\bar tcd}(\beta,\epsilon) = \sum_{i=-4}^{0}
\epsilon^{i} f^{(i)}_{ab\rightarrow t\bar tcd}(\beta) + {\cal O}(\epsilon)\; .
\end{equation}

We sample the functions at twenty equidistant points within the $\beta$
variation range $[0,1]$
\begin{equation}
\beta_i = \frac{2i-1}{40} \; , \;\;\;\; i=1,...,20 \; .
\end{equation}
This is most probably sufficient to obtain a decent fitting
function as has been done in the classic paper \cite{Nason:1987xz}. To this, we add
one point very close to threshold, $\beta = 0.001$, and one very close
to the infinite energy limit $\beta = 0.999$. As we will demonstrate,
the study of the latter limit will require a denser sampling in the case
of gluonic initial states, even though this is not immediately relevant
to phenomenology due to the limited $\beta$ range of current colliders
producing top quarks.

As we are interested in total cross sections at first, we use a
trivial jet function equal to one. For our final results, we use a
cutoff of $\Delta = 10^{-7}$. This implies a missing volume of 
\begin{equation}
\delta \Phi_4(10^{-7}) = 8.4 \times 10^{-5} \; .
\end{equation}
In order to test the independence of the results from the cutoff, we
will also use a higher value of $\Delta = 10^{-6}$, which amounts to
\begin{equation}
\delta \Phi_4(10^{-6}) = 5.5 \times 10^{-4} \; .
\end{equation}
These numbers can only be used as estimates of the relative
integration errors implied by the cutoff in the non-singular case as
discussed in Section~\ref{sec:implementation}. Nevertheless, the
improvement obtained by lowering the cutoff by an order of magnitude
should amount to a factor of three, which will allow us to provide
realistic estimates of the quality of the final results. In fact, we
will aim at a situation, in which we will be restricted by the
integration error rather than by the cutoff.

For each cross section and each value of $\beta$, our simulations are
performed with a total sample of 10 000 000 generated Monte Carlo
events. This number is very close to the number of accepted events,
since the only restriction on the phase space is given by the small
cutoffs. The quality of the obtained results with this sample is
discussed in the following. Nevertheless, we justify its fixed size by
the purpose of this publication, which is to prove the usefulness of our
approach and provide relevant numbers, while not necessarily giving the
highest quality estimates. The latter exercise is left for a future
publication containing complete cross sections, and not only the
double-real contribution. We stress already here, that there is always
a risk of underestimated errors with complicated integrands and low
relative errors. Therefore, our results can probably only be trusted
up to an additional factor multiplying the quoted errors. A safe
bet with no special justification would be a factor of two. This
is unavoidable, as practice shows, and the only way not to have to
worry about the errors is to substantially increase the statistics to
the point, where the estimated precision will vastly surpass the
practical requirements. This will of course be done in the mentioned
future publications.

Finally, we re-stress that all computations have been performed in
quadruple precision, in order to remove at least one source of
concern, namely numerical instabilities. Using higher precision is by
itself not yet reassuring enough, but two calculations with two different
cutoffs and agreement within expectations should be.

We start our presentation with the $q\bar q \rightarrow t\bar tgg$
channel, as it contains all the complications as far as the 
singularity structure is concerned, but can be evaluated in a shorter
time in comparison with the cutting edge $gg \rightarrow t\bar tgg$
process, due to the faster per phase space point computation. The
results for the expansion coefficients can be found in
Tabs.~\ref{tab:qqttgg1} and \ref{tab:qqttgg2}. Notice that we have not given
the values for the leading singularity $1/\epsilon^4$ as it is known
analytically from Section~\ref{sec:leading}. Striking are, of course,
the very large values close to threshold, which were, however,
expected from the leading logarithmic behavior determined in the very
same Section~\ref{sec:leading}. At this point, we can comment on the
integration errors. It is interesting that, while the relative error
varies substantially, the absolute error stays more or less the same up to
some factor. The reason for this behavior that we will see in all
subsequent results is that the dominant, logarithmic contribution in
$\beta$ has much higher precision, due to simpler functional
dependence (neither logarithms nor inverse square roots in the
integration parameters). The remaining functional dependence is due to
cancellations between contributions of the different sectors, which in
turn have all more or less a similar absolute error. If we now assume
that the error is a constant $2\times 10^{-2}$ (not exactly an upper
bound, but rather a realistic estimator for integration with the
partonic flux), we can obtain the implied relative error on the top
quark pair production cross section at the TeVatron, where
this channel dominates. It turns out, that the error would be $0.2\%$,
which is more than acceptable.

\begin{table}[ht]
  \begin{center}
    \begin{tabular}{c|c|c|c}
&& \\ $\beta$ & $f^{(0)}|_{10^{-7}}-f^{(0)}|_{10^{-6}}$ &
$\Delta f^{(0)}|_{10^{-7}}+\Delta f^{(0)}|_{10^{-6}}$ & $f^{(\epsilon,
\; 0)}$ \\ &&& \\ \hline
0.001 & $+1.9\times 10^{\text{-2}}$ & $4.8\times 10^{\text{-2}}$ & $-3.7\times 10^{\text{-8}}$\\
0.025 & $+6.5\times 10^{\text{-2}}$ & $8.8\times 10^{\text{-2}}$ & $-3.0\times 10^{\text{-5}}$\\
0.075 & $+8.5\times 10^{\text{-2}}$ & $7.1\times 10^{\text{-2}}$ & $-3.9\times 10^{\text{-4}}$\\
0.125 & $+8.8\times 10^{\text{-2}}$ & $6.6\times 10^{\text{-2}}$ & $-1.3\times 10^{\text{-3}}$\\
0.175 & $+7.9\times 10^{\text{-2}}$ & $5.5\times 10^{\text{-2}}$ & $-2.7\times 10^{\text{-3}}$\\
0.225 & $+6.3\times 10^{\text{-2}}$ & $4.8\times 10^{\text{-2}}$ & $-4.8\times 10^{\text{-3}}$\\
0.275 & $+5.9\times 10^{\text{-2}}$ & $3.8\times 10^{\text{-2}}$ & $-6.6\times 10^{\text{-3}}$\\
0.325 & $+6.4\times 10^{\text{-2}}$ & $3.2\times 10^{\text{-2}}$ & $-9.5\times 10^{\text{-3}}$\\
0.375 & $+5.6\times 10^{\text{-2}}$ & $3.1\times 10^{\text{-2}}$ & $-1.2\times 10^{\text{-2}}$\\
0.425 & $+1.6\times 10^{\text{-2}}$ & $2.7\times 10^{\text{-2}}$ & $-1.5\times 10^{\text{-2}}$\\
0.475 & $+3.2\times 10^{\text{-2}}$ & $2.7\times 10^{\text{-2}}$ & $-1.6\times 10^{\text{-2}}$\\
0.525 & $+3.1\times 10^{\text{-2}}$ & $2.2\times 10^{\text{-2}}$ & $-1.7\times 10^{\text{-2}}$\\
0.575 & $+1.3\times 10^{\text{-2}}$ & $2.3\times 10^{\text{-2}}$ & $-1.7\times 10^{\text{-2}}$\\
0.625 & $-1.6\times 10^{\text{-4}}$ & $1.9\times 10^{\text{-2}}$ & $-1.5\times 10^{\text{-2}}$\\
0.675 & $-5.1\times 10^{\text{-3}}$ & $1.6\times 10^{\text{-2}}$ & $-1.0\times 10^{\text{-2}}$\\
0.725 & $-1.5\times 10^{\text{-2}}$ & $1.5\times 10^{\text{-2}}$ & $-3.1\times 10^{\text{-3}}$\\
0.775 & $-1.3\times 10^{\text{-2}}$ & $1.1\times 10^{\text{-2}}$ & $+4.9\times 10^{\text{-3}}$\\
0.825 & $-1.3\times 10^{\text{-2}}$ & $8.9\times 10^{\text{-3}}$ & $+1.5\times 10^{\text{-2}}$\\
0.875 & $-1.4\times 10^{\text{-2}}$ & $7.6\times 10^{\text{-3}}$ & $+2.2\times 10^{\text{-2}}$\\
0.925 & $-1.3\times 10^{\text{-2}}$ & $5.6\times 10^{\text{-3}}$ & $+1.8\times 10^{\text{-2}}$\\
0.975 & $-6.0\times 10^{\text{-3}}$ & $3.0\times 10^{\text{-3}}$ & $-3.6\times 10^{\text{-2}}$\\
0.999 & $-2.6\times 10^{\text{-3}}$ & $2.7\times 10^{\text{-3}}$ & $-8.6\times 10^{\text{-2}}$\\
    \end{tabular}
  \end{center}
  \caption{\label{tab:cutoff} A comparison between the finite parts of the
    $f_{q\bar q \rightarrow t\bar tgg}$ function evaluated at two
    different values of the cutoff $\Delta$ defined in Eq.~(\ref{eq:cutoff}),
    $\Delta=10^{-7}$ and $\Delta=10^{-6}$. The second column
    represents the difference between the values, whereas the third
    the sum of the integration errors. The fourth column contains the
    $\epsilon$-contribution of the two-particle phase space.}
\end{table}

Let us now study the cutoff dependence of the results. The latter is
illustrated in Tab.~\ref{tab:cutoff}, where we give the difference
between the finite parts of the cross section evaluated at both
cutoffs, $10^{-7}$ and $10^{-6}$, compared to the sum of the
integration errors. For many values of $\beta$, we notice that the
difference is larger than the errors, albeit not by a large
factor. Taking into account that we can consider the difference to be
close to the actual variation with the cutoff, when changing from
$10^{-6}$ to $10^{-7}$, and that we expect a variation smaller by a
factor of three, when stepping to $10^{-8}$, we expect that the actual
error due to finite cutoff is lower than the integration error, as
certified by dividing all numbers by three and comparing to
Tab.~\ref{tab:qqttgg2}. The table also contains the
$\epsilon$-contribution to the two-particle phase space. We note, that
it is at the level of the current integration error for most points.

\begin{figure}[ht]
\begin{center}
\includegraphics[width=.50\textwidth]{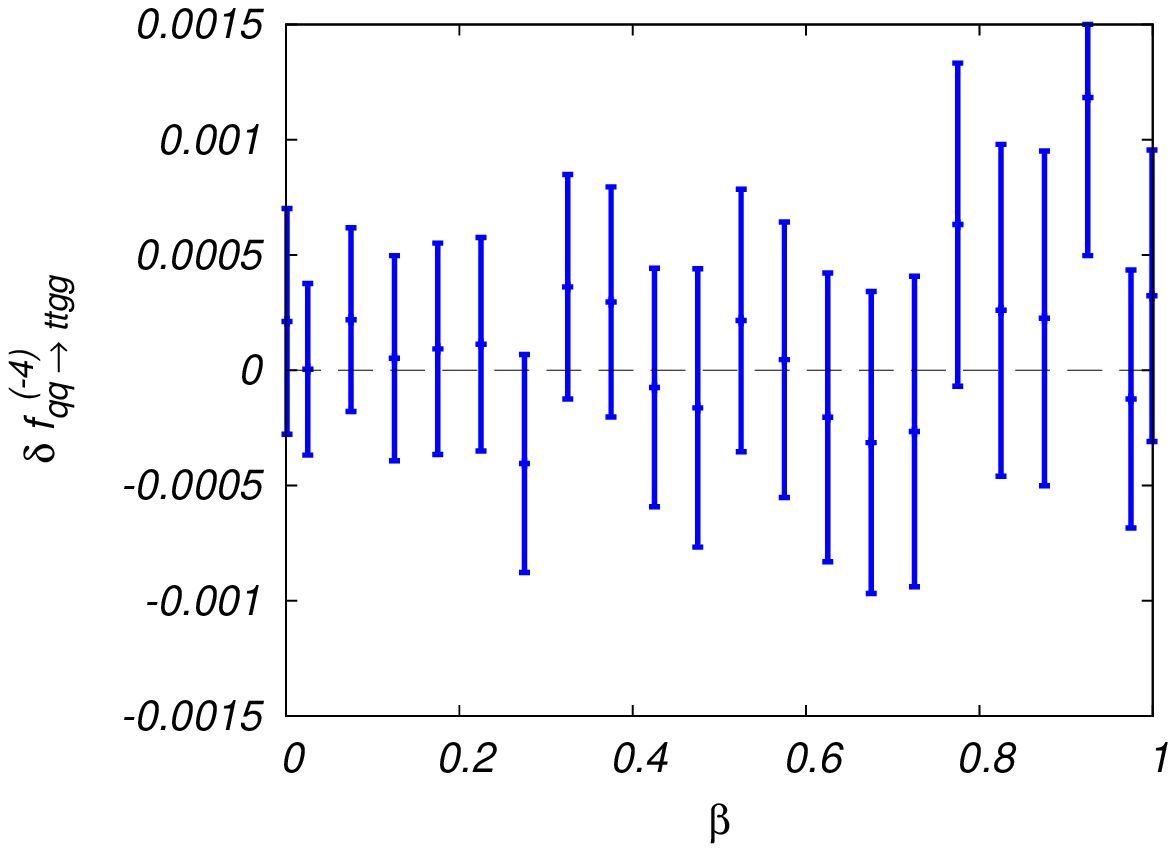}\includegraphics[width=.50\textwidth]{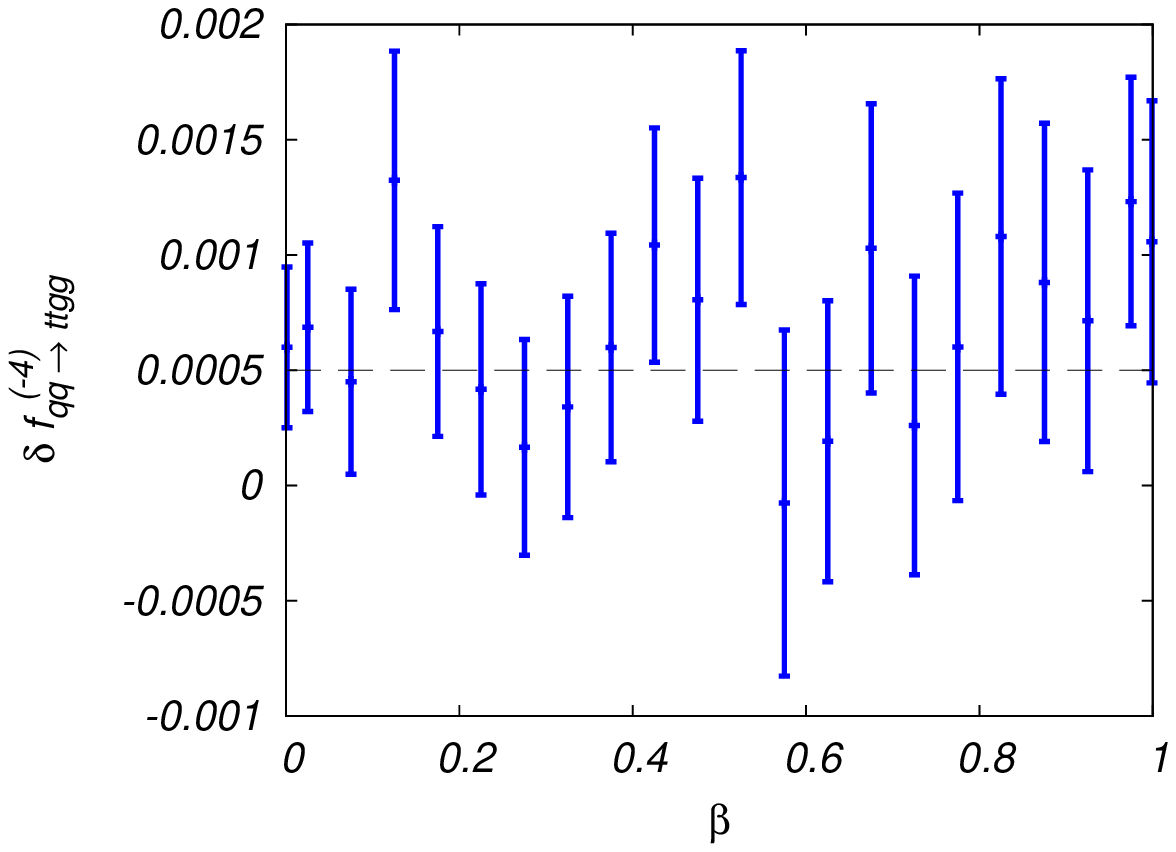}
\end{center}
\caption{Difference between the value of the coefficient of the
  leading divergence of $f_{q\bar q \rightarrow t\bar tgg}$
  obtained by numerical integration and the exact expression
  Eq.~(\ref{eq:divqq}) normalized to the value of the latter. On the left panel,
  the lower cutoff $\Delta = 10^{-7}$ has been used, whereas on the
  right, $\Delta = 10^{-6}$. Notice the shifted scale on the vertical
  axis between the left and right panels.}
\label{fig:leading}
\end{figure}

In order to check the normalization of our results, we can also
compare the numerical estimate of the leading singularity,
$1/\epsilon^4$, with the prediction from
Section~\ref{sec:leading}. This is done in Fig.~\ref{fig:leading} for
both values of the cutoff, where we plot
\begin{equation}
\delta f^{(-4)}_{q\bar q \rightarrow t\bar tgg} =
\frac{(4\pi)^2 f^{(-4)}_{q\bar q \rightarrow t\bar tgg}-2C_F(C_A+4C_F)
f^{(0)}_{q\bar q\rightarrow t\bar t}}{2C_F(C_A+4C_F)
f^{(0)}_{q\bar q\rightarrow t\bar t}} \; ,
\end{equation}
where $f^{(0)}_{q\bar q\rightarrow t\bar t}$ is defined by the
Born cross section in \ref{sec:born}. We notice that in the
case of the higher cutoff, $\Delta = 10^{-6}$, the cutoff dependence
is noticeable due to the tiny numerical errors. What is slightly more
worrisome is that a few errors are indeed underestimated. In this
case, this is due to the fact that the integration errors are very
small, below permille, but the function has integrable singularities
in $\zeta$ at the integration boundaries, and is thus not entirely
well behaved. Based on this, we can only expect worse from the much
more singular finite parts of the cross section.

\begin{figure}[ht]
\begin{center}
\includegraphics[width=.60\textwidth]{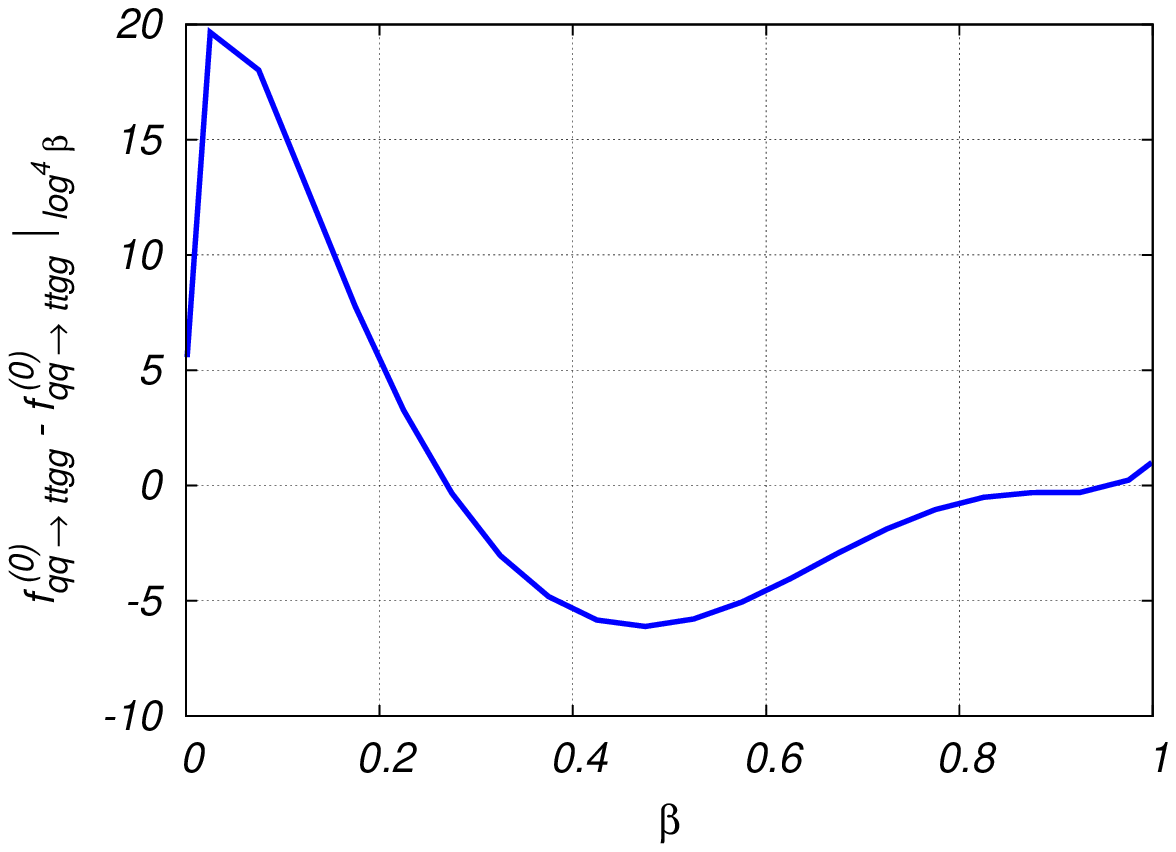}
\end{center}
\caption{Finite part of $f_{q\bar q \rightarrow t\bar tgg}$ after
  removing the dominant logarithmic term Eq.~(\ref{eq:logqq}).}
\label{fig:qqttgg}
\end{figure}

Finally, in Fig.~\ref{fig:qqttgg}, we show the cross section after
subtracting the leading logarithm in $\beta$. The latter would make the
plot span an order of magnitude more. Notice that on the scale of this
plot, the integration errors would not be noticeable.

\begin{table}[ht]
  \begin{center}
    \begin{tabular}{c|ll|ll}
&&&& \\ $\beta$ && $\epsilon^{-3}$ && $\epsilon^{-2}$  \\ &&&& \\ \hline
0.001 & $+4.534\times 10^{\text{-3}}$ & $\pm \;\;\;\;1.8\times 10^{\text{-6}}$ & $+1.568\times 10^{\text{-1}}$ & $\pm \;\;\;\;6.3\times 10^{\text{-5}}$\\
0.025 & $+6.086\times 10^{\text{-2}}$ & $\pm \;\;\;\;2.4\times 10^{\text{-5}}$ & $+1.119\times 10^0$ & $\pm \;\;\;\;4.3\times 10^{\text{-4}}$\\
0.075 & $+1.299\times 10^{\text{-1}}$ & $\pm \;\;\;\;5.4\times 10^{\text{-5}}$ & $+1.664\times 10^0$ & $\pm \;\;\;\;6.8\times 10^{\text{-4}}$\\
0.125 & $+1.776\times 10^{\text{-1}}$ & $\pm \;\;\;\;7.7\times 10^{\text{-5}}$ & $+1.817\times 10^0$ & $\pm \;\;\;\;7.8\times 10^{\text{-4}}$\\
0.175 & $+2.146\times 10^{\text{-1}}$ & $\pm \;\;\;\;9.7\times 10^{\text{-5}}$ & $+1.836\times 10^0$ & $\pm \;\;\;\;8.3\times 10^{\text{-4}}$\\
0.225 & $+2.451\times 10^{\text{-1}}$ & $\pm \;\;\;\;1.1\times 10^{\text{-4}}$ & $+1.798\times 10^0$ & $\pm \;\;\;\;8.4\times 10^{\text{-4}}$\\
0.275 & $+2.707\times 10^{\text{-1}}$ & $\pm \;\;\;\;1.2\times 10^{\text{-4}}$ & $+1.727\times 10^0$ & $\pm \;\;\;\;8.7\times 10^{\text{-4}}$\\
0.325 & $+2.918\times 10^{\text{-1}}$ & $\pm \;\;\;\;1.4\times 10^{\text{-4}}$ & $+1.631\times 10^0$ & $\pm \;\;\;\;8.9\times 10^{\text{-4}}$\\
0.375 & $+3.084\times 10^{\text{-1}}$ & $\pm \;\;\;\;1.4\times 10^{\text{-4}}$ & $+1.516\times 10^0$ & $\pm \;\;\;\;8.8\times 10^{\text{-4}}$\\
0.425 & $+3.191\times 10^{\text{-1}}$ & $\pm \;\;\;\;1.6\times 10^{\text{-4}}$ & $+1.374\times 10^0$ & $\pm \;\;\;\;9.3\times 10^{\text{-4}}$\\
0.475 & $+3.228\times 10^{\text{-1}}$ & $\pm \;\;\;\;1.7\times 10^{\text{-4}}$ & $+1.204\times 10^0$ & $\pm \;\;\;\;1.0\times 10^{\text{-3}}$\\
0.525 & $+3.175\times 10^{\text{-1}}$ & $\pm \;\;\;\;1.8\times 10^{\text{-4}}$ & $+1.007\times 10^0$ & $\pm \;\;\;\;9.8\times 10^{\text{-4}}$\\
0.575 & $+3.016\times 10^{\text{-1}}$ & $\pm \;\;\;\;1.8\times 10^{\text{-4}}$ & $+7.777\times 10^{\text{-1}}$ & $\pm \;\;\;\;9.8\times 10^{\text{-4}}$\\
0.625 & $+2.721\times 10^{\text{-1}}$ & $\pm \;\;\;\;2.1\times 10^{\text{-4}}$ & $+5.218\times 10^{\text{-1}}$ & $\pm \;\;\;\;1.2\times 10^{\text{-3}}$\\
0.675 & $+2.277\times 10^{\text{-1}}$ & $\pm \;\;\;\;1.7\times 10^{\text{-4}}$ & $+2.543\times 10^{\text{-1}}$ & $\pm \;\;\;\;1.0\times 10^{\text{-3}}$\\
0.725 & $+1.657\times 10^{\text{-1}}$ & $\pm \;\;\;\;2.0\times 10^{\text{-4}}$ & $-1.360\times 10^{\text{-2}}$ & $\pm \;\;\;\;1.3\times 10^{\text{-3}}$\\
0.775 & $+8.359\times 10^{\text{-2}}$ & $\pm \;\;\;\;1.8\times 10^{\text{-4}}$ & $-2.575\times 10^{\text{-1}}$ & $\pm \;\;\;\;1.1\times 10^{\text{-3}}$\\
0.825 & $-2.003\times 10^{\text{-2}}$ & $\pm \;\;\;\;2.2\times 10^{\text{-4}}$ & $-4.275\times 10^{\text{-1}}$ & $\pm \;\;\;\;1.2\times 10^{\text{-3}}$\\
0.875 & $-1.461\times 10^{\text{-1}}$ & $\pm \;\;\;\;2.6\times 10^{\text{-4}}$ & $-4.536\times 10^{\text{-1}}$ & $\pm \;\;\;\;1.1\times 10^{\text{-3}}$\\
0.925 & $-2.996\times 10^{\text{-1}}$ & $\pm \;\;\;\;3.3\times 10^{\text{-4}}$ & $-1.805\times 10^{\text{-1}}$ & $\pm \;\;\;\;1.3\times 10^{\text{-3}}$\\
0.975 & $-4.862\times 10^{\text{-1}}$ & $\pm \;\;\;\;4.0\times 10^{\text{-4}}$ & $+9.458\times 10^{\text{-1}}$ & $\pm \;\;\;\;2.3\times 10^{\text{-3}}$\\
0.999 & $-5.970\times 10^{\text{-1}}$ & $\pm \;\;\;\;5.5\times 10^{\text{-4}}$ & $+4.118\times 10^0$ & $\pm \;\;\;\;6.7\times 10^{\text{-3}}$\\
    \end{tabular}
  \end{center}
  \caption{\label{tab:ggttgg1} Coefficients of the Laurent expansion of the
    $f_{gg \rightarrow t\bar tgg}$ function.}
\end{table}
\begin{table}[ht]
  \begin{center}
    \begin{tabular}{c|ll|ll}
&&&& \\ $\beta$ && $\epsilon^{-1}$ && $\epsilon^{0}$  \\ &&&& \\ \hline
0.001 & $+3.601\times 10^0$ & $\pm \;\;\;\;1.5\times 10^{\text{-3}}$ & $+6.175\times 10^1$ & $\pm \;\;\;\;3.0\times 10^{\text{-2}}$\\
0.025 & $+1.349\times 10^1$ & $\pm \;\;\;\;5.4\times 10^{\text{-3}}$ & $+1.197\times 10^2$ & $\pm \;\;\;\;5.2\times 10^{\text{-2}}$\\
0.075 & $+1.374\times 10^1$ & $\pm \;\;\;\;6.0\times 10^{\text{-3}}$ & $+8.129\times 10^1$ & $\pm \;\;\;\;4.5\times 10^{\text{-2}}$\\
0.125 & $+1.175\times 10^1$ & $\pm \;\;\;\;5.8\times 10^{\text{-3}}$ & $+5.280\times 10^1$ & $\pm \;\;\;\;3.8\times 10^{\text{-2}}$\\
0.175 & $+9.741\times 10^0$ & $\pm \;\;\;\;5.3\times 10^{\text{-3}}$ & $+3.461\times 10^1$ & $\pm \;\;\;\;3.3\times 10^{\text{-2}}$\\
0.225 & $+8.015\times 10^0$ & $\pm \;\;\;\;5.1\times 10^{\text{-3}}$ & $+2.300\times 10^1$ & $\pm \;\;\;\;3.2\times 10^{\text{-2}}$\\
0.275 & $+6.574\times 10^0$ & $\pm \;\;\;\;5.4\times 10^{\text{-3}}$ & $+1.557\times 10^1$ & $\pm \;\;\;\;3.4\times 10^{\text{-2}}$\\
0.325 & $+5.350\times 10^0$ & $\pm \;\;\;\;4.9\times 10^{\text{-3}}$ & $+1.062\times 10^1$ & $\pm \;\;\;\;3.4\times 10^{\text{-2}}$\\
0.375 & $+4.290\times 10^0$ & $\pm \;\;\;\;5.3\times 10^{\text{-3}}$ & $+7.186\times 10^0$ & $\pm \;\;\;\;3.3\times 10^{\text{-2}}$\\
0.425 & $+3.334\times 10^0$ & $\pm \;\;\;\;5.3\times 10^{\text{-3}}$ & $+4.774\times 10^0$ & $\pm \;\;\;\;3.1\times 10^{\text{-2}}$\\
0.475 & $+2.453\times 10^0$ & $\pm \;\;\;\;6.2\times 10^{\text{-3}}$ & $+2.944\times 10^0$ & $\pm \;\;\;\;3.6\times 10^{\text{-2}}$\\
0.525 & $+1.649\times 10^0$ & $\pm \;\;\;\;5.7\times 10^{\text{-3}}$ & $+1.595\times 10^0$ & $\pm \;\;\;\;3.2\times 10^{\text{-2}}$\\
0.575 & $+8.930\times 10^{\text{-1}}$ & $\pm \;\;\;\;5.5\times 10^{\text{-3}}$ & $+5.449\times 10^{\text{-1}}$ & $\pm \;\;\;\;3.3\times 10^{\text{-2}}$\\
0.625 & $+2.321\times 10^{\text{-1}}$ & $\pm \;\;\;\;5.6\times 10^{\text{-3}}$ & $-1.188\times 10^{\text{-1}}$ & $\pm \;\;\;\;3.4\times 10^{\text{-2}}$\\
0.675 & $-2.740\times 10^{\text{-1}}$ & $\pm \;\;\;\;5.6\times 10^{\text{-3}}$ & $-4.124\times 10^{\text{-1}}$ & $\pm \;\;\;\;3.5\times 10^{\text{-2}}$\\
0.725 & $-6.121\times 10^{\text{-1}}$ & $\pm \;\;\;\;6.9\times 10^{\text{-3}}$ & $-4.498\times 10^{\text{-1}}$ & $\pm \;\;\;\;3.6\times 10^{\text{-2}}$\\
0.775 & $-7.540\times 10^{\text{-1}}$ & $\pm \;\;\;\;5.6\times 10^{\text{-3}}$ & $-3.971\times 10^{\text{-1}}$ & $\pm \;\;\;\;3.5\times 10^{\text{-2}}$\\
0.825 & $-6.363\times 10^{\text{-1}}$ & $\pm \;\;\;\;6.3\times 10^{\text{-3}}$ & $-3.371\times 10^{\text{-1}}$ & $\pm \;\;\;\;3.9\times 10^{\text{-2}}$\\
0.875 & $-3.241\times 10^{\text{-1}}$ & $\pm \;\;\;\;6.0\times 10^{\text{-3}}$ & $-5.518\times 10^{\text{-1}}$ & $\pm \;\;\;\;3.7\times 10^{\text{-2}}$\\
0.925 & $-1.076\times 10^{\text{-1}}$ & $\pm \;\;\;\;6.3\times 10^{\text{-3}}$ & $-1.462\times 10^0$ & $\pm \;\;\;\;3.7\times 10^{\text{-2}}$\\
0.975 & $-1.867\times 10^0$ & $\pm \;\;\;\;9.1\times 10^{\text{-3}}$ & $-6.540\times 10^{\text{-1}}$ & $\pm \;\;\;\;4.8\times 10^{\text{-2}}$\\
0.999 & $-1.543\times 10^1$ & $\pm \;\;\;\;5.3\times 10^{\text{-2}}$ & $+4.437\times 10^1$ & $\pm \;\;\;\;4.2\times 10^{\text{-1}}$\\
    \end{tabular}
  \end{center}
  \caption{\label{tab:ggttgg2} Coefficients of the Laurent expansion of the $f_{gg
      \rightarrow t\bar tgg}$ function.}
\end{table}

We can, in principle, repeat the same discussion for the most
complicated and computationally intensive channel, $gg \rightarrow
t\bar tgg$. The numerical values are given in Tabs.~\ref{tab:ggttgg1}
and \ref{tab:ggttgg2}, where we have again omitted the leading
singularity. The essential difference to the quark annihilation channel is in
the about twice larger absolute errors of the finite part for most of
the $\beta$ range. One can again estimate that if the error is
consider constant and equal to $4\times 10^{-2}$, then the implied
uncertainty at the LHC would be $0.4\%$, which is also more than
acceptable. As before, we can demonstrate that the cutoff dependence
is lower than the numerical integration error.

\begin{figure}[ht]
\begin{center}
\includegraphics[width=.60\textwidth]{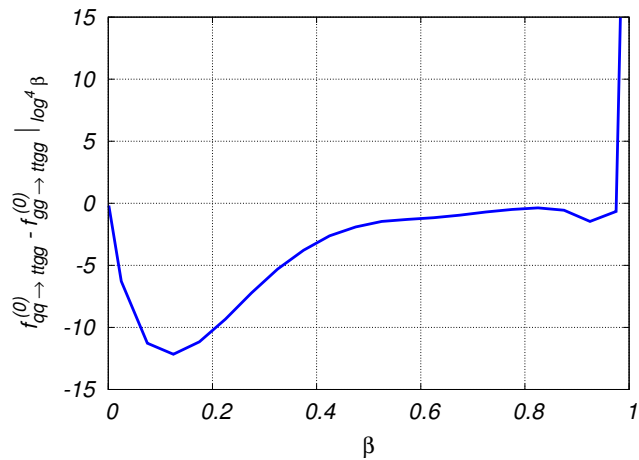}
\end{center}
\caption{Finite part of $f_{gg \rightarrow t\bar tgg}$ after
  removing the dominant logarithmic term Eq.~(\ref{eq:loggg}).}
\label{fig:ggttgg}
\end{figure}

In Fig.~\ref{fig:ggttgg}, we show the finite part of the cross section
after removing the dominant logarithm in $\beta$. We notice the very
steep rise at the end of the range. This phenomenon is well known from
the next-to-leading order cross section in the gluon fusion
channel. For our numerics it has the unpleasant feature of making the
calculation slightly unstable due to the extreme sensitivity to the
value of $\beta$. This also points to an underestimated error at
$\beta=0.999$, anyway quoted to be rather large in
Tab.~\ref{tab:ggttgg2}. None of this is relevant to phenomenology at
present, but we will try to get a better handle of this problem in the
future.

\begin{table}[ht]
  \begin{center}
    \begin{tabular}{c|ll|ll}
&&&& \\ $\beta$ && $\epsilon^{-3}$ && $\epsilon^{-2}$  \\ &&&& \\ \hline
0.001 & $-1.452\times 10^{\text{-6}}$ & $\pm \;\;\;\;2.9\times 10^{\text{-9}}$ & $-1.027\times 10^{\text{-4}}$ & $\pm \;\;\;\;1.9\times 10^{\text{-7}}$\\
0.025 & $-3.627\times 10^{\text{-5}}$ & $\pm \;\;\;\;2.5\times 10^{\text{-8}}$ & $-1.399\times 10^{\text{-3}}$ & $\pm \;\;\;\;9.7\times 10^{\text{-7}}$\\
0.075 & $-1.096\times 10^{\text{-4}}$ & $\pm \;\;\;\;8.7\times 10^{\text{-8}}$ & $-3.029\times 10^{\text{-3}}$ & $\pm \;\;\;\;2.3\times 10^{\text{-6}}$\\
0.125 & $-1.853\times 10^{\text{-4}}$ & $\pm \;\;\;\;1.7\times 10^{\text{-7}}$ & $-4.185\times 10^{\text{-3}}$ & $\pm \;\;\;\;3.3\times 10^{\text{-6}}$\\
0.175 & $-2.644\times 10^{\text{-4}}$ & $\pm \;\;\;\;2.2\times 10^{\text{-7}}$ & $-5.111\times 10^{\text{-3}}$ & $\pm \;\;\;\;4.1\times 10^{\text{-6}}$\\
0.225 & $-3.486\times 10^{\text{-4}}$ & $\pm \;\;\;\;3.1\times 10^{\text{-7}}$ & $-5.901\times 10^{\text{-3}}$ & $\pm \;\;\;\;4.9\times 10^{\text{-6}}$\\
0.275 & $-4.377\times 10^{\text{-4}}$ & $\pm \;\;\;\;4.0\times 10^{\text{-7}}$ & $-6.584\times 10^{\text{-3}}$ & $\pm \;\;\;\;5.7\times 10^{\text{-6}}$\\
0.325 & $-5.323\times 10^{\text{-4}}$ & $\pm \;\;\;\;4.9\times 10^{\text{-7}}$ & $-7.170\times 10^{\text{-3}}$ & $\pm \;\;\;\;6.4\times 10^{\text{-6}}$\\
0.375 & $-6.322\times 10^{\text{-4}}$ & $\pm \;\;\;\;6.0\times 10^{\text{-7}}$ & $-7.656\times 10^{\text{-3}}$ & $\pm \;\;\;\;7.0\times 10^{\text{-6}}$\\
0.425 & $-7.362\times 10^{\text{-4}}$ & $\pm \;\;\;\;7.3\times 10^{\text{-7}}$ & $-8.023\times 10^{\text{-3}}$ & $\pm \;\;\;\;7.9\times 10^{\text{-6}}$\\
0.475 & $-8.422\times 10^{\text{-4}}$ & $\pm \;\;\;\;8.1\times 10^{\text{-7}}$ & $-8.219\times 10^{\text{-3}}$ & $\pm \;\;\;\;8.0\times 10^{\text{-6}}$\\
0.525 & $-9.468\times 10^{\text{-4}}$ & $\pm \;\;\;\;9.6\times 10^{\text{-7}}$ & $-8.196\times 10^{\text{-3}}$ & $\pm \;\;\;\;8.8\times 10^{\text{-6}}$\\
0.575 & $-1.045\times 10^{\text{-3}}$ & $\pm \;\;\;\;1.0\times 10^{\text{-6}}$ & $-7.914\times 10^{\text{-3}}$ & $\pm \;\;\;\;8.4\times 10^{\text{-6}}$\\
0.625 & $-1.136\times 10^{\text{-3}}$ & $\pm \;\;\;\;1.1\times 10^{\text{-6}}$ & $-7.322\times 10^{\text{-3}}$ & $\pm \;\;\;\;8.6\times 10^{\text{-6}}$\\
0.675 & $-1.208\times 10^{\text{-3}}$ & $\pm \;\;\;\;1.2\times 10^{\text{-6}}$ & $-6.310\times 10^{\text{-3}}$ & $\pm \;\;\;\;8.3\times 10^{\text{-6}}$\\
0.725 & $-1.256\times 10^{\text{-3}}$ & $\pm \;\;\;\;1.3\times 10^{\text{-6}}$ & $-4.836\times 10^{\text{-3}}$ & $\pm \;\;\;\;8.0\times 10^{\text{-6}}$\\
0.775 & $-1.264\times 10^{\text{-3}}$ & $\pm \;\;\;\;1.3\times 10^{\text{-6}}$ & $-2.797\times 10^{\text{-3}}$ & $\pm \;\;\;\;7.6\times 10^{\text{-6}}$\\
0.825 & $-1.222\times 10^{\text{-3}}$ & $\pm \;\;\;\;1.3\times 10^{\text{-6}}$ & $-8.582\times 10^{\text{-5}}$ & $\pm \;\;\;\;7.4\times 10^{\text{-6}}$\\
0.875 & $-1.107\times 10^{\text{-3}}$ & $\pm \;\;\;\;1.2\times 10^{\text{-6}}$ & $+3.444\times 10^{\text{-3}}$ & $\pm \;\;\;\;7.9\times 10^{\text{-6}}$\\
0.925 & $-8.799\times 10^{\text{-4}}$ & $\pm \;\;\;\;9.5\times 10^{\text{-7}}$ & $+8.174\times 10^{\text{-3}}$ & $\pm \;\;\;\;9.7\times 10^{\text{-6}}$\\
0.975 & $-4.551\times 10^{\text{-4}}$ & $\pm \;\;\;\;6.3\times 10^{\text{-7}}$ & $+1.546\times 10^{\text{-2}}$ & $\pm \;\;\;\;1.3\times 10^{\text{-5}}$\\
0.999 & $-3.921\times 10^{\text{-5}}$ & $\pm \;\;\;\;5.9\times 10^{\text{-7}}$ & $+2.366\times 10^{\text{-2}}$ & $\pm \;\;\;\;1.9\times 10^{\text{-5}}$\\
    \end{tabular}
  \end{center}
  \caption{\label{tab:ggttqq1} Coefficients of the Laurent expansion of the
    $f_{gg \rightarrow t\bar tq\bar q}$ function.}
\end{table}
\begin{table}[ht]
  \begin{center}
    \begin{tabular}{c|ll|ll}
&&&& \\ $\beta$ && $\epsilon^{-1}$ && $\epsilon^{0}$  \\ &&&& \\ \hline
0.001 & $-3.622\times 10^{\text{-3}}$ & $\pm \;\;\;\;6.2\times 10^{\text{-6}}$ & $-8.500\times 10^{\text{-2}}$ & $\pm \;\;\;\;1.8\times 10^{\text{-4}}$\\
0.025 & $-2.672\times 10^{\text{-2}}$ & $\pm \;\;\;\;1.9\times 10^{\text{-5}}$ & $-3.366\times 10^{\text{-1}}$ & $\pm \;\;\;\;2.7\times 10^{\text{-4}}$\\
0.075 & $-4.108\times 10^{\text{-2}}$ & $\pm \;\;\;\;3.2\times 10^{\text{-5}}$ & $-3.636\times 10^{\text{-1}}$ & $\pm \;\;\;\;3.4\times 10^{\text{-4}}$\\
0.125 & $-4.604\times 10^{\text{-2}}$ & $\pm \;\;\;\;4.1\times 10^{\text{-5}}$ & $-3.275\times 10^{\text{-1}}$ & $\pm \;\;\;\;4.4\times 10^{\text{-4}}$\\
0.175 & $-4.781\times 10^{\text{-2}}$ & $\pm \;\;\;\;4.3\times 10^{\text{-5}}$ & $-2.869\times 10^{\text{-1}}$ & $\pm \;\;\;\;3.7\times 10^{\text{-4}}$\\
0.225 & $-4.806\times 10^{\text{-2}}$ & $\pm \;\;\;\;4.6\times 10^{\text{-5}}$ & $-2.495\times 10^{\text{-1}}$ & $\pm \;\;\;\;3.7\times 10^{\text{-4}}$\\
0.275 & $-4.742\times 10^{\text{-2}}$ & $\pm \;\;\;\;4.8\times 10^{\text{-5}}$ & $-2.173\times 10^{\text{-1}}$ & $\pm \;\;\;\;3.5\times 10^{\text{-4}}$\\
0.325 & $-4.607\times 10^{\text{-2}}$ & $\pm \;\;\;\;5.0\times 10^{\text{-5}}$ & $-1.881\times 10^{\text{-1}}$ & $\pm \;\;\;\;3.6\times 10^{\text{-4}}$\\
0.375 & $-4.404\times 10^{\text{-2}}$ & $\pm \;\;\;\;5.2\times 10^{\text{-5}}$ & $-1.612\times 10^{\text{-1}}$ & $\pm \;\;\;\;3.5\times 10^{\text{-4}}$\\
0.425 & $-4.137\times 10^{\text{-2}}$ & $\pm \;\;\;\;5.5\times 10^{\text{-5}}$ & $-1.364\times 10^{\text{-1}}$ & $\pm \;\;\;\;3.6\times 10^{\text{-4}}$\\
0.475 & $-3.770\times 10^{\text{-2}}$ & $\pm \;\;\;\;5.3\times 10^{\text{-5}}$ & $-1.111\times 10^{\text{-1}}$ & $\pm \;\;\;\;3.5\times 10^{\text{-4}}$\\
0.525 & $-3.296\times 10^{\text{-2}}$ & $\pm \;\;\;\;5.7\times 10^{\text{-5}}$ & $-8.534\times 10^{\text{-2}}$ & $\pm \;\;\;\;3.8\times 10^{\text{-4}}$\\
0.575 & $-2.716\times 10^{\text{-2}}$ & $\pm \;\;\;\;5.9\times 10^{\text{-5}}$ & $-5.918\times 10^{\text{-2}}$ & $\pm \;\;\;\;3.7\times 10^{\text{-4}}$\\
0.625 & $-2.032\times 10^{\text{-2}}$ & $\pm \;\;\;\;5.8\times 10^{\text{-5}}$ & $-3.295\times 10^{\text{-2}}$ & $\pm \;\;\;\;4.0\times 10^{\text{-4}}$\\
0.675 & $-1.234\times 10^{\text{-2}}$ & $\pm \;\;\;\;5.7\times 10^{\text{-5}}$ & $-7.767\times 10^{\text{-3}}$ & $\pm \;\;\;\;3.7\times 10^{\text{-4}}$\\
0.725 & $-3.706\times 10^{\text{-3}}$ & $\pm \;\;\;\;5.2\times 10^{\text{-5}}$ & $+1.446\times 10^{\text{-2}}$ & $\pm \;\;\;\;3.3\times 10^{\text{-4}}$\\
0.775 & $+5.117\times 10^{\text{-3}}$ & $\pm \;\;\;\;5.0\times 10^{\text{-5}}$ & $+3.236\times 10^{\text{-2}}$ & $\pm \;\;\;\;3.1\times 10^{\text{-4}}$\\
0.825 & $+1.331\times 10^{\text{-2}}$ & $\pm \;\;\;\;5.6\times 10^{\text{-5}}$ & $+4.375\times 10^{\text{-2}}$ & $\pm \;\;\;\;2.9\times 10^{\text{-4}}$\\
0.875 & $+1.891\times 10^{\text{-2}}$ & $\pm \;\;\;\;7.8\times 10^{\text{-5}}$ & $+4.700\times 10^{\text{-2}}$ & $\pm \;\;\;\;2.9\times 10^{\text{-4}}$\\
0.925 & $+1.833\times 10^{\text{-2}}$ & $\pm \;\;\;\;4.4\times 10^{\text{-5}}$ & $+4.423\times 10^{\text{-2}}$ & $\pm \;\;\;\;2.4\times 10^{\text{-4}}$\\
0.975 & $-3.231\times 10^{\text{-3}}$ & $\pm \;\;\;\;6.0\times 10^{\text{-5}}$ & $+6.853\times 10^{\text{-2}}$ & $\pm \;\;\;\;2.6\times 10^{\text{-4}}$\\
0.999 & $-6.995\times 10^{\text{-2}}$ & $\pm \;\;\;\;1.9\times 10^{\text{-4}}$ & $+3.312\times 10^{\text{-1}}$ & $\pm \;\;\;\;1.2\times 10^{\text{-3}}$\\
    \end{tabular}
  \end{center}
  \caption{\label{tab:ggttqq2} Coefficients of the Laurent expansion of the $f_{gg
      \rightarrow t\bar tq\bar q}$ function.}
\end{table}

\begin{table}[ht]
  \begin{center}
    \begin{tabular}{c|ll|ll}
&&&& \\ $\beta$ && $\epsilon^{-3}$ && $\epsilon^{-2}$  \\ &&&& \\ \hline
0.001 & $-1.962\times 10^{\text{-6}}$ & $\pm \;\;\;\;8.8\times 10^{\text{-10}}$ & $-1.418\times 10^{\text{-4}}$ & $\pm \;\;\;\;6.5\times 10^{\text{-8}}$\\
0.025 & $-4.907\times 10^{\text{-5}}$ & $\pm \;\;\;\;2.4\times 10^{\text{-8}}$ & $-1.967\times 10^{\text{-3}}$ & $\pm \;\;\;\;9.6\times 10^{\text{-7}}$\\
0.075 & $-1.464\times 10^{\text{-4}}$ & $\pm \;\;\;\;1.4\times 10^{\text{-7}}$ & $-4.255\times 10^{\text{-3}}$ & $\pm \;\;\;\;2.5\times 10^{\text{-6}}$\\
0.125 & $-2.406\times 10^{\text{-4}}$ & $\pm \;\;\;\;1.4\times 10^{\text{-7}}$ & $-5.749\times 10^{\text{-3}}$ & $\pm \;\;\;\;3.1\times 10^{\text{-6}}$\\
0.175 & $-3.297\times 10^{\text{-4}}$ & $\pm \;\;\;\;1.9\times 10^{\text{-7}}$ & $-6.740\times 10^{\text{-3}}$ & $\pm \;\;\;\;3.7\times 10^{\text{-6}}$\\
0.225 & $-4.129\times 10^{\text{-4}}$ & $\pm \;\;\;\;2.5\times 10^{\text{-7}}$ & $-7.342\times 10^{\text{-3}}$ & $\pm \;\;\;\;4.2\times 10^{\text{-6}}$\\
0.275 & $-4.866\times 10^{\text{-4}}$ & $\pm \;\;\;\;3.0\times 10^{\text{-7}}$ & $-7.593\times 10^{\text{-3}}$ & $\pm \;\;\;\;4.5\times 10^{\text{-6}}$\\
0.325 & $-5.508\times 10^{\text{-4}}$ & $\pm \;\;\;\;3.4\times 10^{\text{-7}}$ & $-7.558\times 10^{\text{-3}}$ & $\pm \;\;\;\;4.7\times 10^{\text{-6}}$\\
0.375 & $-6.029\times 10^{\text{-4}}$ & $\pm \;\;\;\;3.9\times 10^{\text{-7}}$ & $-7.250\times 10^{\text{-3}}$ & $\pm \;\;\;\;4.7\times 10^{\text{-6}}$\\
0.425 & $-6.429\times 10^{\text{-4}}$ & $\pm \;\;\;\;4.2\times 10^{\text{-7}}$ & $-6.723\times 10^{\text{-3}}$ & $\pm \;\;\;\;4.7\times 10^{\text{-6}}$\\
0.475 & $-6.682\times 10^{\text{-4}}$ & $\pm \;\;\;\;4.5\times 10^{\text{-7}}$ & $-5.998\times 10^{\text{-3}}$ & $\pm \;\;\;\;4.3\times 10^{\text{-6}}$\\
0.525 & $-6.783\times 10^{\text{-4}}$ & $\pm \;\;\;\;4.6\times 10^{\text{-7}}$ & $-5.117\times 10^{\text{-3}}$ & $\pm \;\;\;\;4.0\times 10^{\text{-6}}$\\
0.575 & $-6.724\times 10^{\text{-4}}$ & $\pm \;\;\;\;4.5\times 10^{\text{-7}}$ & $-4.113\times 10^{\text{-3}}$ & $\pm \;\;\;\;3.5\times 10^{\text{-6}}$\\
0.625 & $-6.503\times 10^{\text{-4}}$ & $\pm \;\;\;\;4.4\times 10^{\text{-7}}$ & $-3.037\times 10^{\text{-3}}$ & $\pm \;\;\;\;3.1\times 10^{\text{-6}}$\\
0.675 & $-6.124\times 10^{\text{-4}}$ & $\pm \;\;\;\;4.2\times 10^{\text{-7}}$ & $-1.945\times 10^{\text{-3}}$ & $\pm \;\;\;\;2.6\times 10^{\text{-6}}$\\
0.725 & $-5.569\times 10^{\text{-4}}$ & $\pm \;\;\;\;3.9\times 10^{\text{-7}}$ & $-8.748\times 10^{\text{-4}}$ & $\pm \;\;\;\;2.2\times 10^{\text{-6}}$\\
0.775 & $-4.869\times 10^{\text{-4}}$ & $\pm \;\;\;\;3.6\times 10^{\text{-7}}$ & $+9.961\times 10^{\text{-5}}$ & $\pm \;\;\;\;2.0\times 10^{\text{-6}}$\\
0.825 & $-4.001\times 10^{\text{-4}}$ & $\pm \;\;\;\;3.1\times 10^{\text{-7}}$ & $+9.114\times 10^{\text{-4}}$ & $\pm \;\;\;\;1.7\times 10^{\text{-6}}$\\
0.875 & $-2.998\times 10^{\text{-4}}$ & $\pm \;\;\;\;2.4\times 10^{\text{-7}}$ & $+1.461\times 10^{\text{-3}}$ & $\pm \;\;\;\;1.5\times 10^{\text{-6}}$\\
0.925 & $-1.874\times 10^{\text{-4}}$ & $\pm \;\;\;\;1.5\times 10^{\text{-7}}$ & $+1.609\times 10^{\text{-3}}$ & $\pm \;\;\;\;1.3\times 10^{\text{-6}}$\\
0.975 & $-6.459\times 10^{\text{-5}}$ & $\pm \;\;\;\;4.3\times 10^{\text{-8}}$ & $+1.052\times 10^{\text{-3}}$ & $\pm \;\;\;\;6.0\times 10^{\text{-7}}$\\
0.999 & $-2.617\times 10^{\text{-6}}$ & $\pm \;\;\;\;2.0\times 10^{\text{-9}}$ & $+1.013\times 10^{\text{-4}}$ & $\pm \;\;\;\;1.1\times 10^{\text{-7}}$\\
    \end{tabular}
  \end{center}
  \caption{\label{tab:qqttbb1} Coefficients of the Laurent expansion of the
    $f_{q\bar q \rightarrow t\bar tq'\bar q'}$ function.}
\end{table}
\begin{table}[ht]
  \begin{center}
    \begin{tabular}{c|ll|ll}
&&&& \\ $\beta$ && $\epsilon^{-1}$ && $\epsilon^{0}$  \\ &&&& \\ \hline
0.001 & $-5.122\times 10^{\text{-3}}$ & $\pm \;\;\;\;2.4\times 10^{\text{-6}}$ & $-1.232\times 10^{\text{-1}}$ & $\pm \;\;\;\;6.2\times 10^{\text{-5}}$\\
0.025 & $-3.933\times 10^{\text{-2}}$ & $\pm \;\;\;\;2.0\times 10^{\text{-5}}$ & $-5.229\times 10^{\text{-1}}$ & $\pm \;\;\;\;2.9\times 10^{\text{-4}}$\\
0.075 & $-6.148\times 10^{\text{-2}}$ & $\pm \;\;\;\;3.6\times 10^{\text{-5}}$ & $-5.894\times 10^{\text{-1}}$ & $\pm \;\;\;\;6.4\times 10^{\text{-4}}$\\
0.125 & $-6.810\times 10^{\text{-2}}$ & $\pm \;\;\;\;3.9\times 10^{\text{-5}}$ & $-5.340\times 10^{\text{-1}}$ & $\pm \;\;\;\;4.0\times 10^{\text{-4}}$\\
0.175 & $-6.802\times 10^{\text{-2}}$ & $\pm \;\;\;\;4.1\times 10^{\text{-5}}$ & $-4.529\times 10^{\text{-1}}$ & $\pm \;\;\;\;3.6\times 10^{\text{-4}}$\\
0.225 & $-6.408\times 10^{\text{-2}}$ & $\pm \;\;\;\;4.1\times 10^{\text{-5}}$ & $-3.670\times 10^{\text{-1}}$ & $\pm \;\;\;\;3.5\times 10^{\text{-4}}$\\
0.275 & $-5.768\times 10^{\text{-2}}$ & $\pm \;\;\;\;4.3\times 10^{\text{-5}}$ & $-2.848\times 10^{\text{-1}}$ & $\pm \;\;\;\;3.8\times 10^{\text{-4}}$\\
0.325 & $-4.992\times 10^{\text{-2}}$ & $\pm \;\;\;\;3.7\times 10^{\text{-5}}$ & $-2.112\times 10^{\text{-1}}$ & $\pm \;\;\;\;2.6\times 10^{\text{-4}}$\\
0.375 & $-4.117\times 10^{\text{-2}}$ & $\pm \;\;\;\;3.7\times 10^{\text{-5}}$ & $-1.453\times 10^{\text{-1}}$ & $\pm \;\;\;\;2.4\times 10^{\text{-4}}$\\
0.425 & $-3.225\times 10^{\text{-2}}$ & $\pm \;\;\;\;3.8\times 10^{\text{-5}}$ & $-9.050\times 10^{\text{-2}}$ & $\pm \;\;\;\;3.4\times 10^{\text{-4}}$\\
0.475 & $-2.344\times 10^{\text{-2}}$ & $\pm \;\;\;\;2.7\times 10^{\text{-5}}$ & $-4.580\times 10^{\text{-2}}$ & $\pm \;\;\;\;1.7\times 10^{\text{-4}}$\\
0.525 & $-1.523\times 10^{\text{-2}}$ & $\pm \;\;\;\;2.9\times 10^{\text{-5}}$ & $-1.203\times 10^{\text{-2}}$ & $\pm \;\;\;\;2.9\times 10^{\text{-4}}$\\
0.575 & $-7.844\times 10^{\text{-3}}$ & $\pm \;\;\;\;2.6\times 10^{\text{-5}}$ & $+1.230\times 10^{\text{-2}}$ & $\pm \;\;\;\;1.7\times 10^{\text{-4}}$\\
0.625 & $-1.715\times 10^{\text{-3}}$ & $\pm \;\;\;\;2.1\times 10^{\text{-5}}$ & $+2.634\times 10^{\text{-2}}$ & $\pm \;\;\;\;1.6\times 10^{\text{-4}}$\\
0.675 & $+2.967\times 10^{\text{-3}}$ & $\pm \;\;\;\;2.0\times 10^{\text{-5}}$ & $+3.260\times 10^{\text{-2}}$ & $\pm \;\;\;\;1.1\times 10^{\text{-4}}$\\
0.725 & $+5.959\times 10^{\text{-3}}$ & $\pm \;\;\;\;1.4\times 10^{\text{-5}}$ & $+3.152\times 10^{\text{-2}}$ & $\pm \;\;\;\;8.3\times 10^{\text{-5}}$\\
0.775 & $+7.193\times 10^{\text{-3}}$ & $\pm \;\;\;\;1.3\times 10^{\text{-5}}$ & $+2.558\times 10^{\text{-2}}$ & $\pm \;\;\;\;9.4\times 10^{\text{-5}}$\\
0.825 & $+6.576\times 10^{\text{-3}}$ & $\pm \;\;\;\;1.2\times 10^{\text{-5}}$ & $+1.680\times 10^{\text{-2}}$ & $\pm \;\;\;\;6.1\times 10^{\text{-5}}$\\
0.875 & $+4.164\times 10^{\text{-3}}$ & $\pm \;\;\;\;7.8\times 10^{\text{-6}}$ & $+8.379\times 10^{\text{-3}}$ & $\pm \;\;\;\;4.3\times 10^{\text{-5}}$\\
0.925 & $+3.782\times 10^{\text{-4}}$ & $\pm \;\;\;\;8.4\times 10^{\text{-6}}$ & $+3.761\times 10^{\text{-3}}$ & $\pm \;\;\;\;4.1\times 10^{\text{-4}}$\\
0.975 & $-3.322\times 10^{\text{-3}}$ & $\pm \;\;\;\;2.9\times 10^{\text{-6}}$ & $+9.508\times 10^{\text{-3}}$ & $\pm \;\;\;\;1.3\times 10^{\text{-5}}$\\
0.999 & $-1.074\times 10^{\text{-3}}$ & $\pm \;\;\;\;8.8\times 10^{\text{-7}}$ & $+7.280\times 10^{\text{-3}}$ & $\pm \;\;\;\;7.3\times 10^{\text{-6}}$\\
    \end{tabular}
  \end{center}
  \caption{\label{tab:qqttbb2} Coefficients of the Laurent expansion of the
    $f_{q\bar q \rightarrow t\bar tq'\bar q'}$ function.}
\end{table}

\begin{figure}[ht]
\begin{center}
\includegraphics[width=.50\textwidth]{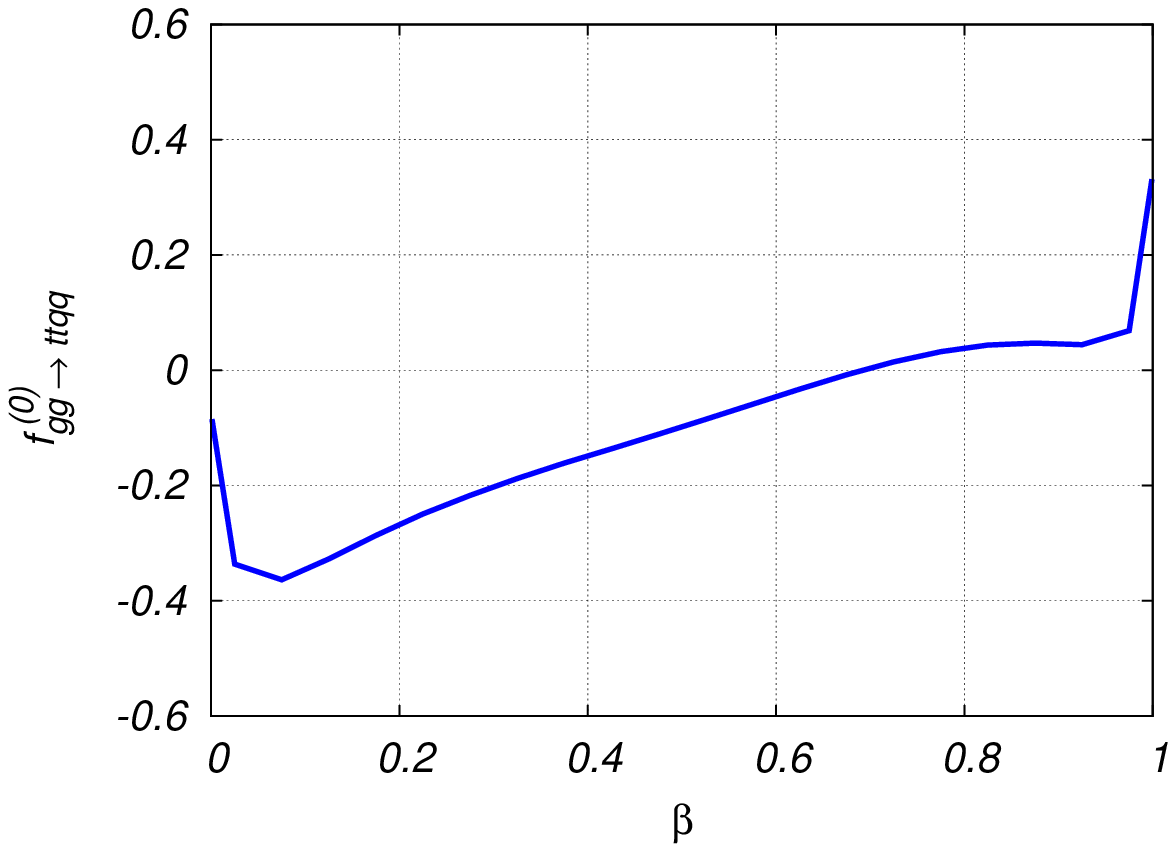}\includegraphics[width=.50\textwidth]{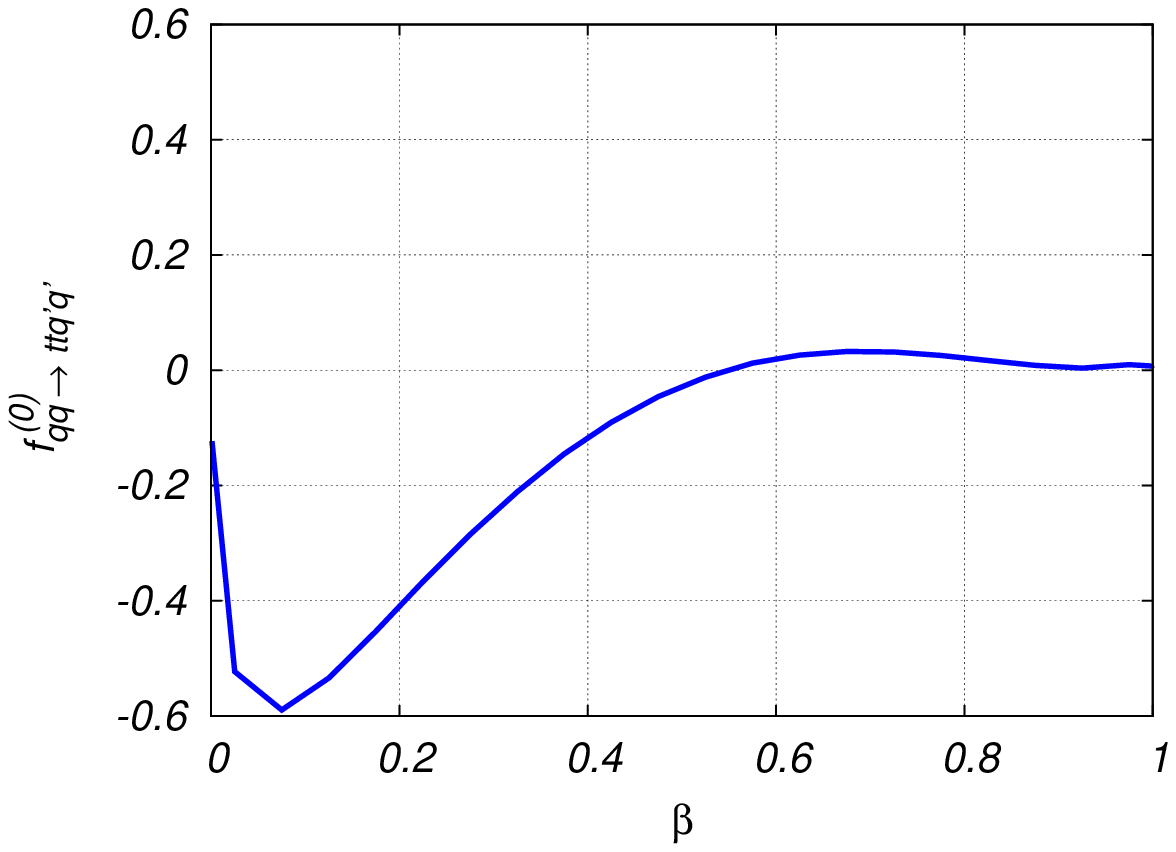}
\end{center}
\caption{Finite parts of $f_{gg \rightarrow t\bar tq\bar q}$ (left)
  and $f_{q\bar q \rightarrow t\bar tq'\bar q'}$ (right).}
\label{fig:remaining}
\end{figure}

The remaining two cross sections are much less interesting. They are
less singular both in $\epsilon$ and $\log\beta$, and are moreover
much smaller even after multiplication by the number of massless quark
species. We give the numbers in Tabs.~\ref{tab:ggttqq1},
\ref{tab:ggttqq2}, \ref{tab:qqttbb1} and \ref{tab:qqttbb2}, and show
the respective finite parts in Fig.~\ref{fig:remaining}.

\section{Conclusions}

The purpose of this work was to prove the usefulness of the
\textsc{Stripper} approach to the problem of double-real radiation. We
have considered the phenomenologically relevant case of top quark pair
production, and evaluated the cross sections for the dominant
channels. We have given most of the formulae needed for the
implementation, and demonstrated pointwise convergence and efficiency. The
immediate consequence is the possibility to evaluate the complete cross
sections for top quark pair production after inclusion of the
double-virtual, and real-virtual contributions. Although this requires
quite some effort, we do not see any conceptual problems, unlike in
the present case of double-real radiation.

We would like to point out that there are three directions of further
development. First, there are many technical improvements of our
implementation that can be studied. The most important are the analysis
of numerical instabilities and implementation of a more efficient
multi-precision library, the latter being almost trivial. Although not
absolutely necessary, this work is always part of software maturation
in the case of higher order calculations. The second direction
involves applications to similar process, in particular removing some
of the subtraction terms is sufficient to treat $e^+ e^- \rightarrow
t\bar t+X$, and $pp(\bar p) \rightarrow W^+W^- + X$ (and other gauge
boson final states) at NNLO. The last, and probably the most
interesting, direction is application of \textsc{Stripper} to final
states with massless particles, such as dijet production. This
requires the specification of the phase space in the case of initial
and final state singularities, but as noticed in \cite{Czakon:2010td}
involves the same treatment of the unresolved partons.


\section*{Acknowledgments}

\noindent

We would like to thank T. Hahn for help with \textsc{FormCalc} and
A. van Hameren for help with \textsc{Parni}.

This work was supported by the Heisenberg and by the Gottfried 
Wilhelm Leibniz programmes of the Deutsche Forschungsgemeinschaft, and
by the DFG Sonderforschungsbereich/Transregio 9 ``Computergest\"utzte
Theoretische Teilchenphysik''.


\appendix


\section{Collinear limits and splitting functions}
\label{sec:splitting}

In this appendix, we will reproduce the splitting functions that have
been used in the derivation of the subtraction terms. The formulae are
taken literally from \cite{Catani:1999ss} (see also
\cite{Campbell:1997hg, Catani:1998nv}). We start by defining the
notation for the matrix elements
\beq
\label{meldef}
\cm^{c_1,c_2,\dots;s_1,s_2,\dots}_{a_1,a_2,\dots}(p_1,p_2,\dots) \;\;,
\eeq
where the $s_i$ indices stand for spin, the $c_i$ for color, and the
$a_i$ for parton flavor. With this object, we define spin correlated
amplitudes squared
\beq
\label{melspindef}
{\cal T}_{a_1,\dots}^{s_1 s'_1}(p_1,\dots) \equiv 
\sum_{{\rm spins} \,\neq s_1,s'_1} \, \sum_{{\rm colors}}
\cm^{c_1,c_2,\dots;s_1,s_2,\dots}_{a_1,a_2,\dots}(p_1,p_2,\dots) \,
\left[ \cm^{c_1,c_2,\dots;s'_1,s_2,\dots}_{a_1,a_2,\dots}(p_1,p_2,\dots)
\right]^\dagger
\;\;.
\eeq

Having defined the matrix elements, we now turn to next-to-leading
order collinear limits of amplitudes. We first define the limits
through auxiliary vectors
\beeq
\label{clim}
&&p_1^\mu = z p^\mu + k_\perp^\mu - \frac{k_\perp^2}{z} 
\frac{n^\mu}{2 p\cdot n} \;\;, \;\;\; p_2^\mu =
(1-z) p^\mu - k_\perp^\mu - \frac{k_\perp^2}{1-z} \frac{n^\mu}{2 p\cdot n}\;\;,
\nonumber \\
&&s_{12} \equiv 2 p_1 \cdot p_2 = - \frac{k_\perp^2}{z(1-z)} \;\;,
\;\;\;\;\;\;\;\; k_\perp \to 0 \;\;,
\eeeq
where $p^2 = n^2 = p\cdot k_\perp = n\cdot k_\perp = 0$. Notice that
all vectors here and below are outgoing. The case we are interested
in, namely some of the vectors being in-going, is recovered by
crossing. In the above collinear limit, the matrix element factorizes
as follows
\beeq
\label{cfac}
| \cm_{a_1,a_2,\dots}(p_1,p_2,\dots) |^2 \simeq \frac{2}{s_{12}} \;
4 \pi \mu^{2\ep} \as 
\;{\cal T}_{a,\dots}^{s s'}(p,\dots) \;
{\hat P}_{a_1 a_2}^{s s'}(z,\kper;\ep) \;\;.
\eeeq
The splitting functions ${\hat P}_{a_1 a_2}^{s s'}$ depend on the
parton flavors. For the general case
\beq
\label{sppro}
a(p) \to a_1(zp + \kper + {\cal O}(\kper^2)) +
a_2((1-z) p - \kper + {\cal O}(\kper^2)) \;\;,
\eeq
they read
\beeq
\label{hpqqep}
{\hat P}_{qg}^{s s'}(z,\kper;\ep) = {\hat P}_{{\bar q}g}^{s s'}(z,\kper;\ep)
= \delta_{ss'} \;C_F
\;\left[ \frac{1 + z^2}{1-z} - \ep (1-z) \right] \;\;,
\eeeq
\beeq
\label{hpqgep}
{\hat P}_{gq}^{s s'}(z,\kper;\ep) = {\hat P}_{g{\bar q}}^{s s'}(z,\kper;\ep)
= \delta_{ss'} \;C_F
\;\left[ \frac{1 + (1-z)^2}{z} - \ep z \right] \;\;,
\eeeq
\beeq
\label{hpgqep}
{\hat P}_{q{\bar q}}^{\mu \nu}(z,\kper;\ep) 
= {\hat P}_{{\bar q}q}^{\mu \nu}(z,\kper;\ep)
= T_F
\left[ - g^{\mu \nu} + 4 z(1-z) \frac{\kper^{\mu} \kper^{\nu}}{\kper^2}
\right] \;\;,
\eeeq
\beq
\label{hpggep}
{\hat P}_{gg}^{\mu \nu}(z,\kper;\ep) = 2C_A
\;\left[ - g^{\mu \nu} \left( \frac{z}{1-z} + \frac{1-z}{z} \right)
- 2 (1-\ep) z(1-z) \frac{\kper^{\mu} \kper^{\nu}}{\kper^2}
\right] \;\;.
\eeq

Let us now turn to the more complicated case of triple-collinear
limits. Consider the set of three vectors
\beq
\label{kin3}
p_i^\mu = x_i p^\mu +k_{\perp i}^\mu - \frac{k_{\perp i}^2}{x_i} 
\frac{n^\mu}{2p \cdot n} \;, \;\;\;\;\;i=1,2,3 \;,
\eeq
where as before $p^2 = n^2 = p\cdot k_{\perp i} = n\cdot k_{\perp i} =
0$. This configuration fulfills no other constraints, but rather the
limits are expressed through derived variables
\beeq
\label{zvar}
z_i &=& \frac{x_i}{\sum_{j=1}^3 \,x_j} \;\;,\\
\label{kvar}
{\ktil}_i^\mu &=& k_{\perp i}^\mu - \frac{x_i}{\sum_{k=1}^3 \,x_k} \;
\sum_{j=1}^3 k_{\perp j}^\mu \;\;.
\eeeq
We also define
\beq
\label{tvar}
t_{ij,k} \equiv 2 \;\f{z_i s_{jk}-z_j s_{ik}}{z_i+z_j} +
\f{z_i-z_j}{z_i+z_j} \,s_{ij} \;\;,
\eeq
with $s_{ij} = (p_i+p_j)^2$.

The factorization formula is now
\beeq
\label{ccfacm}
| \cm_{a_1,a_2,a_3,\dots}(p_1,p_2,p_3,\dots) |^2 \simeq
\left( \frac{8 \pi \mu^{2\ep} \as}{s_{123}}\right)^{2}
 \;{\cal T}_{a,\dots}^{s s'}(xp,\dots) \;
{\hat P}_{a_1 a_2 a_3}^{s s'}
\;\;,
\eeeq
with $s_{123} = (p_1+p_2+p_3)^2$ and $x = x_1+x_2+x_3$.

The complete set of splitting functions is (in the case of spin
conservation, we give only the spin averaged splitting functions $\la
\Ph_{a_1 a_2 a_3} \ra$)
\beq
\label{qqqprimesf}
\la \Ph_{{\bar q}^\prime_1 q^\prime_2 q_3} \ra \, = \f{1}{2} \, 
C_F T_F \,\f{s_{123}}{s_{12}} \left[ - \f{t_{12,3}^2}{s_{12}s_{123}}
+\f{4z_3+(z_1-z_2)^2}{z_1+z_2} 
+ (1-2\ep) \left(z_1+z_2-\f{s_{12}}{s_{123}}\right)
\right] \;\;.
\eeq
Notice that we have omitted the case of identical quarks, which is not
needed in the present paper. The splitting functions for this case can
be found in \cite{Catani:1999ss}. The remaining functions are
\beq
\label{qggsf}
\la \Ph_{g_1 g_2 q_3} \ra \, =
C_F^2 \, \la \Ph_{g_1 g_2 q_3}^{({\rm ab})} \ra \,
+ \, C_F C_A \, \la \Ph_{g_1 g_2 q_3}^{({\rm nab})} \ra  \;\;,
\eeq
with
\beeq
\label{qggabsf}
\la \Ph_{g_1 g_2 q_3}^{({\rm ab})} \ra \, 
&=&\Biggl\{\f{s_{123}^2}{2s_{13}s_{23}}
z_3\left[\f{1+z_3^2}{z_1z_2}-\ep\f{z_1^2+z_2^2}{z_1z_2}-\ep(1+\ep)\right]\nn\\
&+&\f{s_{123}}{s_{13}}\Biggl[\f{z_3(1-z_1)+(1-z_2)^3}{z_1z_2}+\ep^2(1+z_3)
-\ep (z_1^2+z_1z_2+z_2^2)\f{1-z_2}{z_1z_2}\Biggr]\nn\\
&+&(1-\ep)\left[\ep-(1-\ep)\f{s_{23}}{s_{13}}\right]
\Biggr\}+(1\lra 2) \;\;,
\eeeq
\beeq
\label{qggnabsf}
\la \Ph_{g_1 g_2 q_3}^{({\rm nab})} \ra \,
&=&\Biggl\{(1-\ep)\left(\f{t_{12,3}^2}{4s_{12}^2}+\f{1}{4}
-\f{\ep}{2}\right)+\f{s_{123}^2}{2s_{12}s_{13}}
\Biggl[\f{(1-z_3)^2(1-\ep)+2z_3}{z_2}\nn\\
&+&\f{z_2^2(1-\ep)+2(1-z_2)}{1-z_3}\Biggr]
-\f{s_{123}^2}{4s_{13}s_{23}}z_3\Biggl[\f{(1-z_3)^2(1-\ep)+2z_3}{z_1z_2}
+\ep(1-\ep)\Biggr]\nn\\
&+&\f{s_{123}}{2s_{12}}\Biggl[(1-\ep)
\f{z_1(2-2z_1+z_1^2) - z_2(6 -6 z_2+ z_2^2)}{z_2(1-z_3)}
+2\ep\f{z_3(z_1-2z_2)-z_2}{z_2(1-z_3)}\Biggr]\nn\\
&+&\f{s_{123}}{2s_{13}}\Biggl[(1-\ep)\f{(1-z_2)^3
+z_3^2-z_2}{z_2(1-z_3)}
-\ep\left(\f{2(1-z_2)(z_2-z_3)}{z_2(1-z_3)}-z_1 + z_2\right)\nn\\
&-&\f{z_3(1-z_1)+(1-z_2)^3}{z_1z_2}
+\ep(1-z_2)\left(\f{z_1^2+z_2^2}{z_1z_2}-\ep\right)\Biggr]\Biggr\}
+(1\lra 2) \;\;.
\eeeq
Similarly
\beq
\label{gqqsf}
\Ph^{\mu\nu}_{g_1 q_2 {\bar q}_3}  \, =
C_F T_F \, \Ph_{g_1 q_2 {\bar q}_3}^{\mu\nu \,({\rm ab})} \,
+ \, C_A T_F\, \Ph_{g_1 q_2 {\bar q}_3}^{\mu\nu \,({\rm nab})}  \;\;,
\eeq
with
\beeq
\label{gqqabsf}
\Ph^{\mu\nu \,({\rm ab})}_{g_1q_2{\bar q}_3} &=&
-g^{\mu\nu}\Biggl[ -2 
+ \f{2 s_{123} s_{23} + (1-\ep) (s_{123} - s_{23})^2}{s_{12}s_{13}}\Biggr]\nn\\
&+& \f{4s_{123}}{s_{12}s_{13}}\left({\ktil}_{3}^\mu
    {\ktil}_{2}^\nu+{\ktil}_{\hs 2}^\mu
    {\ktil}_{3}^\nu-(1-\ep){\ktil}_{\hs 1}^\mu 
    {\ktil}_{1}^\nu \right)
\;\;,
\eeeq
\beeq
\label{gqqnabsf}
\Ph^{\mu\nu \,({\rm nab})}_{g_1q_2{\bar q}_3} &=& \f{1}{4}
\,\Biggl\{ \f{s_{123}}{s_{23}^2}
\Biggl[ g^{\mu\nu} \f{t_{23,1}^2}{s_{123}}-16\f{z_2^2z_3^2}{z_1(1-z_1)}
\left(\f{{\ktil}_2}{z_2}-\f{{\ktil}_3}{z_3}\right)^\mu
\left(\f{{\ktil}_2}{z_2}-\f{{\ktil}_3}{z_3}\right)^\nu \,\Biggr]\nn\\
&+& \f{s_{123}}{s_{12}s_{13}} \Biggl[ 2 s_{123} g^{\mu\nu}
- 4 ( {\ktil}_2^\mu {\ktil}_3^\nu + {\ktil}_3^\mu {\ktil}_2^\nu
- (1-\ep) {\ktil}_1^\mu {\ktil}_1^\nu ) \Biggr] \nn\\
&-& g^{\mu\nu} \Biggl[ - ( 1 -2 \ep) + 2\f{s_{123}}{s_{12}} 
\f{1-z_3}{z_1(1-z_1)} + 2\f{s_{123}}{s_{23}} 
\f{1-z_1 + 2 z_1^2}{z_1(1-z_1)}\Biggr]\nn\\
&+& \f{s_{123}}{s_{12}s_{23}} \Biggl[ - 2 s_{123} g^{\mu\nu}
\f{z_2(1-2z_1)}{z_1(1-z_1)} - 16 {\ktil}_3^\mu {\ktil}_3^\nu 
\f{z_2^2}{z_1(1-z_1)} 
+ 8(1-\ep) {\ktil}_2^\mu {\ktil}_2^\nu \nn\\
&+& 4 ({\ktil}_2^\mu {\ktil}_3^\nu  + {\ktil}_3^\mu {\ktil}_2^\nu )
\left(\f{2 z_2 (z_3-z_1)}{z_1(1-z_1)}+ (1-\ep) \right)
\Biggr] \Biggr\} + \left( 2 \leftrightarrow 3 \right)  \;\;.
\eeeq
Finally
\beeq
\label{gggsf}
\Ph^{\mu\nu}_{g_1g_2g_3} &=& C_A^2 
\,\Biggl\{\f{(1-\ep)}{4s_{12}^2}
\Biggl[-g^{\mu\nu} t_{12,3}^2+16s_{123}\f{z_1^2z_2^2}{z_3(1-z_3)}
\left(\f{{\ktil}_2}{z_2}-\f{{\ktil}_1}{z_1}\right)^\mu
\left(\f{{\ktil}_2}{z_2}-\f{{\ktil}_1}{z_1}\right)^\nu \;\Biggr]\nn\\
&-& \f{3}{4}(1-\ep)g^{\mu\nu}+\f{s_{123}}{s_{12}}g^{\mu\nu}\f{1}{z_3}
    \Biggl[\f{2(1-z_3)+4z_3^2}{1-z_3}-\f{1-2z_3(1-z_3)}{z_1(1-z_1)}\Biggr]\nn\\
&+& \f{s_{123}(1-\ep)}{s_{12}s_{13}}\Biggl[2z_1\left({\ktil}^\mu_2
    {\ktil}^\nu_2\hs\f{1-2z_3}{z_3(1-z_3)}+
    {\ktil}^\mu_3{\ktil}^\nu_3\hs
    \f{1-2z_2}{z_2(1-z_2)}\right)\nn\\
&+& \f{s_{123}}{2(1-\ep)} g^{\mu\nu}
    \left(\f{4z_2z_3+2z_1(1-z_1)-1}{(1-z_2)(1-z_3)}
    - \f{1-2z_1(1-z_1)}{z_2z_3}\right)\nn\\
&+& \left({\ktil}_2^\mu{\ktil}_3^\nu
   +{\ktil}_3^\mu{\ktil}_2^\nu\right)
    \left(\f{2z_2(1-z_2)}{z_3(1-z_3)}-3\right)\Biggr]\Biggr\}
    + (5\mbox{ permutations}) \;\;.
\eeeq


\section{Soft limits in the presence of massive partons}
\label{sec:soft}

\noindent
While Ref.~\cite{Catani:1999ss} contains a summary of the behavior of
QCD matrix elements in singular limits at next-to-next-to-leading
order, the authors restricted themselves to the case of massless
partons. Since massive partons do not induce collinear singularities,
we need only consider the soft limit. It is well known that the
eikonal current has the same form in both massless and massive
cases. This implies that as long as we describe strongly ordered
limits, no modification of the expressions is needed. Surprisingly,
one observes a difference in the double-soft limit, in which the
energies of both gluons (there is nothing special in the case of a
soft quark pair) vanish at the same rate. To be more specific, we shall
consider two gluons with momenta $q_1$ and $q_2$, which are rescaled
by a factor $\lambda$
\begin{equation}
q_1 \rightarrow \lambda q_1 \; , \;\;\;\; q_2 \rightarrow \lambda q_2
\; ,
\end{equation}
and we will study the limit $\lambda \rightarrow 0$. As explained in
\cite{Catani:1999ss}, the matrix element factorizes as follows
\begin{equation}
\langle a_1,a_2;\mu_1,\mu_2 |
{\cal M}_{g,g,c_1,...,c_n}(q_1,q_2,p_1,...,p_n) \rangle
\simeq g^2\mu^{2\epsilon} J_{\mu_1\mu_2}^{a_1a_2}(q_1,q_2) |
{\cal M}_{c_1,...,c_n}(p_1,...,p_n) \rangle \; ,
\end{equation}
where $g$ is the strong coupling constant, $\mu$ the dimension unit in
dimensional regularization (introduced through the explicit dependence
in the coupling constant), and the two-gluon soft current
$J_{\mu_1\mu_2}^{a_1a_2}(q_1,q_2)$ is given by
\begin{eqnarray}
\label{eq:cur1}
J^{\mu_1\mu_2}_{a_1a_2}(q_1,q_2) &=& \frac{1}{2} \{ J^{\mu_1}_{a_1}(q_1),
  J^{\mu_2}_{a_2}(q_2) \} + i f_{a_1a_2a_3} \sum_{i=1}^n T^{a_3}_i \left\{
  \frac{p_i^{\mu_1}q_1^{\mu_2} - p_i^{\mu_2}q_2^{\mu_1}}{(q_1 \cdot
    q_2)\left[ p_i \cdot (q_1+q_2) \right]} \right. \nonumber \\
 && -\left. \frac{p_i \cdot (q_1-q_2)}{2\left[ p_i \cdot (q_1+q_2) \right]}
  \left[ \frac{p_i^{\mu_1}p_i^{\mu_2}}{(p_i \cdot q_1)(p_i \cdot
    q_2)} +\frac{g^{\mu_1\mu_2}}{q_1 \cdot q_2} \right] \right\} \; ,
\end{eqnarray}
with the eikonal current defined as
\begin{equation}
\label{eq:eikon}
{\bf J}^\mu(q) = \sum_{i=1}^n {\bf T}_i \frac{p_i^\mu}{p_i \cdot
  q} \; .
\end{equation}
The algebra of the colour operators ${\bf T}_i$ has been discussed at
length in \cite{Catani:1996vz}.
The expression Eq.~(\ref{eq:cur1}) is as in \cite{Catani:1999ss} and
can be derived by taking into account all diagrams with the two soft
gluons attached to a hard parton line through an eikonal coupling. The
triple gluon vertex has to be treated exactly, since all momenta are
of the same order. The chosen class of diagrams is shown to be
sufficient by power counting in a physical gauge. Moreover, contraction with
physical polarization vectors has been used to eliminate terms
proportional to the soft gluon momentum.

The difference between massive and massless cases occurs, when
squaring the matrix element. The factorization formula then contains
the factor
\begin{equation}
\label{eq:cur2}
\left[ J_{\mu\nu}^{a_1a_2}(q_1,q_2) \right]^\dagger d^{\mu\sigma}(q_1)
  d^{\nu\rho}(q_2) J_{\sigma\rho}^{a_1a_2}(q_1,q_2) = \frac{1}{2}
  \left\{ {\bf J}^2(q_1) , {\bf J}^2(q_2) \right\}
  - C_A \sum_{i,j=1}^n {\bf T}_i \cdot {\bf T}_j \; {\cal
  S}_{ij}(q_1,q_2) + ... \; ,
\end{equation}
where $d^{\mu\nu}(q)$ is the polarization tensor obtained by summing
over gluon polarizations. Due to current conservation, we can make the
replacement $d^{\mu\nu}(q) \rightarrow -g^{\mu\nu}$. The terms vanishing
when acting on a physical matrix element are denoted by the dots at
the end of the above equation.

In order to recast what is essentially the square of the two-gluon
current in Eq.~(\ref{eq:cur1}) into the form of the right hand side of
Eq.~(\ref{eq:cur2}), some colour algebra is needed (rightfully called
``quite cumbersome'' by the authors of \cite{Catani:1999ss}). The
process is simplified substantially by the use of the following
two identities
\begin{eqnarray}
if^{a_1a_2a_3} \left[ \{ T^{a_1}_i,T^{a_2}_j \},T^{a_3}_k \right] &=&
2C_A \; {\bf T}_i \cdot {\bf T}_j (\delta_{ik}-\delta_{jk}) \; ;
\\ \nonumber \\
\left\{ \{ T^{a_1}_i,T^{a_2}_j \} , \{ T^{a_1}_k,T^{a_2}_l\} \right\} 
&+& \left\{ \{ T^{a_1}_i,T^{a_2}_l \} , \{ T^{a_1}_k,T^{a_2}_j\} \right\} 
\nonumber \\ \nonumber \\ &=& 8 \; \{ {\bf T}_i \cdot {\bf T}_k , {\bf
    T}_j \cdot {\bf T}_l \} + 2 C_A \Big[
  3\delta_{il}\delta_{jk} {\bf T}_i \cdot {\bf T}_j
+ 3\delta_{ij}\delta_{kl} {\bf T}_i \cdot {\bf T}_k
\nonumber \\ \nonumber \\ &&
- 2\delta_{ij}\delta_{jk} {\bf T}_i \cdot {\bf T}_l 
- 2\delta_{ij}\delta_{jl} {\bf T}_i \cdot {\bf T}_k
- 2(\delta_{ik}\delta_{kl}+\delta_{jk}\delta_{kl}) {\bf T}_i \cdot
{\bf T}_j \Big] \; . \nonumber \;
\end{eqnarray}
The result for the ${\cal S}_{ij}(q_1,q_2)$ function can be split into
two parts
\begin{equation}
\label{eq:soft}
{\cal S}_{ij}(q_1,q_2) = {\cal S}^{m=0}_{ij}(q_1,q_2) + \left( m_i^2
\; {\cal S}^{m \neq 0}_{ij}(q_1,q_2) + m_j^2 \; {\cal S}^{m \neq
  0}_{ji}(q_1,q_2) \right) \; ,
\end{equation}
where the first term has already been given in \cite{Catani:1999ss}
and reads
\begin{eqnarray}
\label{eq:sij0}
{\cal S}^{m=0}_{ij}(q_1,q_2) &=& \frac{(1-\epsilon)}{(q_1 \cdot q_2)^2}
\frac{p_i \cdot q_1 \; p_j \cdot q_2 + p_i \cdot q_2 \; p_j \cdot
  q_1}{p_i \cdot (q_1+q_2) \; p_j \cdot (q_1+q_2)} \nonumber
\\ \nonumber \\
&& - \frac{(p_i \cdot p_j)^2}{2 \; p_i \cdot q_1 \; p_j \cdot q_2 \; p_i
  \cdot q_2 \; p_j \cdot q_1} \left[ 2 - \frac{p_i \cdot q_1 \; p_j
    \cdot q_2 + p_i \cdot q_2 \; p_j \cdot q_1}{p_i \cdot (q_1+q_2) \;
    p_j \cdot (q_1+q_2)} \right] \nonumber \\  \nonumber \\
&& + \frac{p_i \cdot p_j}{2 \; q_1 \cdot q_2} \left[ \frac{2}{p_i
    \cdot q_1 \; p_j \cdot q_2} + \frac{2}{p_j \cdot q_1 \; p_i \cdot
    q_2} - \frac{1}{p_i \cdot (q_1+q_2) \; p_j \cdot (q_1+q_2)}
  \right. \nonumber \\ \nonumber \\
&& \times \left. \left( 4 + \frac{(p_i \cdot q_1 \; p_j \cdot q_2 +
    p_i \cdot q_2 \; p_j \cdot q_1)^2}{\; p_i \cdot q_1 \; p_j \cdot
    q_2 \; p_i \cdot q_2 \; p_j \cdot q_1} \right) \right] \; .
\end{eqnarray}
The second contribution in Eq.~(\ref{eq:soft}) is new and represents
the additional terms generated by non-vanishing masses. The relevant
function is
\begin{eqnarray}
\label{eq:sijm}
{\cal S}^{m\neq0}_{ij}(q_1,q_2) &=&  - \frac{1}{4  \; q_1 \cdot q_2 \; p_i \cdot
q_1 \; p_i \cdot q_2} + \frac{p_i \cdot p_j \; p_j
\cdot (q_1+q_2)}{2 \; p_i \cdot q_1 \; p_j \cdot q_2 \; p_i \cdot q_2 \;
  p_j \cdot q_1 \; p_i \cdot (q_1+q_2)} \nonumber \\ \nonumber \\
&& - \frac{1}{2 \; q_1 \cdot q_2 \; p_i \cdot (q_1+q_2) \; p_j \cdot
  (q_1+q_2) } \left( \frac{(p_j \cdot q_1)^2}{p_i \cdot q_1 \; p_j
  \cdot q_2} + \frac{(p_j \cdot q_2)^2}{p_i \cdot q_2 \; p_j \cdot
  q_1} \right) \; . \nonumber \\
\end{eqnarray}
%


\section{Collinear behavior in the double-soft limit}
\label{sec:coll2soft}

\noindent
Due to the particular phase space decomposition introduced in
\cite{Czakon:2010td}, the singular matrix element limits that need
to be considered in the construction of the subtraction terms, are
covered directly by the formulae from  \cite{Catani:1999ss} (aside
from the modification given in the previous appendix for the case of
massive partons). Nevertheless, one case turns out to be slightly
inconvenient. Indeed, the double-soft limit followed by the collinear
limit of the two soft partons, although obtainable with the formulae
of \ref{sec:soft}, requires a careful evaluation, because of
the presence of an apparent quadratic divergence $\sim 1/(q_1\cdot
q_2)^2$ in Eq.~(\ref{eq:sij0}). Of course, the actual leading 
divergence is only logarithmic as can be checked using colour
conservation. To avoid unnecessary complications, we propose 
to use an iterated limit in which the partons become collinear first,
and then produce a soft gluon, which interacts via the usual eikonal
current. This is justified by colour coherence of soft emission in the
collinear limit, which is usually exploited to derive the
soft-collinear limit in which a pair of partons become collinear, and a
gluon, not belonging to the pair, becomes soft. The result for our case
can be written as follows 
\begin{eqnarray}
|{\cal M}_{c_1,c_2,a_1,...,a_n}(q_1,q_2,p_1,...,p_n)|^2 &\simeq&
\frac{2}{s_{12}} g^4\mu^{4\epsilon} \langle
{\cal M}_{a_1,...,a_n}(p_1,...,p_n) |  \\ &\times& {\bf
  J}^\dagger_\mu(q_1+q_2) \hat{P}_{c_1c_2}^{\mu\nu}(z,k_\perp;\epsilon)
{\bf J}_\nu(q_1+q_2) | {\cal M}_{a_1,...,a_n}(p_1,...,p_n) \rangle \;
, \nonumber
\end{eqnarray}
where $c_1c_2=q\bar q$ or $gg$, $s_{12}=(q_1+q_2)^2$, ${\bf J}_\mu$ is
the eikonal current defined in Eq.~(\ref{eq:eikon}), and
$\hat{P}_{c_1c_2}^{\mu\nu}(z,k_\perp;\epsilon)$ is the $d$-dimensional
polarized Altarelli-Parisi splitting function given in
Eqs.~(\ref{hpqqep}, \ref{hpqgep}, \ref{hpgqep}, \ref{hpggep}).

The factorization formula demonstrates the usual spin correlations,
which are transferred here to the eikonal currents and not directly to
the matrix element, since the nearly on-shell gluon is fully
described by the current in the soft limit. One might wonder why the
spin correlations survive the soft limit, since they do not at the
next-to-leading order. The reason is that the double-soft limit cannot
be defined with the momenta of the collinear pair alone, because the
splitting functions depend only on the ratio of the energies of the
two partons. Therefore, as long as this ratio remains constant, from
the point of view of the collinear limit, we are considering two {\it
  hard} partons.


\section{Born level cross sections for top quark pair production}
\label{sec:born}

Although text book material, we reproduce these cross sections here
for convenience of the reader. We have
\begin{equation}
\sigma^{B}_{q\bar q \rightarrow t\bar t}(s,m^2,\alpha_s) =
\frac{\alpha_s^2}{m^2} f^{(0)}_{q\bar q \rightarrow t\bar t}(\beta) \;
, \;\;\;\; \sigma^{B}_{gg \rightarrow t\bar t}(s,m^2,\alpha_s) =
\frac{\alpha_s^2}{m^2} f^{(0)}_{gg \rightarrow t\bar t}(\beta) \; ,
\end{equation}
with
\begin{eqnarray}
f_{q{\bar q} \rightarrow t\bar t}^{(0)}(\beta) &=& \frac{\pi}{6}
\;\frac{T_F C_F}{N} \; \beta \rho \;( 2 + \rho) \; , \\ f_{gg
  \rightarrow t\bar t}^{(0)}(\beta) &=& \frac{\pi}{12}
\;\frac{T_F}{N^2 -1} \; \beta \rho  \left\{ 3 C_F \left[ ( 4 + 4 \rho
  - 2 \rho^2) \; \frac{1}{\beta} \; \ln \frac{1+\beta}{1-\beta} - 4 -
  4 \rho \right] \right. \nonumber \\ &+& C_A \left. \left[ 3 \rho^2
  \;\frac{1}{\beta} \;\ln \frac{1+\beta}{1-\beta} - 4 - 5 \rho \right]
\right\} \; ,
\end{eqnarray}
and $\rho = 1-\beta^2$.


\section{Software}
\label{sec:software}

The results obtained for the present publication have required the use
of numerous software systems. We list them here

\begin{itemize}

\item{\textsc{DiaGen/IdSolver}}, our own private system for diagram
  generation, analysis and evaluation, has been used for the generation
  of the cut diagrams and reduction of the integrals needed to compute
  the volume of the phase space;

\item{\textsc{Fermat} \cite{fermat}}, an algebra system, is the rational
  function algebra library of \textsc{DiaGen/IdSolver}, and has been
  used in the reduction of the phase space integrals;

\item{\textsc{Form} \cite{Vermaseren:2000nd}}, has been used for the
  algebraic simplification of the diagrams, mostly Dirac algebra and
  color factor evaluation, for which the package \textsc{Color.h} has
  proven useful;

\item{\textsc{FormCalc} \cite{Hahn:1998yk}}, the backbone of
  \textsc{FeynArts} \cite{Hahn:2000kx}, has been used for the low
  level formatting of \textsc{Fortran} code generated by
  \textsc{Mathematica};

\item{\textsc{Helac/Phegas} \cite{Kanaki:2000ey, Cafarella:2007pc,
    Papadopoulos:2000tt}}, has been used for tests of the 
  matrix elements at specified phase space points and checks of the
  numerical integration routines;

\item{\textsc{Mathematica}}, has been used for the derivation of the
  subtraction and integrated subtraction terms, convergence tests with
  very high numerical precision, and generation of \textsc{Fortran}
  code;

\item{\textsc{Intel Fortran Compiler}}, although not essential for this
  project, we have used its quadruple precision functionality to spare
  some minor effort in implementing interfaces to external libraries
  (see comments in Section~\ref{sec:implementation});

\item{\textsc{Parni} \cite{vanHameren:2007pt}} adaptive Monte Carlo
  random number generation optimizer, has been used in the numerical
  integration routines;

\item{\textsc{Ranlux} \cite{Luscher:1993dy, James:1993vv}}, the
  classic random number generator.

\end{itemize}



\section*{References}

\begin{thebibliography}{00}

\bibitem{Czakon:2010td}
  M.~Czakon,
  Phys.\ Lett.\  B {\bf 693} (2010) 259;

\bibitem{Catani:1996vz}
  S.~Catani and M.~H.~Seymour,
  Nucl.\ Phys.\  B {\bf 485} (1997) 291
  [Erratum-ibid.\  B {\bf 510} (1998) 503];

\bibitem{Catani:2002hc}
  S.~Catani, S.~Dittmaier, M.~H.~Seymour and Z.~Trocsanyi,
  Nucl.\ Phys.\  B {\bf 627} (2002) 189;

\bibitem{Frixione:1995ms}
  S.~Frixione, Z.~Kunszt and A.~Signer,
  Nucl.\ Phys.\  B {\bf 467} (1996) 399;

\bibitem{Frederix:2009yq}
  R.~Frederix, S.~Frixione, F.~Maltoni and T.~Stelzer,
  JHEP {\bf 0910} (2009) 003;

\bibitem{Nagy:1996bz}
  Z.~Nagy and Z.~Trocsanyi,
  Nucl.\ Phys.\  B {\bf 486} (1997) 189;

\bibitem{Binoth:2000ps}
  T.~Binoth and G.~Heinrich,
  Nucl.\ Phys.\  B {\bf 585} (2000) 741;

\bibitem{Anastasiou:2003gr}
  C.~Anastasiou, K.~Melnikov and F.~Petriello,
  Phys.\ Rev.\  D {\bf 69} (2004) 076010;

\bibitem{Binoth:2004jv}
  T.~Binoth and G.~Heinrich,
  Nucl.\ Phys.\  B {\bf 693} (2004) 134;

\bibitem{GehrmannDeRidder:2005cm}
  A.~Gehrmann-De Ridder, T.~Gehrmann and E.~W.~N.~Glover,
  JHEP {\bf 0509} (2005) 056;

\bibitem{Daleo:2009yj}
  A.~Daleo, A.~Gehrmann-De Ridder, T.~Gehrmann and G.~Luisoni,
  JHEP {\bf 1001} (2010) 118;

\bibitem{Glover:2010im}
  E.~W.~Nigel Glover and J.~Pires,
  JHEP {\bf 1006} (2010) 096;

\bibitem{Boughezal:2010mc}
  R.~Boughezal, A.~D.~Ridder and M.~Ritzmann,
  arXiv:1011.6631 [hep-ph];

\bibitem{Catani:2007vq}
  S.~Catani and M.~Grazzini,
  Phys.\ Rev.\ Lett.\  {\bf 98} (2007) 222002;

\bibitem{Anastasiou:2010pw}
  C.~Anastasiou, F.~Herzog and A.~Lazopoulos,
  arXiv:1011.4867 [hep-ph];

\bibitem{Somogyi:2005xz}
  G.~Somogyi, Z.~Trocsanyi and V.~Del Duca,
  JHEP {\bf 0506} (2005) 024;

\bibitem{Somogyi:2006cz}
  G.~Somogyi and Z.~Trocsanyi,
  arXiv:hep-ph/0609041;

\bibitem{Somogyi:2006db}
  G.~Somogyi and Z.~Trocsanyi,
  JHEP {\bf 0701} (2007) 052;

\bibitem{Somogyi:2006da}
  G.~Somogyi, Z.~Trocsanyi and V.~Del Duca,
  JHEP {\bf 0701} (2007) 070;

\bibitem{Bolzoni:2010bt}
  P.~Bolzoni, G.~Somogyi and Z.~Trocsanyi,
  arXiv:1011.1909 [hep-ph];

\bibitem{Weinzierl:2003fx}
  S.~Weinzierl,
  JHEP {\bf 0303} (2003) 062;

\bibitem{Kilgore:2004ty}
  W.~B.~Kilgore,
  Phys.\ Rev.\  D {\bf 70} (2004) 031501;

\bibitem{Frixione:2004is}
  S.~Frixione and M.~Grazzini,
  JHEP {\bf 0506} (2005) 010;

\bibitem{Denner:2010jp}
  A.~Denner, S.~Dittmaier, S.~Kallweit and S.~Pozzorini,
  arXiv:1012.3975 [hep-ph];

\bibitem{Bevilacqua:2010qb}
  G.~Bevilacqua, M.~Czakon, A.~van Hameren, C.~G.~Papadopoulos and M.~Worek,
  arXiv:1012.4230 [hep-ph];

\bibitem{Bernreuther:2004jv}
  W.~Bernreuther, A.~Brandenburg, Z.~G.~Si and P.~Uwer,
  Nucl.\ Phys.\  B {\bf 690} (2004) 81;

\bibitem{Melnikov:2009dn}
  K.~Melnikov and M.~Schulze,
  JHEP {\bf 0908} (2009) 049;

\bibitem{Beneke:2009ye}
  M.~Beneke, M.~Czakon, P.~Falgari, A.~Mitov and C.~Schwinn,
  Phys.\ Lett.\  B {\bf 690} (2010) 483;

\bibitem{Ahrens:2009uz}
  V.~Ahrens, A.~Ferroglia, M.~Neubert, B.~D.~Pecjak and L.~L.~Yang,
  Phys.\ Lett.\  B {\bf 687} (2010) 331;

\bibitem{Beneke:2009rj}
  M.~Beneke, P.~Falgari and C.~Schwinn,
  Nucl.\ Phys.\  B {\bf 828} (2010) 69;

\bibitem{Czakon:2009zw}
  M.~Czakon, A.~Mitov and G.~F.~Sterman,
  Phys.\ Rev.\  D {\bf 80} (2009) 074017;

\bibitem{Moch:2008qy}
  S.~Moch and P.~Uwer,
  Phys.\ Rev.\  D {\bf 78} (2008) 034003;

\bibitem{Ahrens:2010zv}
  V.~Ahrens, A.~Ferroglia, M.~Neubert, B.~D.~Pecjak and L.~L.~Yang,
  JHEP {\bf 1009} (2010) 097;

\bibitem{Kidonakis:2010dk}
  N.~Kidonakis,
  arXiv:1009.4935 [hep-ph];

\bibitem{Beneke:2010da}
  M.~Beneke, P.~Falgari and C.~Schwinn,
  Nucl.\ Phys.\ B {\bf 842} (2011) 414;

\bibitem{Czakon:2007ej}
  M.~Czakon, A.~Mitov and S.~Moch,
  Phys.\ Lett.\  B {\bf 651} (2007) 147;

\bibitem{Czakon:2007wk}
  M.~Czakon, A.~Mitov and S.~Moch,
  Nucl.\ Phys.\  B {\bf 798} (2008) 210;

\bibitem{Bonciani:2009nb}
  R.~Bonciani, A.~Ferroglia, T.~Gehrmann and C.~Studerus,
  JHEP {\bf 0908} (2009) 067;

\bibitem{Bonciani:2010mn}
  R.~Bonciani, A.~Ferroglia, T.~Gehrmann, A.~von Manteuffel and C.~Studerus,
  arXiv:1011.6661 [hep-ph];

\bibitem{Bonciani:2008az}
  R.~Bonciani, A.~Ferroglia, T.~Gehrmann, D.~Maitre and C.~Studerus,
  JHEP {\bf 0807} (2008) 129;

\bibitem{Czakon:2008zk}
  M.~Czakon,
  Phys.\ Lett.\  B {\bf 664} (2008) 307;

\bibitem{Korner:2008bn}
  J.~G.~Korner, Z.~Merebashvili and M.~Rogal,
  Phys.\ Rev.\  D {\bf 77} (2008) 094011;

\bibitem{Kniehl:2008fd}
  B.~Kniehl, Z.~Merebashvili, J.~G.~Korner and M.~Rogal,
  Phys.\ Rev.\  D {\bf 78} (2008) 094013;

\bibitem{Dittmaier:2007wz}
  S.~Dittmaier, P.~Uwer and S.~Weinzierl,
  Phys.\ Rev.\ Lett.\  {\bf 98} (2007) 262002;

\bibitem{Bevilacqua:2010ve}
  G.~Bevilacqua, M.~Czakon, C.~G.~Papadopoulos and M.~Worek,
  Phys.\ Rev.\ Lett.\  {\bf 104} (2010) 162002;

\bibitem{Melnikov:2010iu}
  K.~Melnikov and M.~Schulze,
  Nucl.\ Phys.\  B {\bf 840} (2010) 129;

\bibitem{Bern:1994zx}
  Z.~Bern, L.~J.~Dixon, D.~C.~Dunbar and D.~A.~Kosower,
  Nucl.\ Phys.\  B {\bf 425} (1994) 217;

\bibitem{Bern:1998sc}
  Z.~Bern, V.~Del Duca and C.~R.~Schmidt,
  Phys.\ Lett.\  B {\bf 445} (1998) 168;

\bibitem{Kosower:1999rx}
  D.~A.~Kosower and P.~Uwer,
  Nucl.\ Phys.\  B {\bf 563} (1999) 477;

\bibitem{Bern:1999ry}
  Z.~Bern, V.~Del Duca, W.~B.~Kilgore and C.~R.~Schmidt,
  Phys.\ Rev.\  D {\bf 60} (1999) 116001;

\bibitem{Catani:2000pi}
  S.~Catani and M.~Grazzini,
  Nucl.\ Phys.\  B {\bf 591} (2000) 435;

\bibitem{Czakon:2008ii}
  M.~Czakon and A.~Mitov,
  Nucl.\ Phys.\  B {\bf 824} (2010) 111;

\bibitem{GehrmannDe Ridder:2003bm}
  A.~Gehrmann-De Ridder, T.~Gehrmann and G.~Heinrich,
  Nucl.\ Phys.\  B {\bf 682} (2004) 265;

\bibitem{Kotikov:1990kg}
  A.~V.~Kotikov,
  Phys.\ Lett.\  B {\bf 254} (1991) 158;

\bibitem{Remiddi:1997ny}
  E.~Remiddi,
  Nuovo Cim.\  A {\bf 110} (1997) 1435;

\bibitem{Remiddi:1999ew}
  E.~Remiddi and J.~A.~M.~Vermaseren,
  Int.\ J.\ Mod.\ Phys.\  A {\bf 15} (2000) 725;

\bibitem{Catani:1999ss}
  S.~Catani and M.~Grazzini,
  Nucl.\ Phys.\  B {\bf 570} (2000) 287;

\bibitem{Ferroglia:2009ii}
  A.~Ferroglia, M.~Neubert, B.~D.~Pecjak and L.~L.~Yang,
  JHEP {\bf 0911} (2009) 062;

\bibitem{Catani:2000ef}
  S.~Catani, S.~Dittmaier and Z.~Trocsanyi,
  Phys.\ Lett.\  B {\bf 500} (2001) 149;

\bibitem{Ellis:1985er}
  R.~K.~Ellis and J.~C.~Sexton,
  Nucl.\ Phys.\  B {\bf 269} (1986) 445;

\bibitem{Kanaki:2000ey}
  A.~Kanaki and C.~G.~Papadopoulos,
  Comput.\ Phys.\ Commun.\  {\bf 132} (2000) 306;

\bibitem{Cafarella:2007pc}
  A.~Cafarella, C.~G.~Papadopoulos and M.~Worek,
  Comput.\ Phys.\ Commun.\  {\bf 180} (2009) 1941;

\bibitem{Czakon:2009ss}
  M.~Czakon, C.~G.~Papadopoulos and M.~Worek,
  JHEP {\bf 0908} (2009) 085;

\bibitem{vanHameren:2007pt}
  A.~van Hameren,
  Acta Phys.\ Polon.\  B {\bf 40}, 259 (2009);

\bibitem{Nason:1987xz}
  P.~Nason, S.~Dawson and R.~K.~Ellis,
  Nucl.\ Phys.\  B {\bf 303}, 607 (1988);

\bibitem{Campbell:1997hg}
  J.~M.~Campbell and E.~W.~N.~Glover,
  Nucl.\ Phys.\  B {\bf 527} (1998) 264;

\bibitem{Catani:1998nv}
  S.~Catani and M.~Grazzini,
  Phys.\ Lett.\  B {\bf 446} (1999) 143;

\bibitem{fermat}
  R.~H.~Lewis, Fermat, http://www.bway.net/\verb!~!lewis/;

\bibitem{Vermaseren:2000nd}
  J.~A.~M.~Vermaseren,
  arXiv:math-ph/0010025;

\bibitem{Hahn:1998yk}
  T.~Hahn and M.~Perez-Victoria,
  Comput.\ Phys.\ Commun.\  {\bf 118} (1999) 153;

\bibitem{Hahn:2000kx}
  T.~Hahn,
  Comput.\ Phys.\ Commun.\  {\bf 140} (2001) 418;

\bibitem{Papadopoulos:2000tt}
  C.~G.~Papadopoulos,
  Comput.\ Phys.\ Commun.\  {\bf 137} (2001) 247;

\bibitem{Luscher:1993dy}
  M.~Luscher,
  Comput.\ Phys.\ Commun.\  {\bf 79} (1994) 100;

\bibitem{James:1993vv}
  F.~James,
  Comput.\ Phys.\ Commun.\  {\bf 79} (1994) 111
  [Erratum-ibid.\  {\bf 97} (1996) 357].

\end{thebibliography}
\end{document}